\documentclass[fleqn,usenatbib]{mnras}

\usepackage{newtxtext,newtxmath}

\usepackage[T1]{fontenc}
\usepackage{threeparttable}
\DeclareRobustCommand{\VAN}[3]{#2}
\let\VANthebibliography\thebibliography
\def\thebibliography{\DeclareRobustCommand{\VAN}[3]{##3}\VANthebibliography}

\usepackage{graphicx}	
\usepackage{amsmath}	
\usepackage{pdflscape}

\title[High-$z$ radio quasars in the eRASS:1 ]{The most extreme high-$z$ radio quasars in the eRASS:1 X-ray survey}

\author[L. Ighina]{
L. Ighina,$^{1,2}$\thanks{E-mail: luca.ighina@cfa.harvard.edu}
L. Palmieri,$^{2,3}$
L. N. Martínez-Ramírez,$^{4}$
A. Caccianiga,$^{2}$
T. Connor,$^{1}$
A. Moretti,$^{2}$
\newauthor
B. Arsioli,$^{5,6}$
Y. Beletsky,$^{7}$
E. Marini,$^{8}$
A. Rossi$^{9}$
\\
$^{1}$Center for Astrophysics | Harvard \& Smithsonian, 60 Garden St., Cambridge, MA 02138, USA\\
$^{2}$INAF, Osservatorio Astronomico di Brera, via Brera 28, 20121, Milano, Italy \\
$^{3}$Dipartimento di Fisica e Astronomia, Alma Mater Studiorum, Università degli Studi di Bologna, Via Gobetti 93/2, 40129 Bologna, Italy\\
$^{4}$Hamburger Sternwarte, Universität Hamburg, Gojenbergsweg 112, D-21029 Hamburg, Germany\\
$^{5}$Instituto de Astrofísica e Ciências do Espaço, Universidade de Lisboa, OAL, Tapada da Ajuda, Lisboa, Portugal\\
$^{6}$Departamento de Física, Faculdade de Ciências, Universidade de Lisboa, Lisbon, Portugal\\
$^{7}$Carnegie Science Observatories, 813 Santa Barbara Street, Pasadena, CA 91101, USA\\
$^{8}$INAF, Observatory of Rome, Via Frascati 33, 00078, Monte Porzio Catone, RM, Italy\\
$^{9}$INAF - Osservatorio di Astrofisica e Scienza dello Spazio, Via Piero Gobetti 93/3, 40129, Bologna, Italy
}
\date{Accepted XXX. Received YYY; in original form ZZZ}

\pubyear{2026}

\begin{document}
\label{firstpage}
\pagerange{\pageref{firstpage}--\pageref{lastpage}}
\maketitle

\begin{abstract}
We present the selection and identification of high-$z$ radio quasars from a combination of X-ray (eROSITA All-Sky Survey), optical/NIR (DECam Local Volume Exploration), and radio (Rapid ASKAP Continuum Survey) surveys. After building a sample of 68 new high-$z$ quasar candidates, we performed follow-up spectroscopic observations on 46 sources, confirming 39 to have redshifts $z>3.5$. For the subset of radio quasars at $z>4$, the 14 new sources identified here represent an increase of $\sim65$\% in the known population of objects above the same flux limits in the same area.
Using X-ray-to-optical relative intensity, we identified blazars in our sample, which we then used to compare the total number of $z>4$, X-ray selected blazars to the predictions of a fractional IC/CMB model previously proposed in the literature. Overall we find a relatively good agreement, even considering the potential biases in the redshift distribution from the incompleteness of the spectroscopic follow-up and the blazar classification criteria. The sources presented in this work represent a sample well suited for investigating the evolution of the most extreme systems in the early Universe, including their relativistic jets and accretion properties.

\end{abstract}




\section{Introduction}

Blazars are a subset of jetted quasars and active galactic nuclei (AGN) defined such that the axis of their relativistic jets is oriented close to our line of sight ($\theta_{\rm view}\lesssim1/\Gamma$, where $\Gamma$ is the bulk Lorentz factor of the jet). Due to their orientation, the emission produced within the jets is relativistically boosted, amplifying these sources' radio, X-ray and $\gamma$-ray emission by orders of magnitudes \cite[e.g.][]{Urry1995}. At the same time, every blazar detection implies the existence of hundreds of similar sources at the same redshift, but with jets oriented in different directions. In particular, in a volume complete sample, we expect the total number of such sources to be $N_{\rm tot} \approx N_{\rm blazars} \times 2\Gamma^2$, with typical values of $\Gamma\sim5-10$ assumed \citep[e.g.][]{Ghisellini2014,Spingola2020}.
Combined, these two features make blazars unique tools to perform detailed studies on relativistic jet emission \citep[e.g.][]{An2020,Ighina2022a} and statistical studies on the jetted AGN population \citep[e.g.][]{Ghisellini2013,Diana2022}, even at high redshift.

Several recent studies found a redshift evolution of the X-ray properties of very radio-powerful (i.e. likely blazars) quasars, with higher redshift sources being X-ray brighter compared to their lower redshift counterparts with similar optical \citep[e.g.][]{Zhu2019} and radio \citep[e.g.][]{Ighina2019} properties. One of the most natural explanations for the excess of high-energy radiation is the Inverse Compton interaction of the Cosmic Microwave Background (CMB) photons with the jetted electrons (IC/CMB; \citealt{Wu2013}). Even if only a small fraction of the entire X-ray emission in blazars is produced through IC/CMB at low redshift ($\lesssim0.2\%$ at $z\lesssim1$; \citealt{Ighina2021b}), this fraction will increase at higher redshift following the CMB energy density evolution, which increases in proportion to $(1+z)^4$ \citep[e.g.][]{Schwartz2019}. Such scaling means that the overall X-ray luminosity could potentially increase by a factor $\sim2$ at $z\sim4$ \citep[e.g.][]{Zuo2024}.

Most of these studies are based on follow-up X-ray observations of already known $z>4$ quasars. For such studies, it is difficult to disentangle observational bias from population evolution, and to constrain how the potential IC/CMB enhancement affects the observed number density of blazars in the X-rays.
Furthermore, when it comes to statistical studies focused on the luminosity function and space density evolution of blazars, the analysis is limited by the scarce number of known objects at high redshift. For example, only one $z>4$ blazar is detected by the SWIFT-BAT telescope \citep{Oh2018}, which was used by both \cite{Ighina2021b} and \cite{Marcotulli2022} to constrain the space density evolution of this population across redshifts.

The recent release of the eROSITA All-Sky Survey (eRASS; \citealt{Merloni2024}) offers the perfect opportunity to advance past these issues and to build a statistical sample of X-ray selected blazars at $z>4$. eRASS provides both a large area of coverage (1/2 the sky in the German release) and relatively deep sensitivity, with sources detected up to $z\sim6$ \citep{Khorunzhev2021,Medvedev2020}. Unfortunately, most of the known $z>4$ blazars detected in the first data release of the eRASS (eRASS:1) belong to a wide variety of samples (see e.g. \citealt{Haemmerich2025,Sbarrato2026}) with significantly different selection functions, making them unsuited for statistical studies. While eRASS:1 has the potential to enable a search for high-redshift blazars, a dedicated multi-wavelength search is needed to reliably identify and confirm high-$z$ sources. 

In this work, we present the selection of $z>4$ blazar candidates through the combination of eRASS:1 with optical and radio wide-area surveys, along with spectroscopic observations of photometric candidates. 
In Sec. \ref{sec:selection}, we describe the selection criteria we used to uncover high-$z$ jetted quasars; in Sec. \ref{sec:spectroscopy} we describe spectroscopic observations for a subset of the final candidate list; in Sec. \ref{sec:comparison} we compare the properties of the selected high-$z$ quasars to the broader quasar population; in Sec. \ref{sec:IC/CMB} we identify the blazars within the sample and use them to test previous models describing their redshift evolution; and in Sec. \ref{sec:conclusions} we summarise our results. \\
Throughout the paper we assume a flat $\Lambda$ cold dark matter cosmology with $H_{0}$=70 km s$^{-1}$ Mpc$^{-1}$, $\Omega_m$=0.3, and $\Omega_{\Lambda}$=0.7. Spectral indices are given assuming $S_{\nu}\propto \nu^{-\alpha}$, and all uncertainties are reported at 68\% confidence unless otherwise specified.

\section{Sample selection}
\label{sec:selection}
In order to efficiently select high-$z$ blazars, we combined multi-wavelength datasets together. In this section we discuss the surveys and the criteria that we adopted in the selection of radio and X-ray bright quasars at high redshift.

\subsection{eRASS and RACS cross-match}
The starting point of our selection is the first data release of the eRASS survey \citep{Merloni2024}, which covers the entire western Galactic hemisphere ($359.9442^\circ > l >  179.9442^\circ$). The corresponding catalogue contains about 930,000 X-ray sources and is 50\% complete at a flux limit of $F_{\rm 0.5-2.0~keV} > 5 \times 10^{-14}$~erg~s$^{-1}$~cm$^{-2}$. The point spread function of the eROSITA telescope is $\sim30''$ in the 0.2--2.3~keV energy band and remains constant across the entire sky.
Since jetted high-$z$ quasars are compact on arcsec scales, or may only show faint extended emission up to $\sim5''$ \citep[e.g.][]{Worrall2020,Connor2021,Ighina2022a}, we only considered the point-like sources in the eRASS:1 catalogue, reducing the total number of X-ray sources to $\sim$900,000.

To only select radio-powerful quasars (i.e. likely blazars) and to increase the positional accuracy of the X-ray sources, we also considered radio information. In particular, we used the second data release of the Rapid ASKAP Continuum survey (RACS; \citealt{Duchesne2023,Duchesne2024}) at 1.37~GHz. This survey covers the entire sky south of Dec=+49$^{\circ}$ and, therefore, the vast majority of the German eRASS:1 footprint. The median synthesised beam is $\sim12''$ and the median typical RMS is $\sim$180~\textmu Jy~beam$^{-1}$. The primary catalogue presented in \cite{Duchesne2024} has a $>95$\% completeness for fluxes above 1.6~mJy. We also note that in this project we are mainly focusing on lower declinations (Dec~$<+0^{\circ}$) and outside the Galactic plane, where the quality of the observations is typically better than the median values reported for the entire survey.

When cross matching the X-ray and the radio sources, we only considered the extra-galactic sky at $\rm |b|>20^\circ$, to avoid contamination from Galactic sources. To reduce potential spurious associations to the minimum we adopted different crossing match radii based on the eRASS X-ray flux. In particular, we  chose three 0.2-2.3~keV flux bins: $f>10^{-12}$~erg~s$^{-1}$~cm$^{-1}$; $2\times10^{-13}$~erg~s$^{-1}$~cm$^{-1}<f<10^{-12}$~erg~s$^{-1}$~cm$^{-1}$; $f<2\times10^{-13}$~erg~s$^{-1}$~cm$^{-1}$. To find an appropriate radius for each flux bin, we first performed a cross match between the catalogues adopting a large radius (30$''$) meant to include all the associations. After simulating a sample with the eRASS characteristics, but with random sky positions, and cross-matching it again to the RACS-mid, we also estimated the number of spurious associations expected in each bin. Finally, we selected the radius that, for each flux bin, contains 90\% of all the true associations. The radii obtained, in order of decreasing X-ray flux bin, are the following: 3.25$''$, 5.43$''$ and 14.45$''$.
The total number of sources recovered from the combination of the X-ray and radio catalogues is 45,350.

\subsection{DELVE selection}
\label{ssec:DELVE}
In order to select good $z>4$ candidates we also considered optical and near-infrared photometric data from the second data release of the DECam Local Volume Exploration (DELVE; \citealt{Drlica-Wagner2022}) survey. This is a wide-area photometric survey in the $griz$ filters aimed at covering the entire sky south of DEC$<$+30$^\circ$. In the second data release the area covered by each filter is slightly different (see \citealt{Drlica-Wagner2022}), with $\sim17000$~deg$^2$ covered in all four bands. The $5\sigma$ AB median depth in each filter are: $g=24.3$, $r=23.9$, $i=23.5$, and $z=22.8$. 

We cross-matched the catalogue of X-ray and radio sources built from eRASS and RACS-mid to the DELVE catalogue by using a radius of $r=1.7''$ from the radio position, which corresponds to more than 90\% confidence interval of the radio positions. From this cross-match, we recover a total of 30,849 optical/NIR counterparts. 

We then applied the well tested Ly$\alpha$ dropout technique (for similar approaches, see e.g. \citealt{Banados2016,Caccianiga2019,Gloudemans2022,Ighina2025}) to select high-$z$ sources. This technique consists of selecting objects showing a significant drop in two adjacent photometric colours, which, in high-$z$ sources is caused by the absorption of the UV emission by the Intergalactic medium. In this work we focused on objects with a drop on the $g-r$, $r-i$ and $i-z$ colours, corresponding to a redshift range of $3.6<z<6.4$. {\bf To select only point-like sources, we considered the \texttt{EXTENDED\_CLASS} parameter, which takes into account both the likelihood that an object is point-like in a specific band and the associated uncertainty (see sec.~4.7 of \citealt{Drlica-Wagner2022} for further details)}. Finally, we also included IR photometric measurements from the Wide-field Infrared Survey Explorer Catalogue (CATWISE; \citealt{Eisenhardt2020}), by applying a cross-matching radius of $2.5''$ between the optical-IR counterparts (see \citealt{Caccianiga2019} for a similar approach). We applied three criteria to the optical/NIR colours of the sources and we used the $\sim$1800 $z>4$ quasars currently known in the DELVE area to compute the completeness of the selection. We applied three sets of criteria based on the type of dropout, i.e. on the redshift, of the candidates:

\begin{itemize}

    \item $g-$drop, $3.7\lesssim z\lesssim4.7$:\\
    \indent mag$\_r<21.5$\,\\
    \indent $g-r>1$, \\
    \indent \texttt{EXTENDED\_CLASS\_$r$}$\, \le$ 1,\\
    \indent $r-i<1.14$, $-0.83<z-W2<1.92$\\
    
    \item $r-$drop, $4.7\lesssim z\lesssim5.7$:\\
    \indent mag$\_i<21.5$, \\
    \indent $r-i>1$, $g-i>2.8$, \\
    \indent $i-z<0.97$, $-0.33<z-W2<1.66$, \\
    \indent\texttt{EXTENDED\_CLASS\_$i$}\, $\le$ 1\\

    \item $i-$drop, $5.7\lesssim z\lesssim6.4$:\\
    \indent mag$\_z<21.5$,\\
    \indent $i-z>1$, non-detection $g$-band,\\
    \indent $-0.33<z-W2<1.66$,\\
    \indent \texttt{EXTENDED\_CLASS\_$z$}\,$\le$ 1\\

\end{itemize}

By applying these criteria to all the known $z>4$ quasars detected in DELVE (radio and not), we recover 90\%, 76\% and 58\% of the total quasars in the $g-$, $r-$ and $i-$dropouts, respectively. In section \ref{sec:IC/CMB}, we use these values to correct for the completeness of our selection. The total number of candidates selected with these criteria is 486.

We note that, out of all the $z>4$ radio quasars in the literature that have been detected in the eRASS:1 and that fall in the RACS--mid+DELVE area, we only missed four known $z>4$ quasars in our selection: J032444.2$-$291821 at $z=4.63$ \citep{Hook2002}, was missed because the X-ray source was associated with a second component of the RACS-mid radio source, which was $\sim4''$ away from the optical/NIR counterpart; J111004.2+263522 at $z=4.12$ \citep{Lyke2020} and J140850.9+020522 at $z=4.00$ \citep{Caccianiga2024} were missed because the radio and optical/NIR counterparts are $>1.7''$; J132206.4$-$132354 at $z=4.70$ \citep{Ighina2025,Belladitta2025} would be selected as a $g$-dropout ($g-r>1$), however its mag$\_r=21.56$ is above the threshold we adopted. We also note that a potential quasar has been found recently at $z=4.71$ \citep{Davies2026} based on the Spectro-Photometer for the History of the Universe, Epoch of Reionization, and Ices Explorer \citep[SPHEREx;][]{Bock2026} low-resolution spectrum in the NIR.


\subsection{Additional optical/NIR surveys}
\label{ssec:adtl_opt_srvys}
To further improve our selection, we also considered data from additional optical/NIR surveys. We started by considering the sources detected in the GAIA data release 3 \citep{Gaia2023}. Out of the 438 sources detected in GAIA, we removed all the objects that showed a significant proper motion or parallax measurement (i.e., with the ratio between the proper motion or parallax value and its uncertainty above $>4$; see \citealt{Wolf2018} for a similar approach). 

We then considered data from the Panoramic Survey Telescope \& Rapid Response System (Pan-STARRS; \citealt{Chambers2016}), the Dark Energy Survey (DES; \citealt{Abbott2021}) and SkyMapper \citep{Onken2024}. If detected in either of these surveys, we only kept sources that still showed a dropout $>0.7$. This criteria allowed us to discard objects with unreliable values in the DELVE catalogue and/or highly variable systems\footnote{While the emission produced by relativistic jets in blazars is highly variable, the optical/UV emission of  flat spectrum radio quasas (FSRQs) is dominated by the radiation produced by accretion disc, which is expected to be only $\Delta{\rm mag}~\sim0.1-0.2$ \citep[e.g.][]{King2004}.}. After the GAIA, Pan-STARRS, DES and SkyMapper criteria, the final number of sources in the sample is 153. Among these objects, 43 already have a redshift estimate from the literature and 30 (18) are at $z>3.4$ ($>4$), leaving 110 new candidates.

\subsection{Optical/IR Spectral Energy Distribution}
\begin{figure}
\centering
    \includegraphics[width=\hsize]{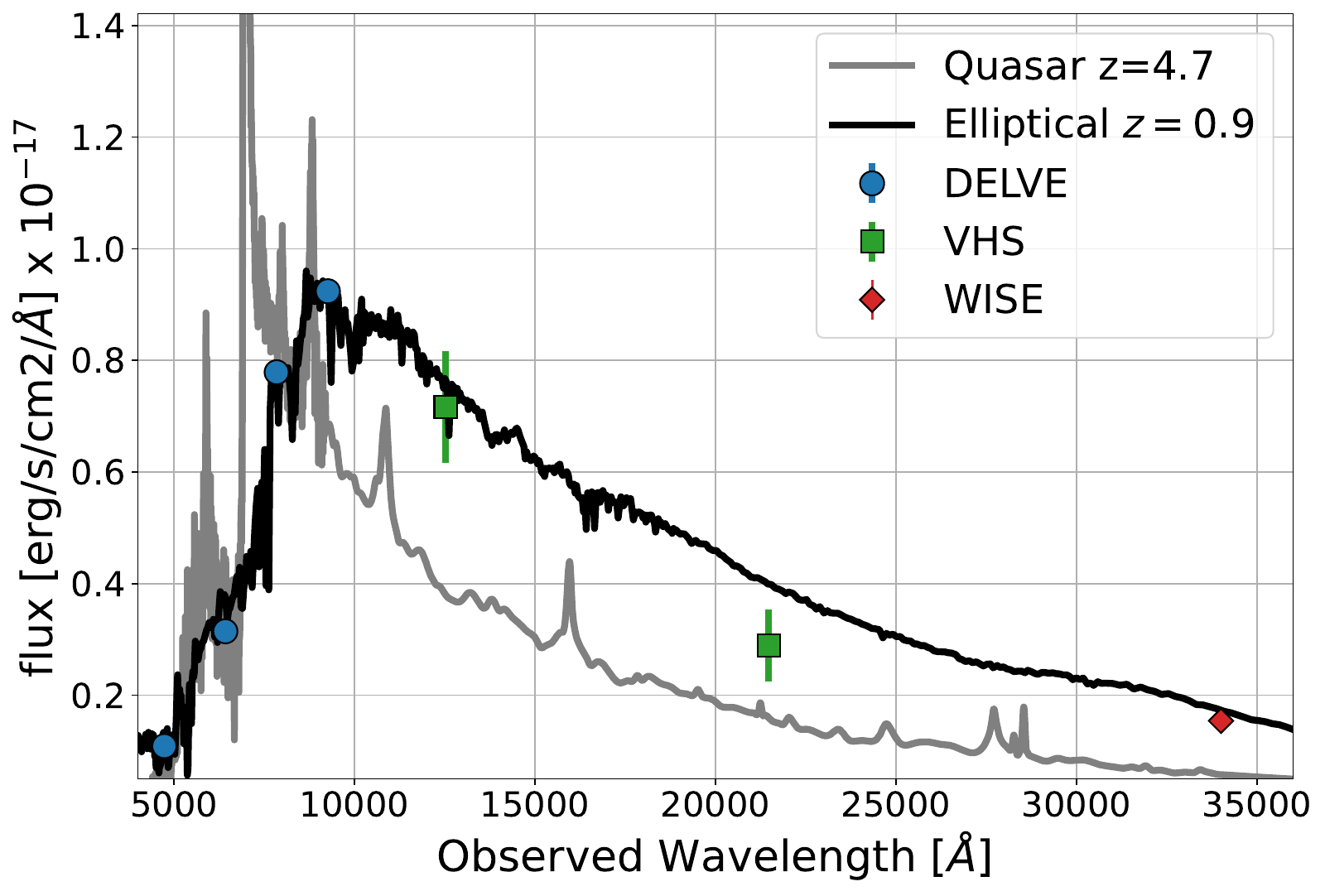}
    
    \caption{Example of the optical/NIR SED of an $r$-dropout candidate selected with the criteria described in the text. As shown with the templates overlaid, the observed optical emission can be explained by both an elliptical galaxy at $z\sim0.9$ (black line; from {\protect \citealt{Polletta2007}}) and a $z\sim4.7$ quasar (gray line; from {\protect \citealt{Polletta2007}}). Based on the additional IR coverage provided by the VHS and CatWISE surveys, we excluded this source from the observable sample. }
    \label{fig:contaminant}
\end{figure}

As a final step, we also visually checked the optical-NIR spectral energy distributions (SEDs) of the selected candidates. We considered all the available optical and NIR photometric data available\footnote{Mainly from DELVE, DES, SkyMapper, PanSTARRS, and CatWISE catalogues, as well as the VISTA Hemisphere Survey \citep[VHS; ][]{McMahon2021} and the VISTA Kilo-degree Infrared Galaxy \citep[VIKING; ][]{Edge2013}}. In this way we were able to discard sources with NIR colours consistent with low-$z$ elliptical galaxies. We show such an example in Fig. \ref{fig:contaminant}.
In this way we discarded 42 further candidates that we did not consider for follow-up observations.
From the remaining 68 candidates, we spectroscopically observed 46. In the next section, we describe the observing setups and the data reductions for this sample.

\begin{figure}
\centering
    \includegraphics[width=\hsize]{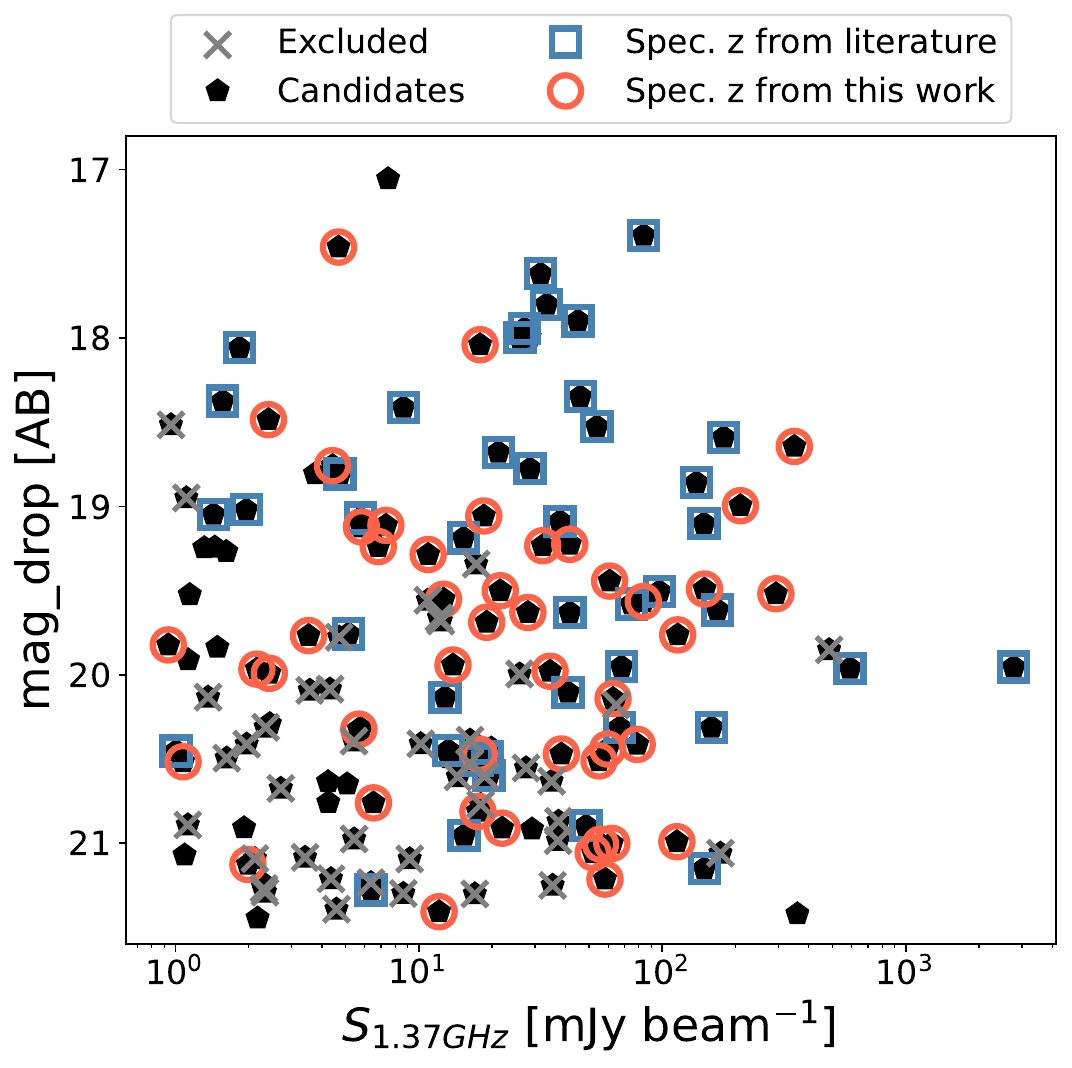}
    \caption{Apparent magnitude from the DELVE survey as a function of the $S_{\rm 1.37GHz}$ radio surface brightness from RACS-mid for the high-$z$ candidates selected starting from the eRASS:1 survey. The magnitude corresponds to the filter after the dropout of the given candidate: $r$-filter for $z<4.5$, the $i$-filter for $4.7 \leq z \leq 5.7$ and the $z$-filter for $z>5.7$. Black pentagons indicate all the sources selected after optical and NIR colour criteria. Additional symbols indicate objects that already had a redshift measurement from the literature (blue squares), were excluded from the final sample because of their optical/NIR SED, have been observed as part of this work (red circles). As clear from the plot, the majority of the optically and radio-bright target have been identified.}
    \label{fig:all_candidates}
\end{figure}

We show in Fig. \ref{fig:all_candidates} the apparent optical/NIR magnitude as a function of the observed radio flux for all the candidates we selected (black pentagons). We highlight the objects with a spectroscopic redshift available from the literature (43; blue squares), the candidates we discarded (42; gray crosses), and the targets we observed in this work (46; red circles).

\section{Observations and data reduction}
\label{sec:spectroscopy}

To confirm the quasar classification of our high-$z$ candidates, we performed a campaign of spectroscopic observations using multiple telescopes and instruments, covering 2024 October through 2026 January. We targeted candidate blazars from the above sample based on their radio, optical, and/or X-ray brightness. Individual candidates were prioritised for different observing runs based on telescope latitude, seasonal observability, source brightness and weather conditions.

For all observations, we used long slit observations, with a spectroscopic setup sensitive to the $\sim5000-9000$~\AA \, wavelength range. For $z\sim3.5-5.5$ sources, this setup covers the Ly$\alpha$ emission line and the associated break, which is at $\sim5500-8000$~\AA. We report the telescope and the on-source time of each target in Tab. \ref{tab:obs_targets}. Details of each telescope setup and corresponding data reduction steps are as follows:

\begin{itemize}
    \item \textbf{Telescopio Nazionale Galileo}. We observed using the Device Optimized for the LOw RESolution (DOLORES) in 2025 February under the project AOT50\_46 (P.I.: Ighina). We used the LR-R grism (covering $\sim$5000--10,000~\AA) and a long-slit with a width between $1\farcs0$ and $2\farcs00$, depending on the seeing conditions. The slit was oriented along the parallactic angle or oriented along a pivot star for fainter targets. In total, we observed 9 targets. For the data reduction, we followed the \texttt{easyspec} tutorials \citep{deMenezes2025} for long-slit observations of point-like targets\footnote{see \url{https://github.com/ranieremenezes/easyspec}.}. The relative flux calibration was obtained by observing multiple spectrophotometric standard stars each night.
\item \textbf{Large Binocular Telescope}. We used the Multi-Object Double Spectrograph (MODS; \citealt{Pogge2010}) to observe 4 targets between 2024 October and 2025 January under the project IT-2024B-039 (P.I.: Ighina). Observations were conducted with the red grating G670L coupled with the GG495 filter, covering the $\sim$5000--10,000~\AA \, wavelength range. A spectrophotometric standard star was observed for each target. We performed data reduction with the \texttt{SIPGI} software \citep{Gargiulo2022} following a standard long-slit data reduction\footnote{see \url{http://pandora.lambrate.inaf.it/sipgi/}.}.
\item \textbf{New Technology Telescope}. Using the ESO Faint Object Spectrograph and Camera (EFOSC2, \citealt{Buzzoni1984}), we observed 23 targets under project 115.28BJ.001 (P.I.: Ighina) in 2025 September. Our observations were conducted with the grism $\#5$, covering the $\sim$5200--9300~\AA \, wavelength range, and with a long slit oriented along the parallactic angle. Depending on the seeing conditions, we used a long slit of width between $1\farcs0$ and $2\farcs0$. We used \texttt{pypeit} \citep{Prochaska2020b,Prochaska2026} to reduce the data following a standard data reduction for long-slit spectra\footnote{see \url{https://pypeit.readthedocs.io/en/stable/tutorials/tutorials.html}.}.
\item \textbf{Gemini South}. We observed 4 targets in 2025 May--June using the Gemini Multi-Object Spectrographs (GMOS, \citealt{Hook2004}) under the project GS-2025A-Q-414 (P.I.: Ighina). All observations were conducted with a $2\farcs0$ slit and with the R150 grism. Half of the exposures were centred at 5200~\AA \, and half at 5500~\AA \, in order to cover the wavelength gap in the detector. The final wavelength coverage is $\sim$4000--10,000~\AA. To reduce the observations we used the DRAGONS pipeline \citep{Labrie2023} following a standard long-slit data reduction\footnote{see \url{https://dragons.readthedocs.io/projects/gmosls-drtutorial/en/v4.1.0/}.}.
\item \textbf{Magellan Clay}. Our final observations were conducted with the Low Dispersion Survey Spectrograph 3 \citep[LDSS-3;][]{Stevenson2016} in 2026 January under the project 814 (P.I. Ighina).  All five targets were observed with the $0\farcs75$ slit and the VPH-ALL grism, covering the 6000--10,000~\AA \, wavelength range. As before, we used \texttt{pypeit} \citep{Prochaska2020b,Prochaska2026} for the data reduction, following \citet{Martinez2026}.

\end{itemize}

\begin{figure}
\centering
    \includegraphics[width=\hsize]{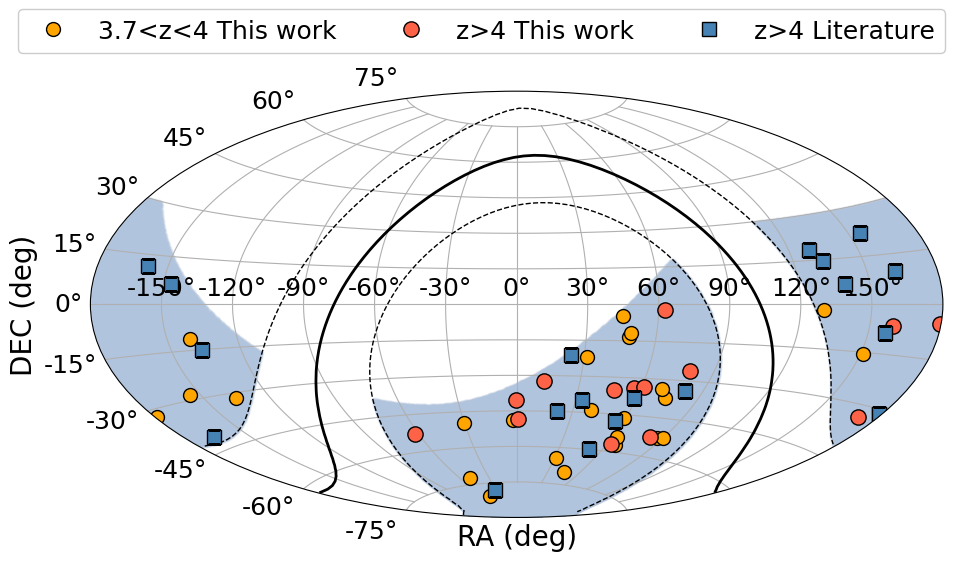}
    \caption{Sky distribution of the high-$z$ quasars selected from the eRASS:1 survey. Objects already reported in the literature are marked by blue squares, while quasars newly discovered in this work are shown with orange ($z<4$) and red ($z>4$) circles. The shaded blue area corresponds to the area covered by eRASS:1 below Dec.=$+30^\circ$, where most of the sky is covered by DELVE.}
    \label{fig:sky_dist}
\end{figure}

We report in Figs. \ref{fig:optical_spectra1}--\ref{fig:optical_spectra5} the final spectra obtained for the newly identified high-$z$ quasars, and we show in Fig. \ref{fig:contaminants} the spectra of the objects not confirmed to be at high redshifts. All spectra were re-normalised to their $i$-band magnitude from DELVE. Out of the 46 targets we observed, we confirmed the high-$z$ nature of 39\footnote{We note that one of the targets observed, J115839.3-052221, was also independently identified as part of the Dark Energy Spectroscopic Instrument (DESI; \citealt{Yang2023}) a few months before our observations.}, 14 of which are at $z>4$, corresponding to an increase of 65\% in the number of $z>4$ radio quasars detected in the same eRASS:1 area. We show in Fig. \ref{fig:sky_dist} the sky distribution of the newly discovered high-$z$ quasars and the $z>4$ already known in the literature.  The high faction of confirmed candidates ($\sim85\%$) is mainly due to the X-ray and radio associations, which remove most stars as potential contaminants. Additionally, given the bright optical nature of the observed targets and the large number of photometric data-points available, we were able to screen out most lower-redshift contaminants, significantly improving the efficiency of the spectroscopic confirmation. Indeed, most of the observed objects that were not confirmed at high-$z$ are targets for which only DELVE photometry was available (cf. Fig. \ref{fig:contaminant}).

\section{Properties of the eRASS:1 high-\texorpdfstring{\lowercase{$z$}}{z} radio-quasar sample}
\label{sec:comparison}

To compare our sample with the larger quasar population, we computed optical/UV and X-ray properties for the new blazars, parametrised by the monochromatic luminosity at 2500~\AA \, (L$_{\rm 2500\, \textup{\footnotesize \AA}}$) and the UV--to--X-ray spectral slope ($\alpha_{\rm ox}$\footnote{$\alpha_{\rm ox}=0.384 \times \log(L_{\rm 2500\, \textup{\footnotesize \AA}}$/L$_{\rm 2~keV}$); we follow the convention that a larger value of $\alpha_{\rm ox}$ corresponds to a higher relative UV brightness.}).
To compute the monochromatic luminosity at 2500~\AA \, we assumed a power-law spectrum normalized to the reddest optical band available, typically $z$-band for sources covered by the DELVE survey. We then computed an optical spectral index from the $z$ and $W1$-band magnitudes. 
For computing the monochromatic X-ray luminosity (L$_{\rm 2~keV}$), we started with the 0.2--2.3~keV flux reported in the eRASS catalogue and assumed a simple power-law model with $\Gamma=2.0$, as used to compute fluxes from count rates in the eRASS:1 catalogue \citep{Merloni2024}. Uncertainties on L$_{\rm 2~keV}$ are computed including both the eRASS flux uncertainties as well as an adopted $\sigma(\Gamma_{\rm X})=0.2$.
Rest-frame luminosities in radio, optical, and X-ray bands, as well as the radio-loudness parameter and $\alpha_{\rm ox}$, are presented for the full sample in Tabs. \ref{tab:lum_sample} and \ref{tab:lum_sample2}.

\begin{figure}
\centering
    \includegraphics[width=\hsize]{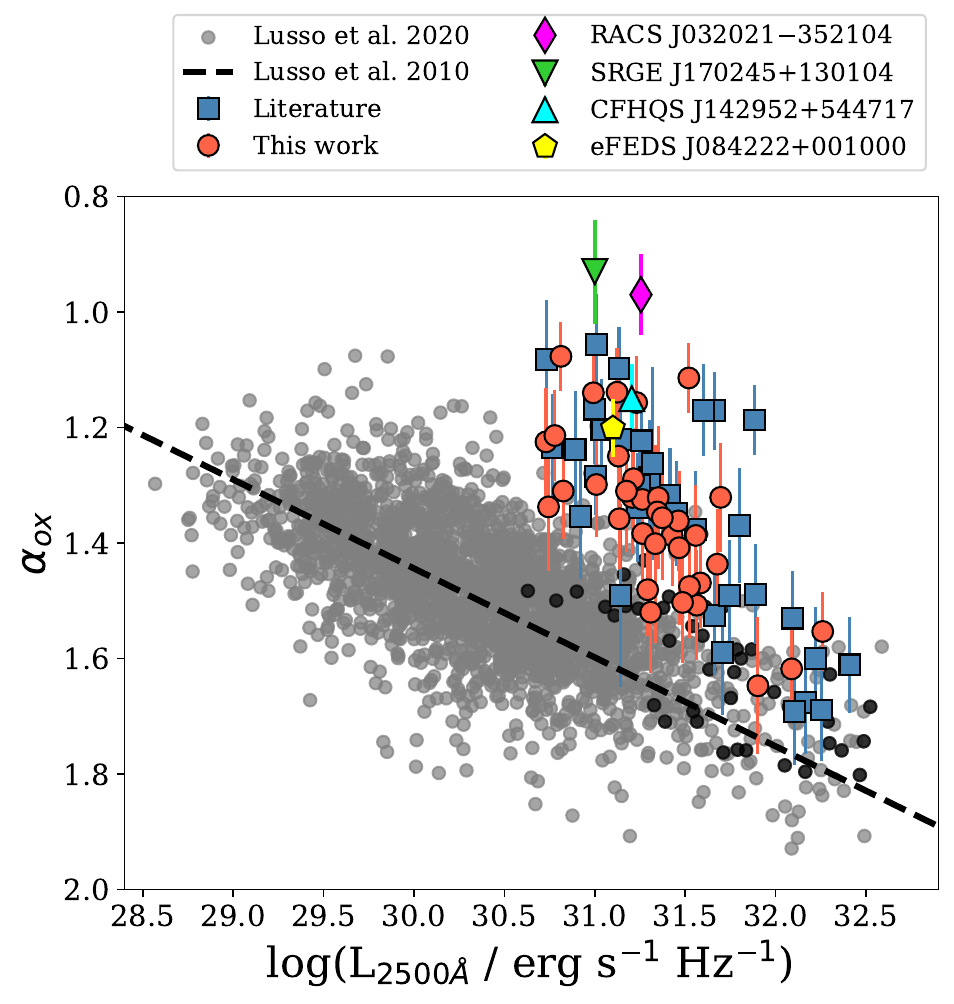}
    \caption{Distribution of the $\alpha_{\rm ox}$ parameter as a function of the UV luminosity at 2500~\AA \, for the high-$z$ quasars selected in this work. As a reference we also report the $0<z<7$ quasars discussed in \protect\citealt{Lusso2020} ($z>3.5$ sources are highlighted with black circles) as well as the best-fit $\alpha_{\rm ox}$--L$_{\rm UV}$ relation derived by \protect\cite{Lusso2010}. We highlight two blazars recently discovered in the Russian half of the eRASS. SRGE~J170245.3+130104 at $z=5.47$ (green triangle; \protect\citealt{Khorunzhev2021}) and CFHQS~J142952+544717 at $z=6.18$ (cyan triangle, \protect\citealt{Medvedev2020}). We also show two highly accreting radio quasars, one at $z=3.4$ in the eFEDS field, eFEDS~J084222.9+001000 (yellow pentagon; \protect\citealt{Obuchi2026}), and one at $z=6.1$, RACS~J032021.44$-$352104.1 (magenta diamond; \protect\citealt{Ighina2025b}).
    Most of the sources selected in this work are significantly above the typical $\alpha_{\rm ox}$--L$_{\rm UV}$ relation and are located in the same region of the parameter space as other high-$z$ blazars and super-Eddington accreting quasars}.
    
    \label{fig:aox_comparison}
\end{figure}

In Fig. \ref{fig:aox_comparison} we show the L$_{\rm 2500\, \textup{\footnotesize \AA}}$ and $\alpha_{\rm ox}$ distribution of the high-$z$ quasars selected in this work, both newly discovered (red circles) and already known in the literature (blue squares). In the same plot, we also show the distribution of the $0<z<7$ quasars discussed in \cite{Lusso2020} and the best-fit L$_{2500~\textup{\footnotesize \AA}}$--$\alpha_{\rm ox}$ relation derived by \cite{Lusso2010}. Sources from \cite{Lusso2020} at $z>3.5$ are highlighted with black data points.

Based on Fig. \ref{fig:aox_comparison}, the sources selected in this work sampled the brightest X-ray and UV part of the general quasar population distribution. Indeed, all of our targets have intrinsic UV luminosities above $\sim4\times 10^{30}$~erg~sec$^{-1}$~Hz$^{-1}$ and lie above the mean L$_{\rm 2500~\textup{\footnotesize \AA}}$--$\alpha_{\rm ox}$ relation from \cite{Lusso2010}. This is a consequence of the optical and X-ray selection limits. At the same time, the X-ray excess observed, compared to the relation, can also be attributed to the emission from the relativistic jets. Indeed, by requiring a radio counterpart in our selection, we are restricting the selection to jetted quasars. 

Given the high X-ray flux limit we are most likely to select blazars, for which relativistically boosted radiation from the jets can significantly affect the observed X-ray emission, or sources with a very soft X-ray emission, given low-energy sensitivity of eRASS ($\lesssim2$~keV).

An example of a high-$z$ blazar previously discovered in the eRASS is the source SRGE~J170245.3+130104 at $z=5.48$ selected from the first scan in the Russian half of the sky (light green triangle in Fig. \ref{fig:aox_comparison}; \citealt{Khorunzhev2021}). This corresponds to the X-ray brightest source known at $z\gtrsim5.5$ and, based on its multi-wavelength properties, including radio VLBI \citep{Liu2024}, this source is a blazar. Similarly, the quasar CFHQS~J142952+544717 at $z=6.18$ was also detected in the Russian half of the eRASS:1 (dark green triangle in Fig. \ref{fig:aox_comparison}; \citealt{Medvedev2020}). Although complementary radio observations \citep{Frey2011} seem to discard a blazar scenario, the strong X-ray emission together with its rapid variability \citep{Medvedev2021,Migliori2023,Marcotulli2025} suggest that its high-energy emission is relativistically boosted.

At the same time, sources with very soft X-ray emission are more likely to be detected in the eRASS. For example, highly accreting quasars \citep[e.g.][]{Tortosa2022} are expected to have a very steep X-ray spectrum ($\Gamma_{\rm X}>2.4$; \citealt{Madau2024,Pacucci2024,Inayoshi2024}) and therefore most of their high energy emission is produced in the soft X-rays ($\lesssim2$~keV). Interestingly, two examples of fast accreting, super-Eddington\footnote{Super-Eddington accretion occurs when $\lambda_{\rm Edd} = \dot{M}/\dot{M}_{\mathrm{Edd}} > 1$, where $\dot{M}_{\mathrm{Edd}} \approx 2.2  M_{\rm BH} \epsilon^{-1} \times 10^{-9}\ {\rm yr}^{-1}$, $\epsilon$ is the radiative efficiency, $\dot{M}$ is the mass accretion rate, and  $M_{\rm BH}$ is the mass of the BH.} sources have been recently found at high redshift: eFEDS~J084222.9+001000 at $z=3.4$ \citep{Obuchi2026} and RACS~J032021.44$-$352104.1 at $6.13$ \citep{Ighina2025b}, reported as yellow pentagon and magenta diamond in Fig. \ref{fig:aox_comparison}, respectively. The high-energy properties of eFEDS~J084222.9+001000 were uncovered by eROSITA as part of the equatorial deep field (eROSITA Final Equatorial-Depth Survey, eFEDS; \citealt{Brunner2022}). Based on the strong (L$_{\rm 2-10keV}$ = $9.8_{-0.3}^{+0.5}\times10^{45}$~erg~s$^{-1}$) and soft ($\Gamma_{\rm X}=2.4\pm0.2$) X-ray emission, \cite{Obuchi2026} argued that the high-energy emission in this system is a consequence of super-Eddington accretion, even when accounting for the potential contribution of the relativistic jet, whose presence was inferred from a strong radio emission. Similarly, the high luminosity (L$_{\rm 2-10~keV}=1.8^{+1.1}_{-0.7} \times 10^{46}$~erg~sec$^{-1}$) and the large photon index value ($\Gamma_{\rm X}=3.3\pm0.4$) of RACS~J032021.44$-$352104.1 revealed by dedicated {\it Chandra} observations together with the absence of a strong compact radio emission on VLBI scales suggest that this quasar is accreting above its Eddington limit. Complementary JWST observations support these findings (Gloudemans et al. in prep.).

As clearly shown in Fig. \ref{fig:aox_comparison}, the sources selected in this work occupy a similar position in the L$_{\rm 2500~\textup{\footnotesize \AA}}$--$\alpha_{\rm ox}$ plot as both blazars and super-Eddington sources. Both classes of objects are very rare and extremely valuable to study the evolution of SMBH across cosmic times.

\section{Redshift evolution of blazars}
\label{sec:IC/CMB}

One of the scientific drivers of this project was to identify all the high-$z$ blazars in the eRASS:1 survey. This is because the specific orientation of blazars with respect to our line of sight ($\theta_{\rm v}\lesssim1/\Gamma$) makes them powerful tools to constrain the evolution of the entire population of AGN with relativistic jets oriented in different directions (see, e.g., \citealt{Ghisellini2013,Diana2022}). In particular, previous studies focused on the X-ray selected blazars found that this population has a unique space density evolution peaking at $z\sim4$ \citep{Ajello2009,Marcotulli2022}. If confirmed, this result would have strong implications on the cosmological evolution of jetted systems compared to the overall quasar population. 

In addition, the X-ray brightest blazars are the most suited sources to resolve the emission produced by kpc-scale jets both at low- and high-energies. These systems offer a unique opportunity to constrain the evolution of the emission mechanism in relativistic jets (see e.g. \citealt{Connor2021,Ighina2022a,Breiding2023}). 

In this section we first identify the bona-fide blazars among the eRASS:1 sample discussed above and we then use these blazars to constrain models describing the redshift evolution of this class of objects as a function of redshift \citep[see][]{Ighina2021b}.

\subsection{Identification of blazars}
\label{sec:blazar_class}

While blazars are significantly different from an observational point of view to other quasars, when it comes to high-$z$ systems, it can be challenging to have a solid blazar classification \citep[e.g.][]{Krezinger2026}. Typically, brightness temperature measurement from VLBI observations (e.g. \citealt{Coppejans2016,Krezinger2022}) or the intensity and the shape of the X-ray emission \citep[e.g.][]{Ighina2019,Ighina2024b,Sbarrato2022} are used to identify relativistic boosting and, therefore, blazars. However, the faint fluxes of high-$z$ sources make these classification harder and more time consuming.

\begin{figure}
\centering
    \includegraphics[width=\hsize]{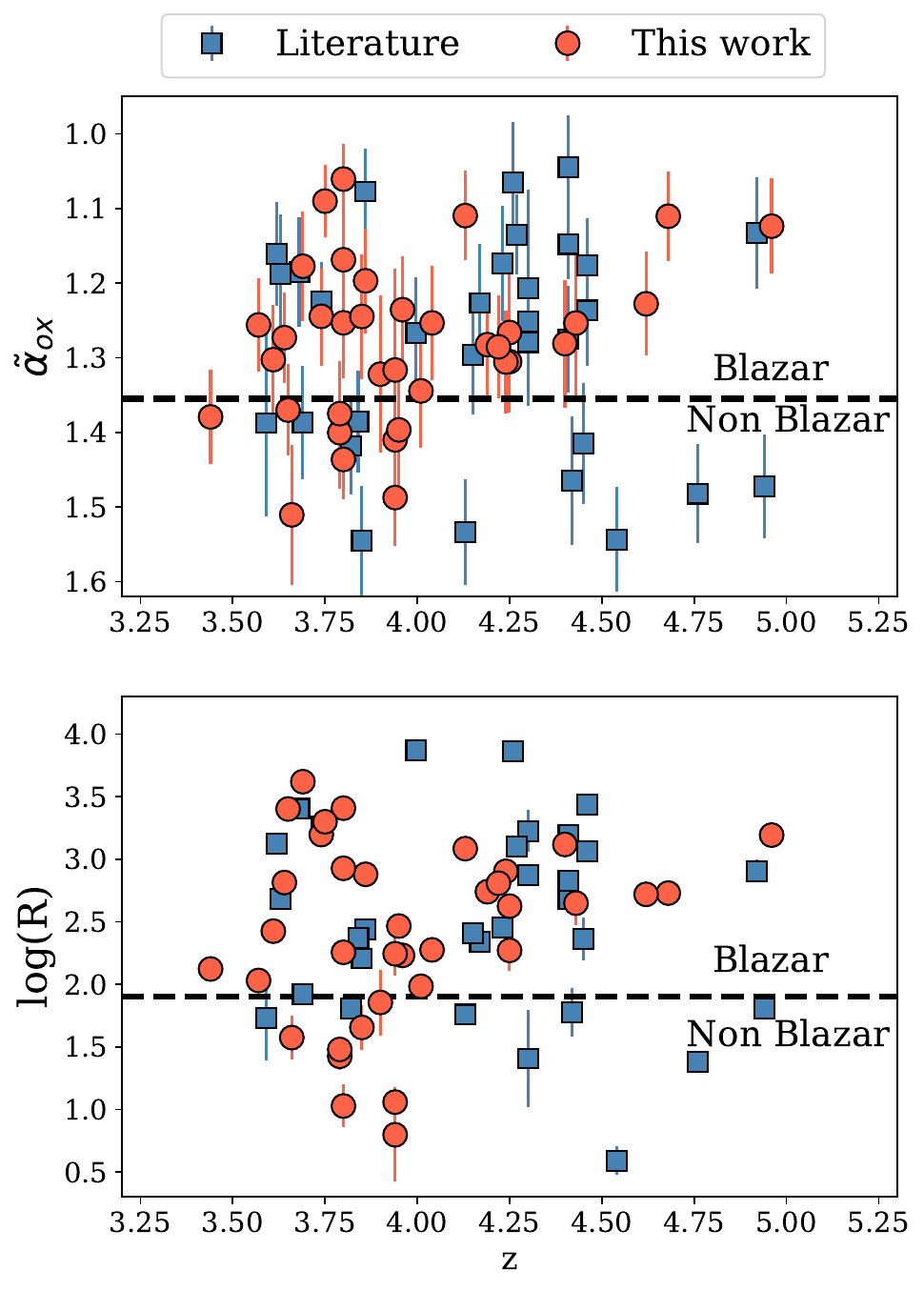}
    \caption{Distribution of the $\tilde{\alpha}_{\rm ox}$ (top panel) and the radio-loudness (in logarithm scale; bottom panel) parameters as a function of redshift. Red circles indicate the new sources discovered in this work, while blue squares show sources already reported in the literature in the same area. The horizontal black lines indicate the typical thresholds used to identify blazars: $\tilde{\alpha}_{\rm ox}$=1.355 and R=80. Based on these thresholds, most of the sources selected in this work are consistent with being blazars.}
    \label{fig:aox_R_dist}
\end{figure}

In this work, we use the high-energy information to classify sources as blazars, since this is available for all the targets in the sample. In particular, we considered the classification proposed by \cite{Ighina2019} based on the shape (photon index, $\Gamma_{\rm X}<1.8$) and the X-ray-to-UV intensity ratio ($\tilde{\alpha}_{\rm ox}=0.303 \times {\rm log}(L_{\rm 10keV}/L_{\rm 2500~\textup{\footnotesize \AA}}<1.355$). Even though by definition we have X-ray information for all the quasars, in most cases the statistics is not sufficient to get significant constraints on the photon index value. For this reason, we only consider the $\tilde{\alpha}_{\rm ox}<1.355$ criterion for the blazar classification. We stress that, while a classification based on the $\tilde{\alpha}_{\rm ox}$ only cannot be considered fully reliable for each individual object, especially at faint X-ray fluxes, we expect the final number of blazar in a statistical sample to be reliable \citep[e.g.][]{Krezinger2026}. Dedicated X-ray and VLBI observations will be useful to have a ore secure classification of individual sources in the sample. To compute the rest-frame luminosities at 10~keV and the $\tilde{\alpha}_{\rm ox}$ parameter, we followed the same prescription described in the previous section for the $\alpha_{\rm ox}$ parameter\footnote{For simplicity, we also note that the two parameters are related by the following equation: $\tilde{\alpha}_{\rm ox} =0.79\alpha_{\rm ox} + 0.21(\Gamma_{\rm X} - 1)$.}. 
In Fig. \ref{fig:aox_R_dist} (top panel) we show the $\tilde{\alpha}_{\rm ox}$ parameter as a function of redshift. The threshold for the blazar classification is indicated with a dashed black line ($\tilde{\alpha}_{\rm ox}=1.355$; \citealt{Ighina2019}). Out of all the high-$z$ quasars selected, 52 present X-ray emission consistent with a blazar nature. Among these, 26 sources are at $z>4$. We note that the analysis of deeper eRASS data of the objects discussed in this work, including up to five scans, agrees with our blazar classification reported here. In particular, out of the 10 sources in common with \cite{Sbarrato2026}, the blazar classification is consistent for 9 of them. The remaining object (1eRASS~J141209.5+062411) is not considered a blazar here due to the high $\alpha_{\rm ox}$ value. Interestingly, based on dedicated X-ray observations and adopting a similar blazar classification, \cite{Ighina2019} classified this same source as a potential blazar. We note however, that the two $\alpha_{\rm ox}$ values are consistent. This highlights how variability and low statistics can play a role in the proposed classification.

In Fig. \ref{fig:aox_R_dist} (bottom panel) we also show the radio loudness parameter (R$=S_{\rm 5GHz}/S_{4400\AA}$ in the rest frame; \citealt{Kellerman1989}) as a function of redshift. Also in this case we show a threshold often used in the literature to pre-select blazars (R=80; e.g. \citealt{Sbarrato2022,Caccianiga2024}). We note that even by adopting a blazar classification based on the radio loudness, the total number of blazars at $z>4$ does not change and only two sources would be classified differently compared to the X-ray method. 

In order to compute the radio-loudness parameter we considered radio information from the literature. In particular, we integrated the information at 1.37~GHz from RACS-mid with surveys at both low frequencies, such as the TIFR GMRT Sky Survey (TGSS, at 150~MHz; \citealt{Intema2017}) and the GaLactic and Extragalactic All-sky Murchison Widefield Array survey eXtended (GLEAM-X, at 200~MHz; \citealt{Hurley-Walker2022,Ross2024}), and higher frequencies, such as RACS-high  (at 1.67~GHz; \citealt{Duchesne2025}) and the Very Large Array Sky Survey (VLASS, at 3~GHz; \citealt{Lacy2020}).
We then fit the available data points with a simple power law and used the corresponding spectral index to compute the rest-frame flux at 5~GHz. While we expect more complex spectral shapes \citep[e.g.][]{Shao2022,Ighina2022b} and variability \citep[e.g.][]{Sotnikova2021} to play a role in the observed radio properties, the relatively small uncertainty on the spectral indices indicate that these effects, if present, are not significant. There are only two cases where the uncertainty is large ($\sigma_{\alpha_r}0.5$), but this is due to the very limited radio coverage available, only from the RACS scans (i.e. 0.888--1.67~GHz). In these cases we fixed the uncertainty to $\sigma_{\alpha_r}=0.5$.

\subsection{Comparison to IC/CMB model predictions}

In order to compare the total number of blazars detected in the eRASS:1 survey to the number expected in the fractional IC/CMB model, we proceeded in a similar way to \citet[see both their Sec. 5 and Fig. 5]{Ighina2021b}. In the following we provide a summary of the assumptions and steps used.

We started from the radio luminosity function (RLF) of blazars derived by \cite{Mao2017}. By integrating this LF at different redshift and luminosity bins, we built a mock sample of blazars with known redshift and radio luminosity. We then computed their X-ray luminosity assuming two scenarios: a constant X-ray-to-radio luminosity (X/R) ratio as a function of redshift; a redshift evolution of the X/R ratio given by the IC/CMB model:

\begin{equation}
\frac{L_X}{L_R}(z)\: = \: \frac{L_X}{L_R}(z=0) \times \,  \left[(1 - A_{0}) + A_0 (1 + z)^4\right]
    \label{eq:enhace}
\end{equation}

where $L_x/L_r(z=0)$ is the distribution of X/R ratios of blazars at $z=0$ and the term $A_0$ is the fraction of the X-ray emission produced through IC/CMB at $z=0$.
In \cite{Ighina2021b} we assumed that both X/R and $A_0$ follow a logarithmic Gaussian distribution and the corresponding best-fit  we derived are: log(X/R)=$1.95$, $\sigma_{\rm X/R}=0.35$, log($A_0$)=$-2.74$, and $\sigma_{A_0}=0.66$. 
We stress that these values were computed from  the comparison of radio-selected samples of blazars at $z\sim1$ and $z\sim4.5$. In the following computations we use these values for the X/R($z=0$) and $A_0$ parameter distributions.

\begin{figure}
\centering
    \includegraphics[width=\hsize]{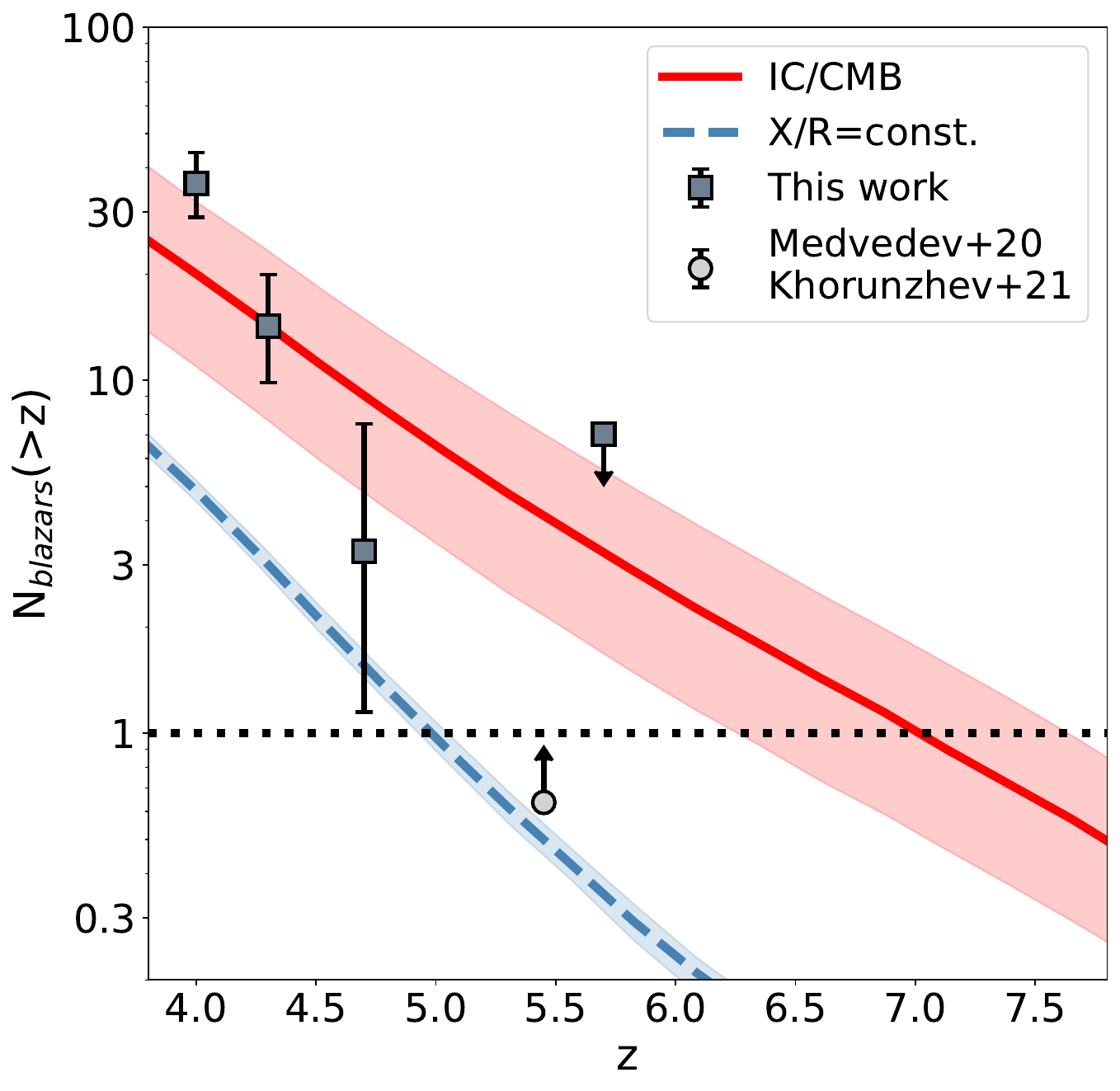}
    \caption{Number of blazars expected to be detected eRASS:1 in the $\sim12000$~deg$^2$ covered by this analysis as a function of redshift. In the plot we show expectations for both constant X/R ratios as a function of redshift (dashed blue line) and for X/R evolution following the IC/CMB model (solid red line; see Eq. \ref{eq:enhace} and \citealt{Ighina2021b}). The black data points represent the observed number of blazars in this work above a given redshift value. The upper limit at $z\sim5.5$ is based on the detection of two blazars in the Russian half of the sky {\protect \citep{Medvedev2020,Khorunzhev2021}}.}
    \label{fig:ICCMB_pred}
\end{figure}

To convert the X-ray luminosity into a flux we assumed a photon index $\Gamma_{\rm X}=1.5$, typical of blazars. We then computed the number of sources detected in the eRASS:1 by considering the sky-dependent sensitivity over the extra-galactic sky ($|b|>20^\circ$) as derived in \citet[see their Fig. 11]{Merloni2024}. We then re-normalised the number of objects over the fraction of the sky we considered for our selection ($\sim1/3$ of the total sky area). We show the expected number of blazars detected based on our selection in Fig. \ref{fig:ICCMB_pred}, assuming both the constant (dashed blue line) and the IC/CMB (solid red line) X/R evolving models. The shaded areas around the curves represent the expected values assuming a photon index in the range $\Gamma_{\rm X}=1.3-1.7$ and the A$_0$ distribution parameters with a p-value $>0.7$ (see \citealt{Ighina2021b}). In the same plot, we also show the number of blazars we detected in this work (dark gray data points), which we corrected for the optical and radio selection completeness described in Sec. \ref{sec:selection}. Each data-point represents the number of blazars observed above a given redshift value. The upper limit set at $z=5.7$ represents the non confirmation of any $i$-dropout candidates in the sample. The light gray lower limit is given by the detection of two $z\gtrsim5.5$ quasars in the Russian half of the eRASS survey (see \citealt{Medvedev2020, Khorunzhev2021})\footnote{While the source SRGE~J170245.3+130104 presents multi-wavelength properties typical of a blazar (see \citealt{Khorunzhev2021, An2023, Liu2024}), the nature of CFHQS~J142952+544717 is unclear. High resolution radio observations did not find strong signs of relativistic boosting \citep{Frey2011}; however, the very strong \citep{Medvedev2021,Migliori2023} and the rapidly variable X-ray emission \citep{Marcotulli2025} point towards a blazar nature. For consistency with our classification based on the $\tilde{\alpha}_{\rm ox}<1.355$, we consider it a blazar in this work.}. All the errors in the plot are based on Poisson statistics.

Based on the observed number of sources, we can rule out a scenario where the X/R ratios in blazars does not evolve as a function of redshift. In particular, the data points at $z=4-4.3$ are $\sim5$ times larger than the expectations, while the agreement with the IC/CMB evolution scenario is significantly better. However, also in this case the total number of $z>4$ blazars is larger compared to the model, although the discrepancy is much smaller  with respect to the constant X/R case. Considering the Poisson error on the observed number of blazars, data are $\sim2.6\sigma$ away from the best-fit model derived in \cite{Ighina2021b}. The discrepancy reduces if we also consider the uncertainty on the model. 
Interestingly, while the IC/CMB model under-predicts the total number of blazars  in eRASS at $z>4$, it over-predicts the number of $z>4.7$ blazars, although, the uncertainty is still very large on $z>4.7$ estimates, due to the very limited statistics (only two sources). While this could be due to the incompleteness in the spectroscopy follow-up of the candidate sample, as mentioned below, it could also indicate that the redshift evolution observed in the X-ray properties of blazars is associated to another process rather than the IC/CMB, with a different redshift dependence.


We do note that there are a few caveats in the total number of blazars we considered when constraining the models. 
We expect to have selected and identified most of the eRASS:1 blazars in this work. However, there are still some not observed candidates in our sample, which could still be high-$z$ blazars. In particular, sources in the higher redshift bins ($z>4.7$) are more likely to have fainter optical and radio fluxes, for which our spectroscopic follow-up is highly incomplete. This effect would result in a sharp decrease of the number of blazars as a function of redshift compared to the expectations, similar to what is observed in Fig. \ref{fig:ICCMB_pred}. 

At the same time, the blazar classification presented in Sec. \ref{sec:blazar_class} should not be considered conclusive. For example we could have classified as blazars more sources due to the X-ray enhancement of the IC/CMB interaction. As shown in \cite{Krezinger2026}, even with a complementary VLBI and X-ray observations it may be hard to identify relativistic boosting at high redshift.

Adopting the same identification efficiency of $z>4$ given by the targets already observed ($\sim$30\%), we expect $\sim4-5$ sources from the candidates not observed yet to be $z>4$ blazars (i.e. $\alpha_{\rm ox}<1.355$; assuming that they are at $z\approx4$). These candidates have photometric redshifts in between $z\sim3.7-5.7$, therefore their identification could potentially modify the shape of number of blazars detected as a function of redshift in Fig. \ref{fig:ICCMB_pred}.

\section{Summary and Conclusions}
\label{sec:conclusions}

In this work we presented the selection and the identification of new high-$z$ blazars from the first scan of the eRASS survey in the German half of the sky. Our first selection targeted sources detected in eRASS:1 with a radio counterpart from the RACS-mid survey and an optical/NIR counterpart in DELVE that is point-like and exhibits a dropout in at least one observed band. The total sky area covered by this selection amounts to approximately $\sim 14,000$ deg$^2$.

We subsequently refined the candidate list by incorporating additional optical and NIR information from complementary surveys, including Gaia, Pan-STARRS, DES, SkyMapper, VHS, VIKING, and CatWISE. This multi-wavelength filtering resulted in a final sample of 153 high-$z$ blazar candidates, of which 20 are quasars at $z>4$ already known in the literature.

We performed spectroscopic observations on 46 of the remaining candidates and identified 39 new $z>3.4$ quasars, 14 of which at $z>4$. The newly discovered $z>4$ consists in a $\sim65$\% of the total number of quasars with similar properties in the same sky area. The quasars uncovered in this work correspond to the most extreme sources currently known. In particular, when compared to the wider population of quasars (from \citealt{Lusso2020}), the sources in the high-$z$ eRASS sample discussed in this work present the highest X-ray-to-UV emission ratios ($\alpha_{\rm ox}$). This is a consequence of their X-ray selection and the excess high-energy radiation observed can be attributed to the relativistically boosted radiation from the relativistic jets, that is, most sources in the sample are blazars. At the same time, given the soft X-ray sensitivity of the eRASS observatory, the excess in the high-energy radiation can also be due super-Eddington accretion, as observed in a few recent examples \citep[e.g.][]{Ighina2025b,Obuchi2026}.
X-ray and optical/NIR follow-up observations on this eRASS-selected sample will be crucial to constrain both the jet orientations and the accretion properties of these extreme high-$z$ quasars. 

We used the available X-ray information to provide a first classification of the selected sources.  Following \cite{Ighina2019}, we classified objects with $\tilde{\alpha}_{\rm ox}<1.355$ as blazars. This criterion yields 26 blazars in our sample, which we then used to constrain the X-ray space density evolution models described in \cite{Ighina2021b}. By comparing the observed number of blazars as a function of redshift with model predictions, we rule out scenarios in which the X-ray-to-radio luminosity ratio of blazars does not evolve with redshift. Observations are broadly consistent with the predictions from the fractional IC/CMB model. Nevertheless, some discrepancies remain, particularly in the total number of $z>4$ blazars and at the highest redshifts ($z>4.7$). We emphasize that these differences may arise from the incomplete spectroscopic follow-up of the candidate sample and from the blazar classification being based solely on X-ray properties.

Once the full candidate sample is spectroscopically confirmed, we will be able to place significantly tighter constraints on IC/CMB and alternative models describing the cosmological evolution of blazars. Moreover, the additional eRASS scans will provide progressively deeper X-ray coverage, enabling improved constraints on the blazar luminosity function and space density at even higher redshifts.

\section*{Acknowledgements}
{\small
We thank the anonymous referee for their useful comments. L.I. thanks all the staff and the astronomers at the Roque de los Muchachos and La Silla observatories for the great experience during the observing runs.
Support for L.I.'s  work was provided by the National Aeronautics and Space Administration through Chandra Award Number GO3-24069X issued by the Chandra X-ray Observatory Center, which is operated by the Smithsonian Astrophysical Observatory for and on behalf of the National Aeronautics Space Administration under contract NAS8-03060.
T.C. acknowledges support from NASA Contract NAS8-03060 to the \textit{Chandra} X-ray Center.
L.I., L. P., A.C. and A.M. acknowledge financial support from INAF under the projects ``Quasar jets in the early Universe'' (Ricerca Fondamentale 2022) and ``Testing the obscuration in the early Universe'' (Ricerca Fondamentale 2023).

Based on observations obtained at the international Gemini Observatory, a program of NSF NOIRLab (ID GS-2025A-Q-414), which is managed by the Association of Universities for Research in Astronomy (AURA) under a cooperative agreement with the U.S. National Science Foundation on behalf of the Gemini Observatory partnership: the U.S. National Science Foundation (United States), National Research Council (Canada), Agencia Nacional de Investigaci\'{o}n y Desarrollo (Chile), Ministerio de Ciencia, Tecnolog\'{i}a e Innovaci\'{o}n (Argentina), Minist\'{e}rio da Ci\^{e}ncia, Tecnologia, Inova\c{c}\~{o}es e Comunica\c{c}\~{o}es (Brazil), and Korea Astronomy and Space Science Institute (Republic of Korea).

This paper includes data gathered with the 6.5 meter Magellan Telescopes located at Las Campanas Observatory, Chile.

We acknowledge the support from the LBT-Italian Coordination Facility for the execution of the observations. The LBT is an international collaboration among institutions in the United States, Italy and Germany. LBT Corporation partners are: The University of Arizona on behalf of the Arizona Board of Regents; Istituto Nazionale di Astrofisica, Italy; LBT Beteiligungsgesellschaft, Germany, representing the Max-Planck Society, The Leibniz Institute for Astrophysics Potsdam, and Heidelberg University; The Ohio State University, representing OSU, University of Notre Dame, University of Minnesota and University of Virginia. This research used the facilities of the Italian Center for Astronomical Archive (IA2) operated by INAF at the Astronomical Observatory of Trieste.

Based on observations made with ESO Telescopes at the La Silla Paranal Observatory under programme ID 115.28BJ.001.

This research made use of Astropy (\url{http://www.astropy.org}) a community-developed core Python package for Astronomy \citep{astropy2018}.

}
\section*{Data Availability}

 Reprocessed data are available upon reasonable request to the corresponding author.



\bibliographystyle{mnras}
\bibliography{ref} 




\appendix

\section{Optical Spectra of Candidates}

In this section we report the discovery spectra of the new high-$z$ quasars selected as part of this work. The spectra are shown in Figs. \ref{fig:optical_spectra1}--\ref{fig:optical_spectra5}. Additionally, we show the spectra of candidate objects that we could not confirm to be high-redshift sources in Fig. \ref{fig:contaminants}.

\begin{figure*}
    \includegraphics[width=0.49\hsize]{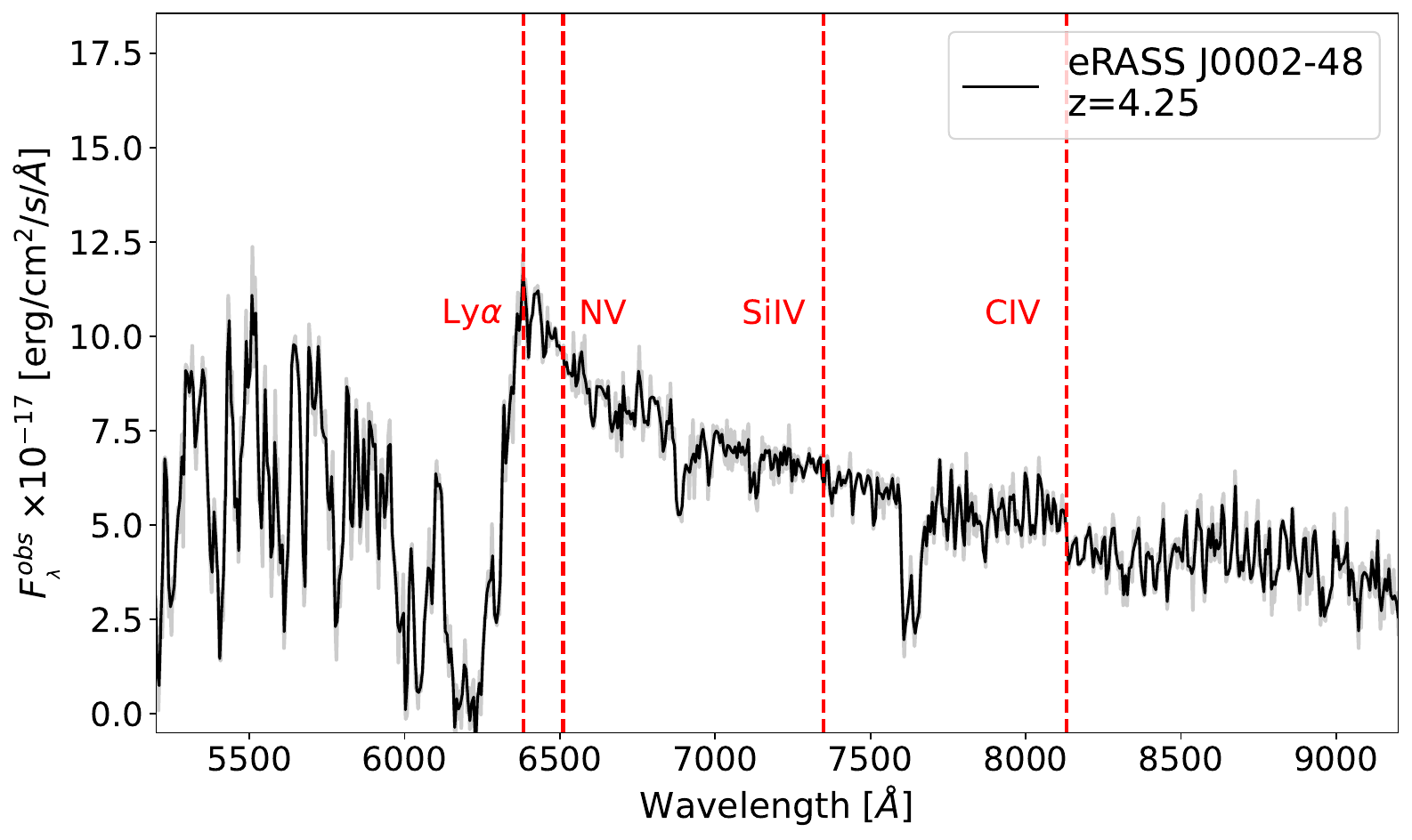}
    \includegraphics[width=0.49\hsize]{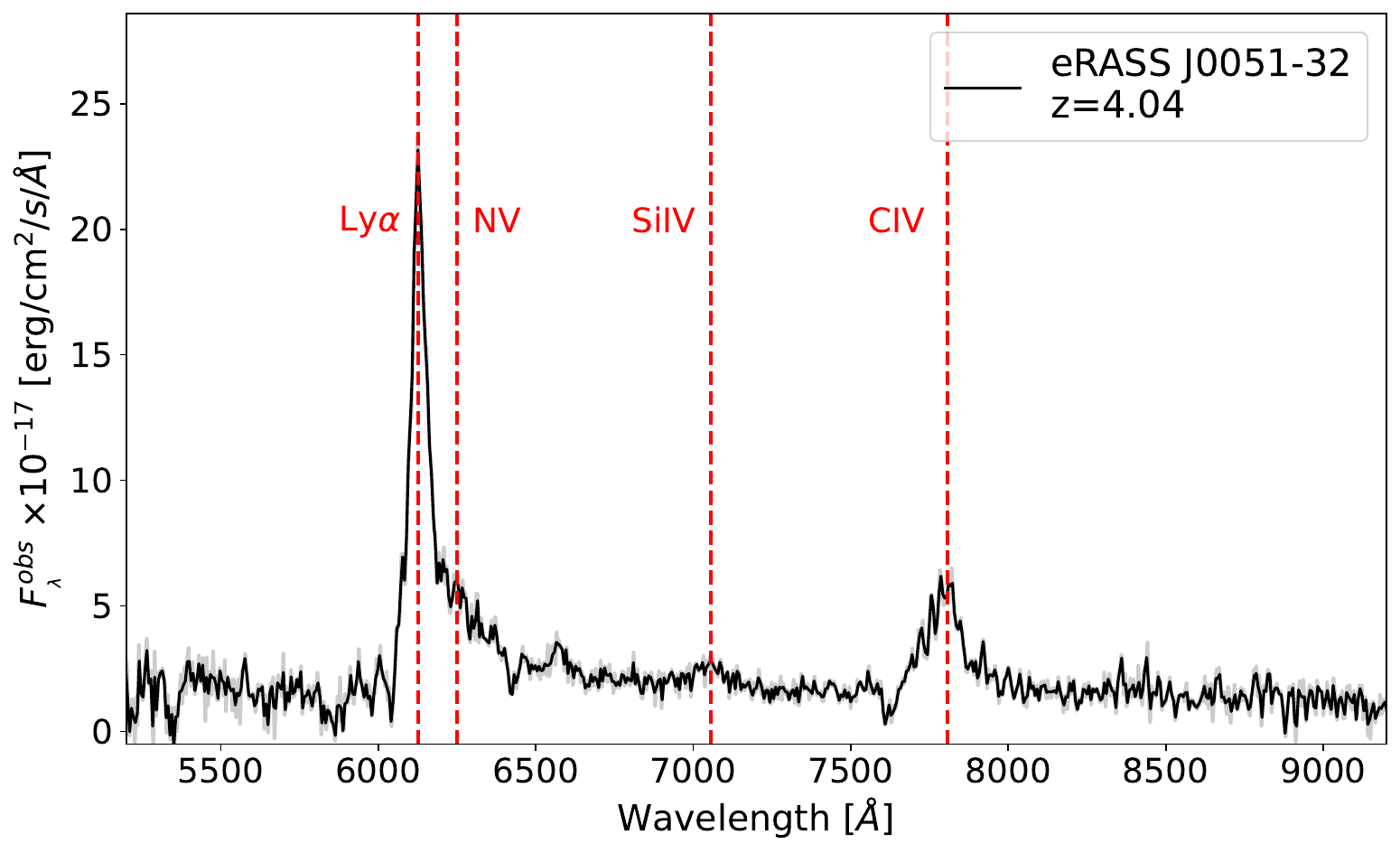}
	\includegraphics[width=0.49\hsize]{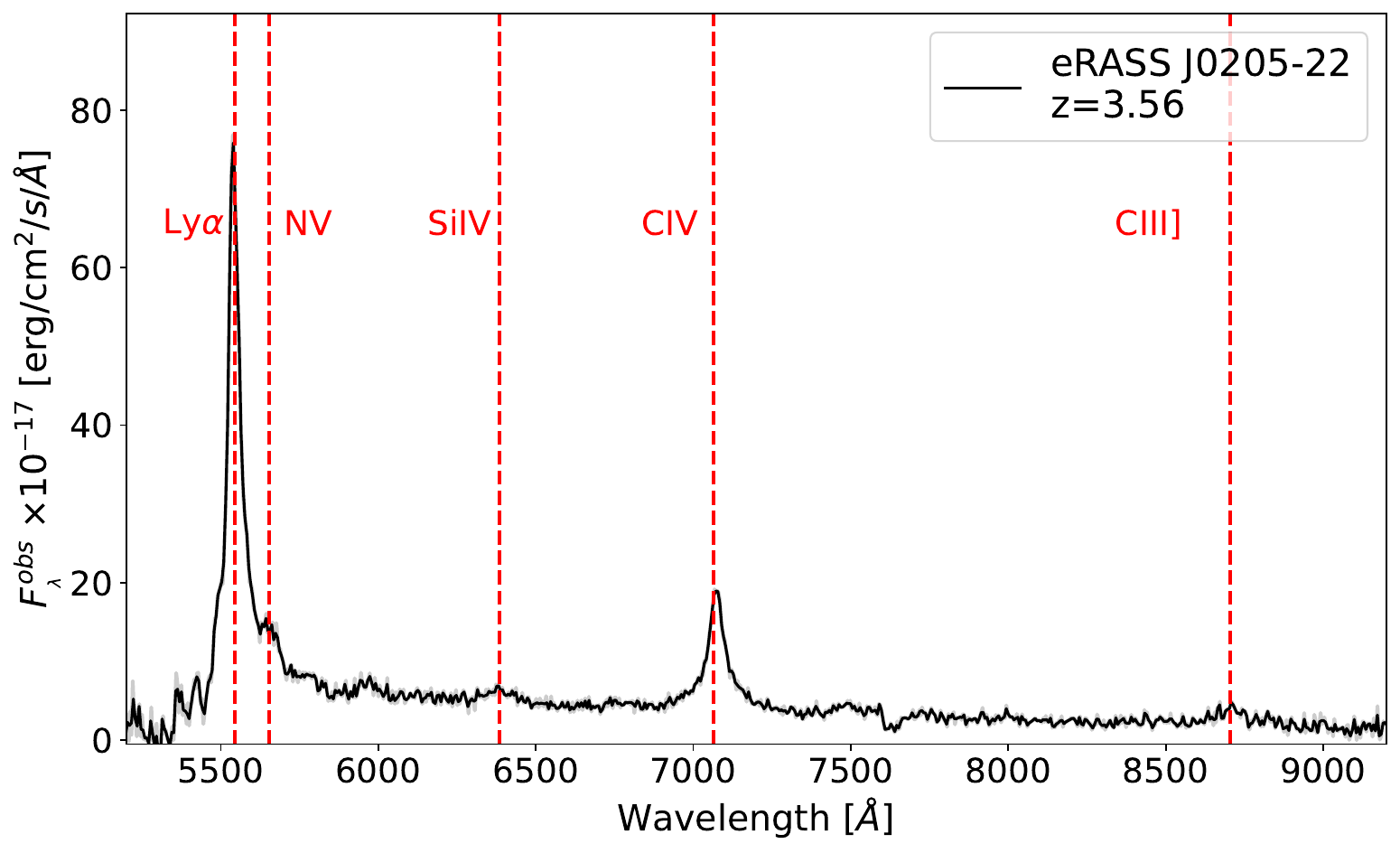}
	\includegraphics[width=0.49\hsize]{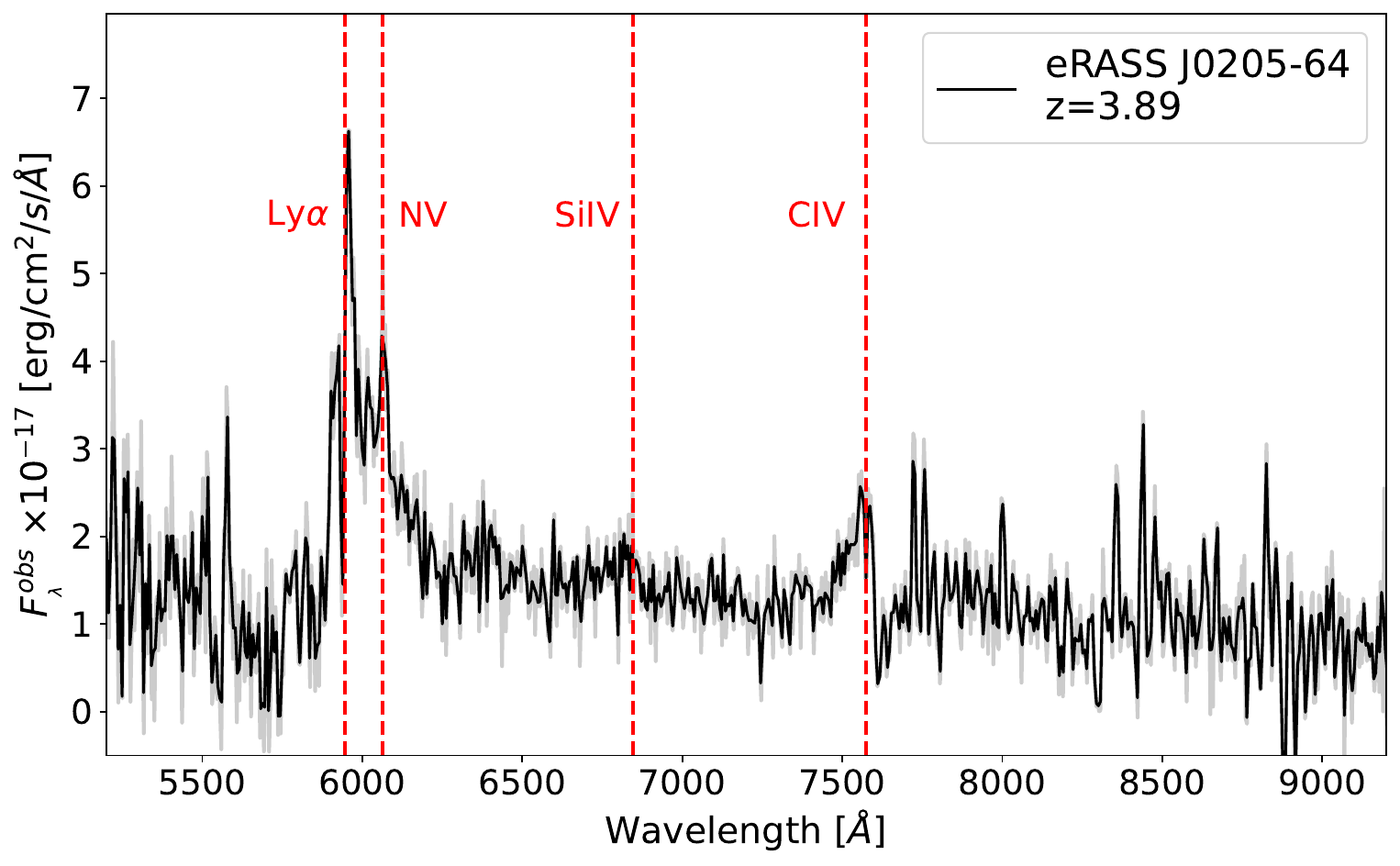}
	\includegraphics[width=0.49\hsize]{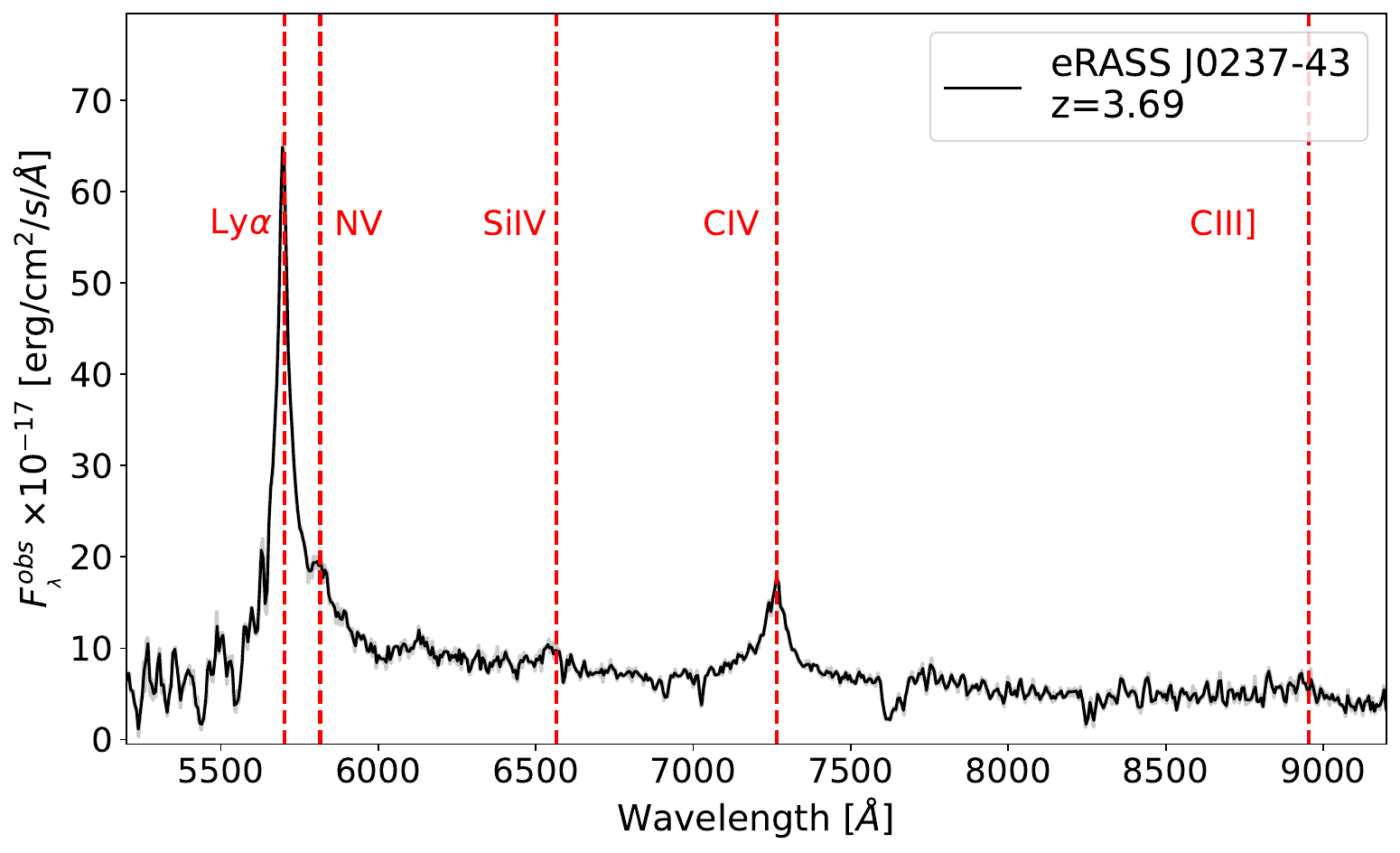}
	\includegraphics[width=0.49\hsize]{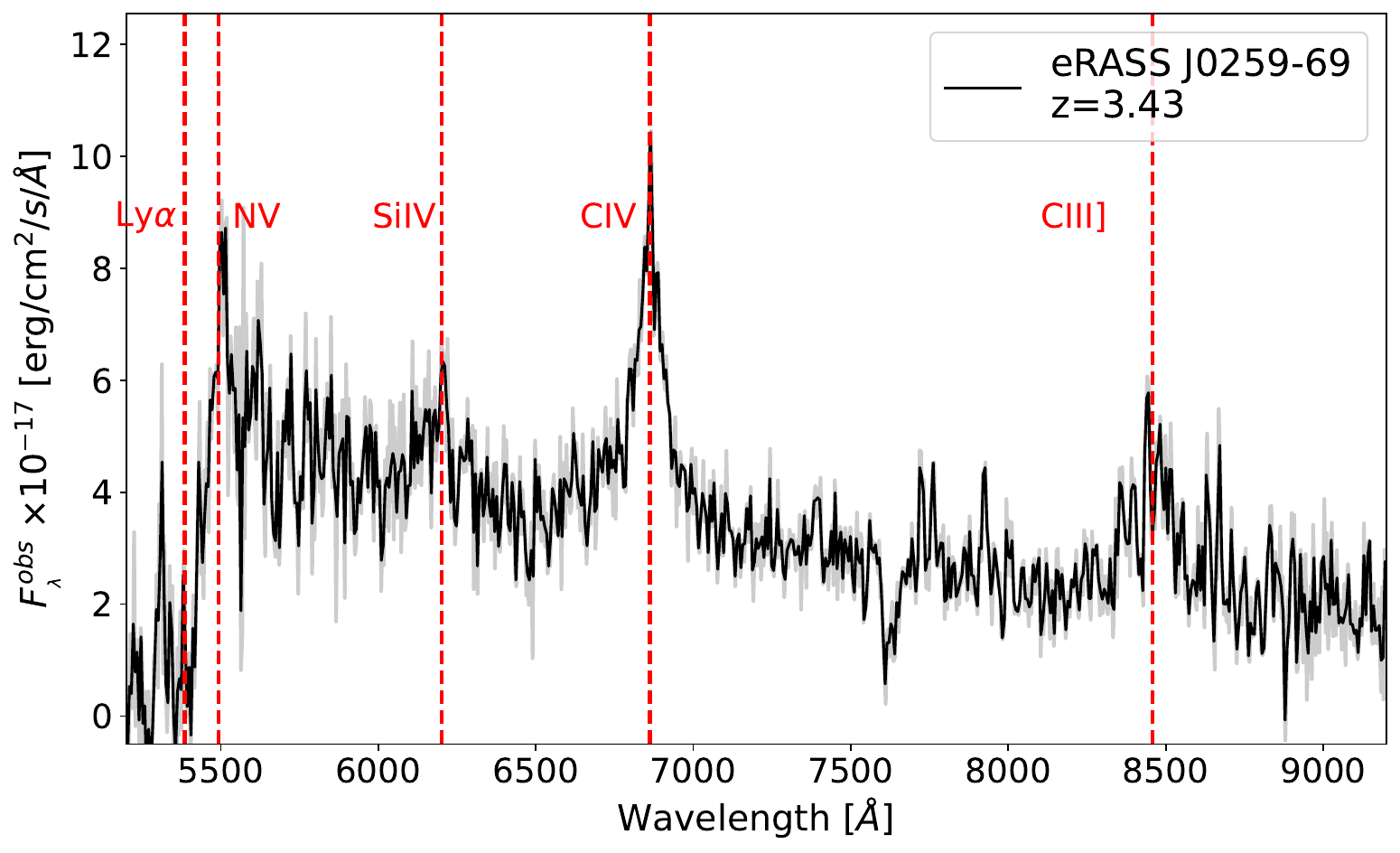}
	\includegraphics[width=0.49\hsize]{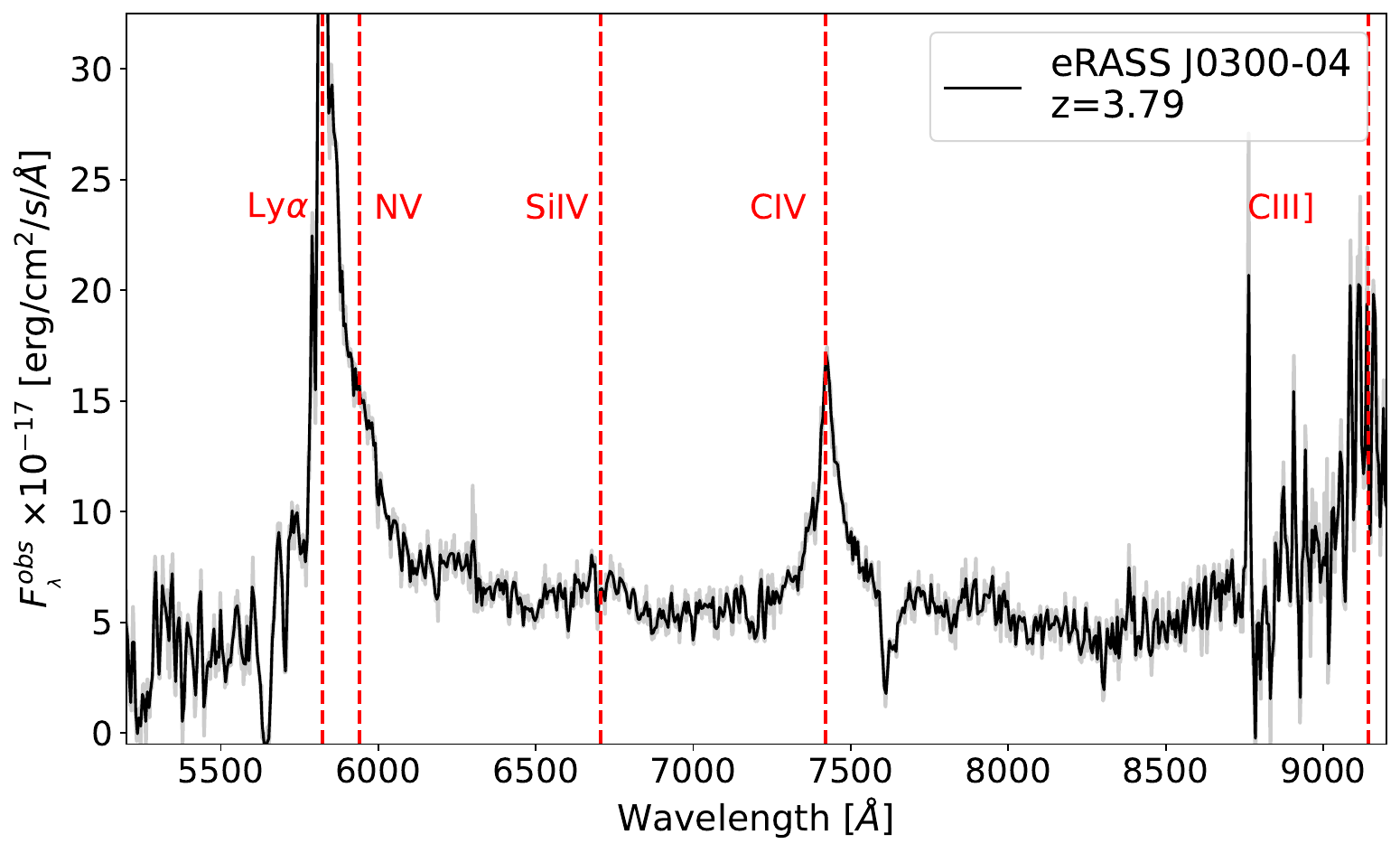}
	\includegraphics[width=0.49\hsize]{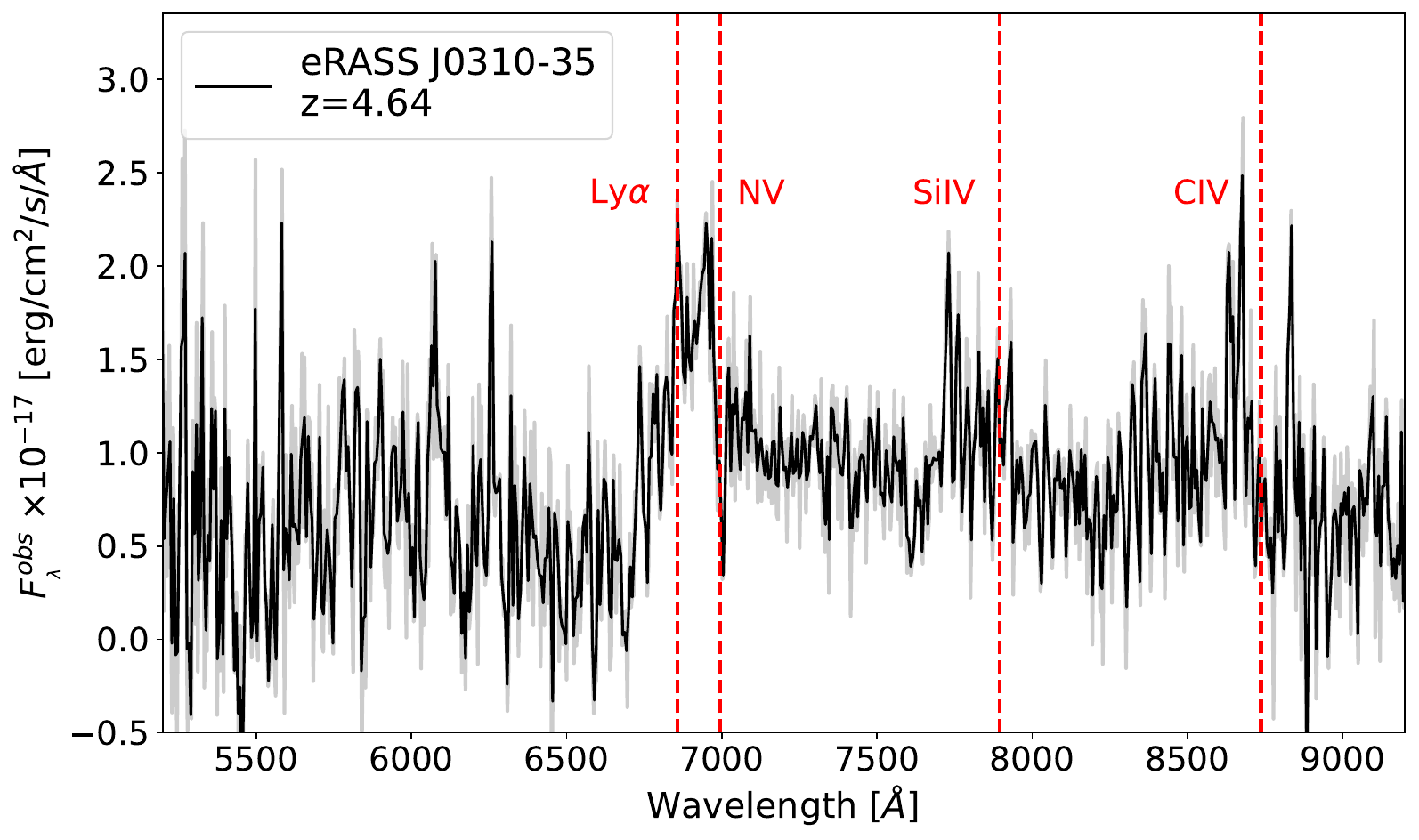}
    
    \caption{Discovery spectra of the high-$z$ quasars newly identified in this work. Vertical dashed lines indicate the expected wavelength of redshifted emission lines in the observed range. }
    \label{fig:optical_spectra1}
\end{figure*}

\begin{figure*}
	\includegraphics[width=0.49\hsize]{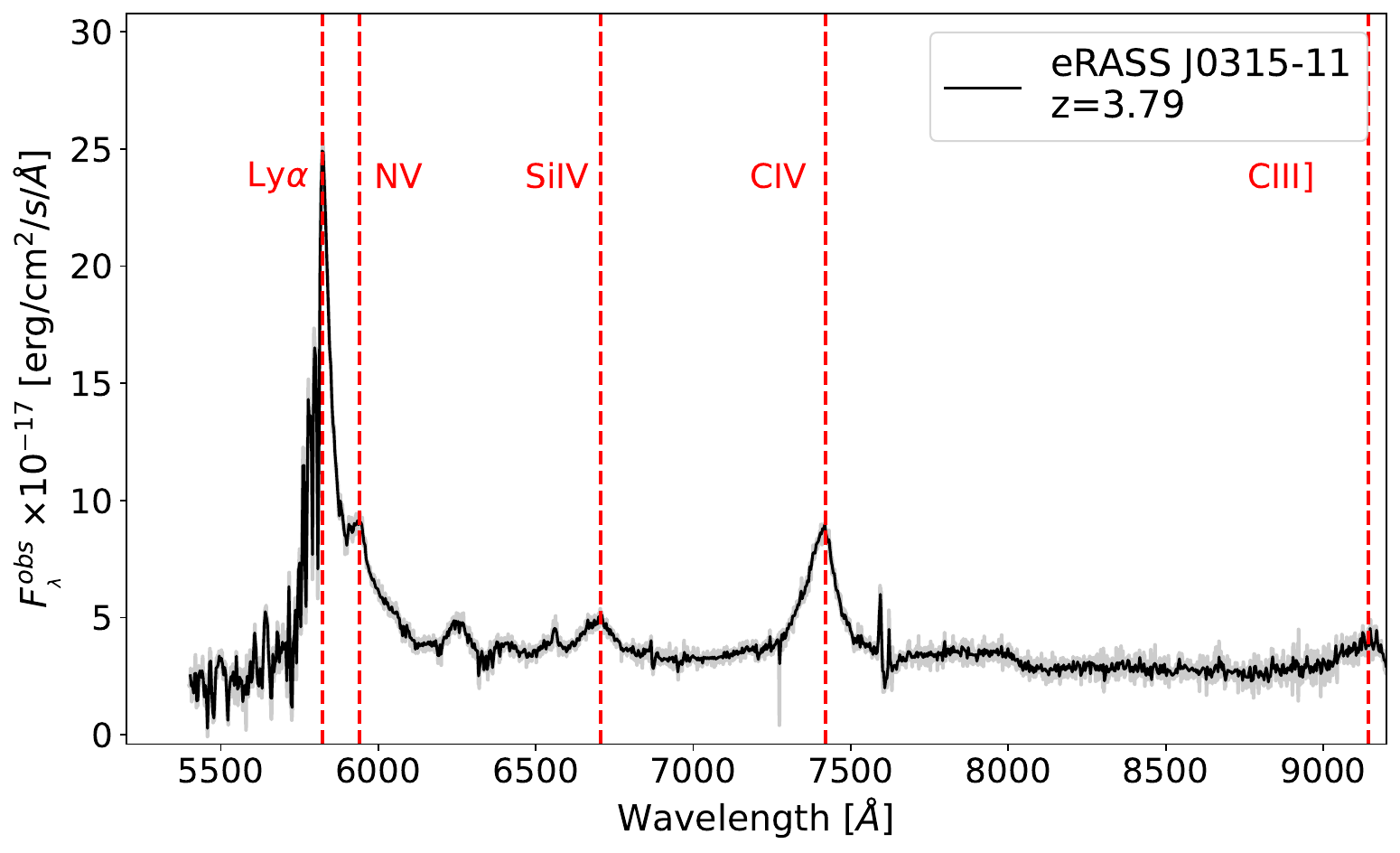}
	\includegraphics[width=0.49\hsize]{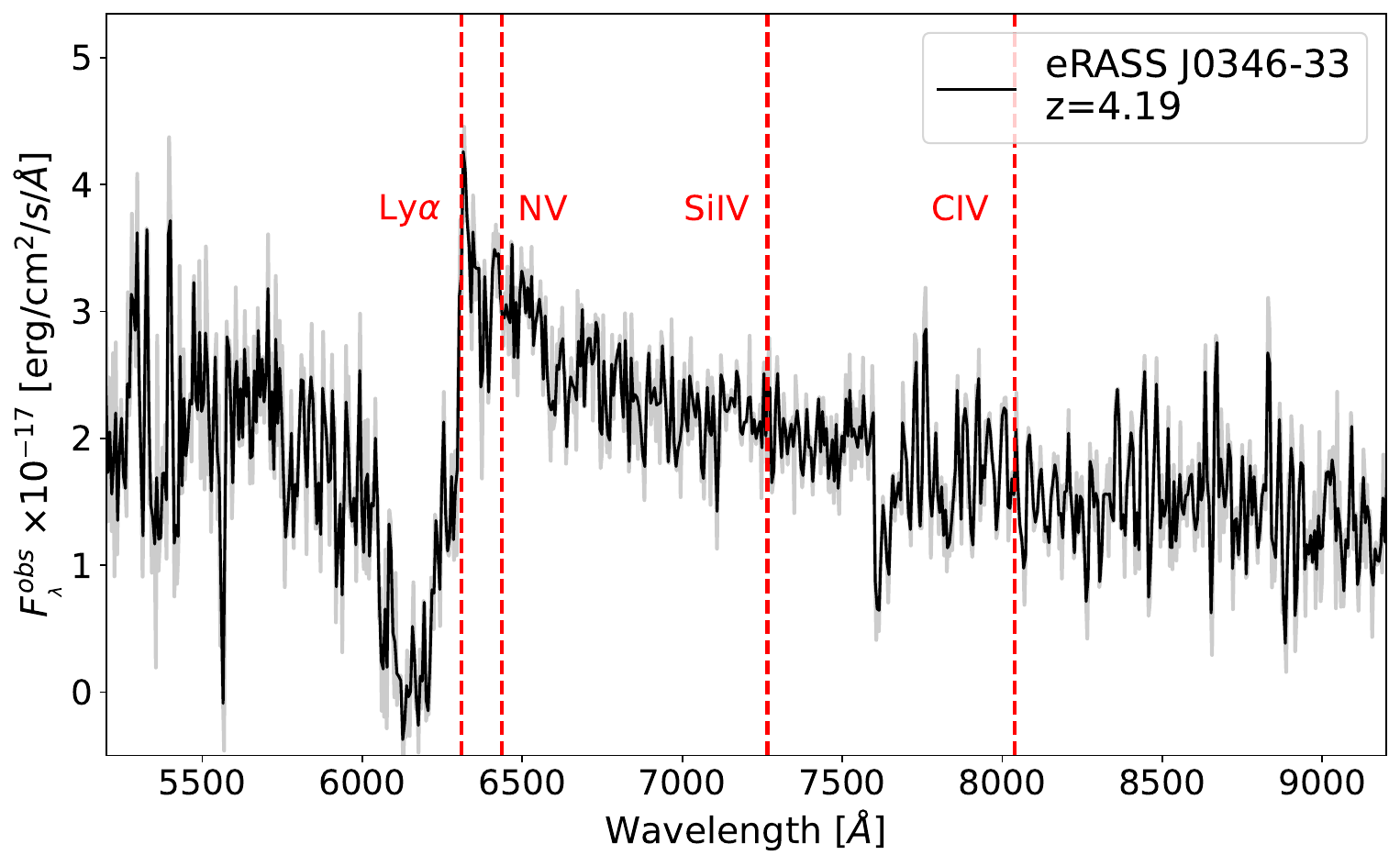}
	\includegraphics[width=0.49\hsize]{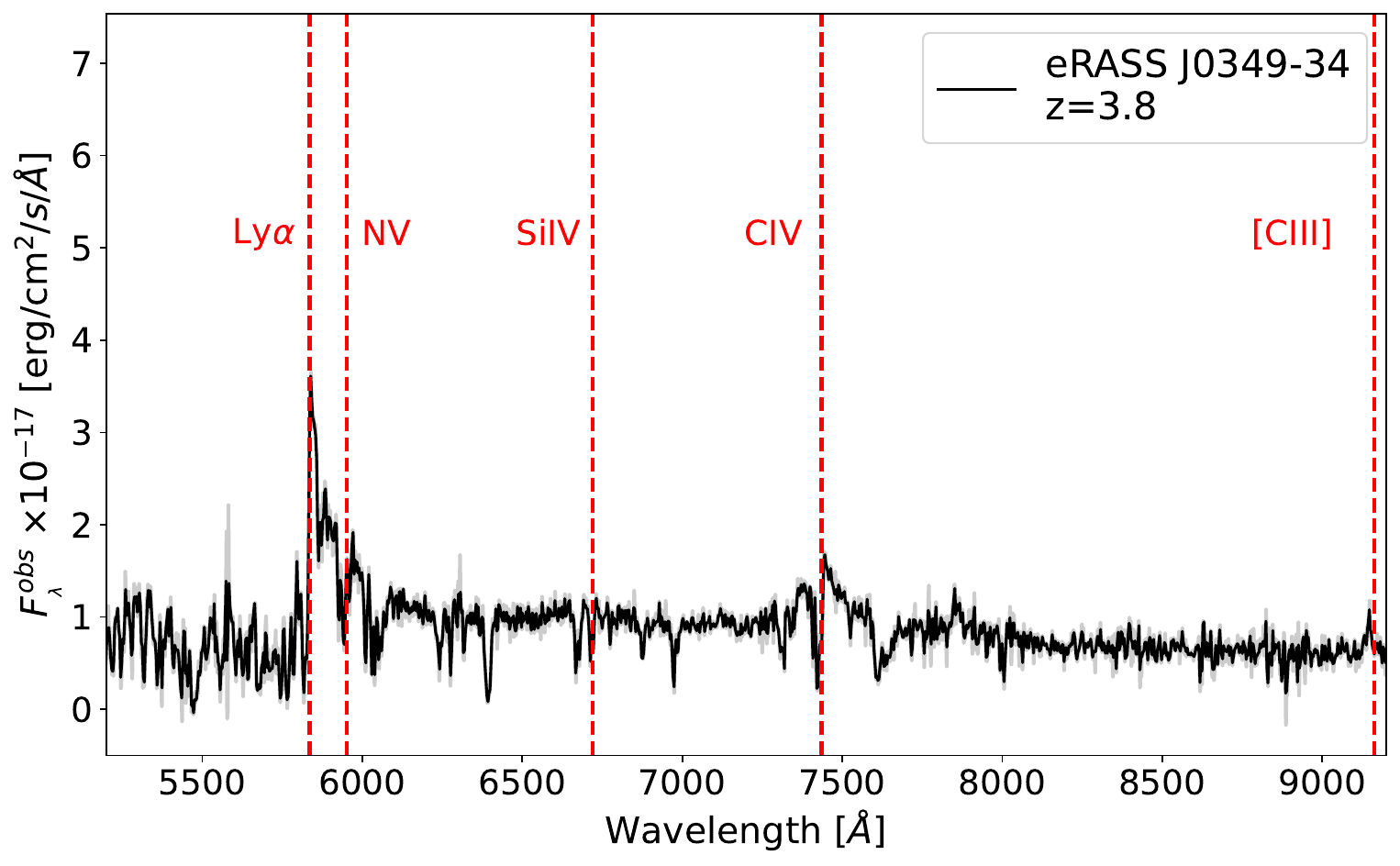}
	\includegraphics[width=0.49\hsize]{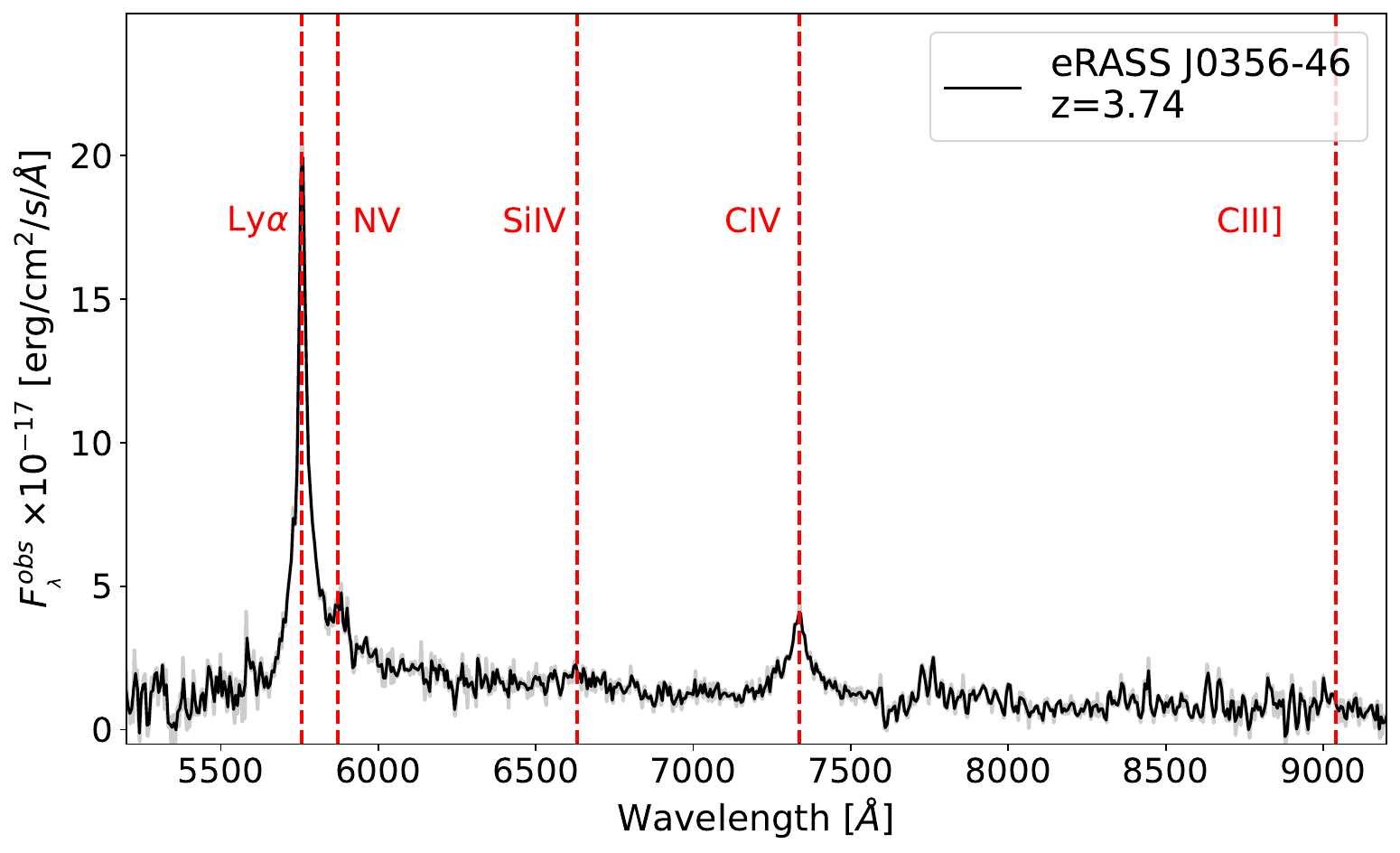}
	\includegraphics[width=0.49\hsize]{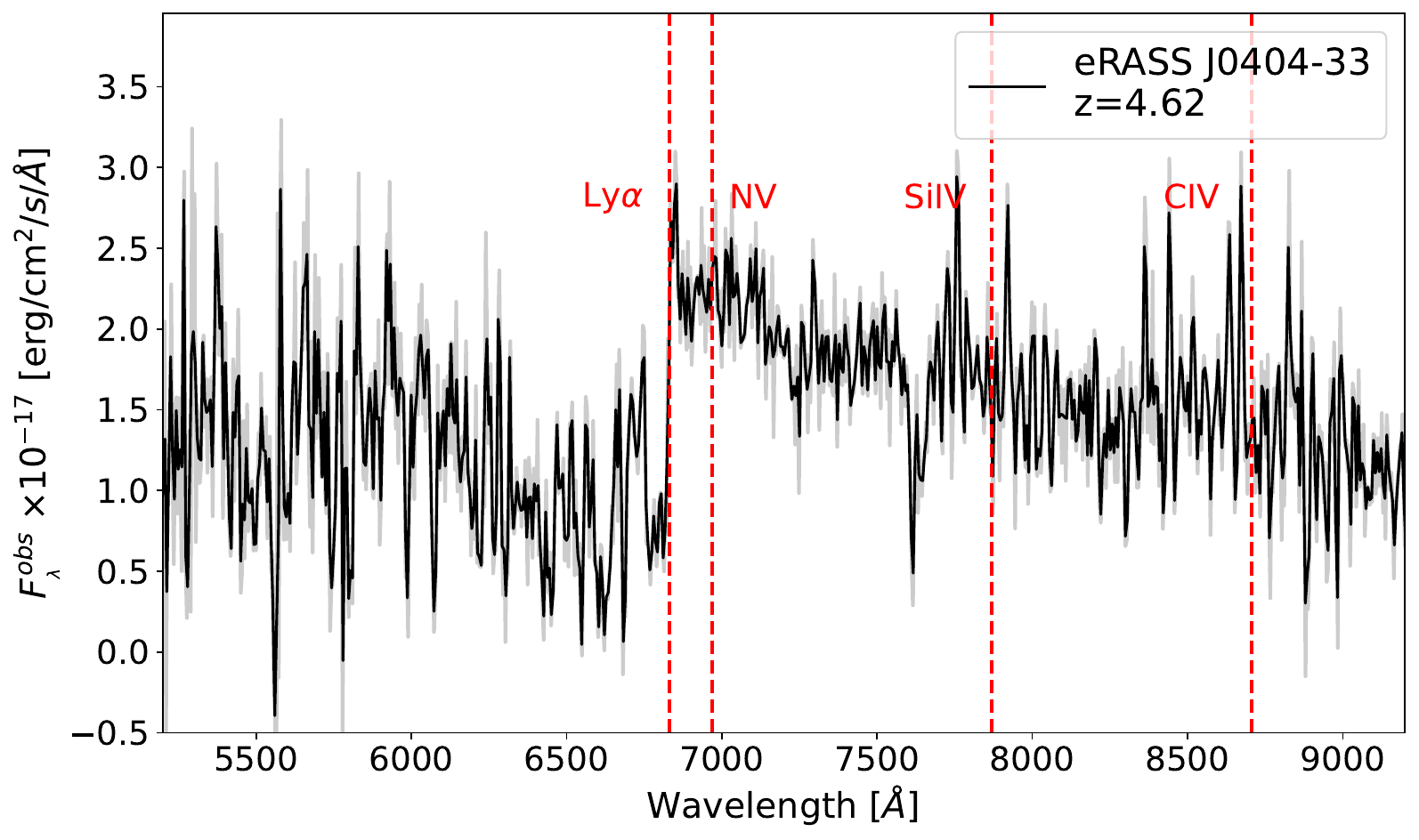}
	\includegraphics[width=0.49\hsize]{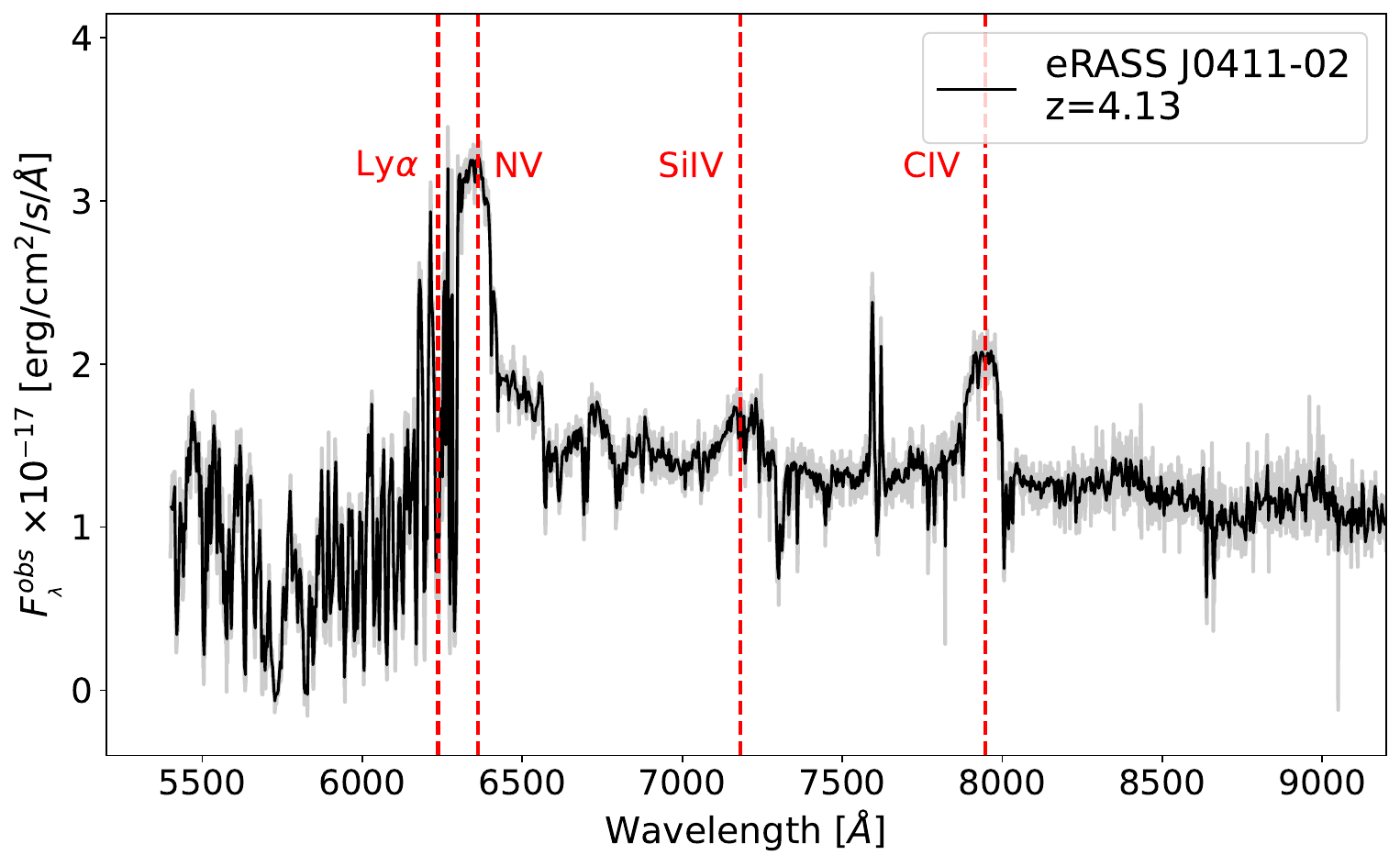}
	\includegraphics[width=0.49\hsize]{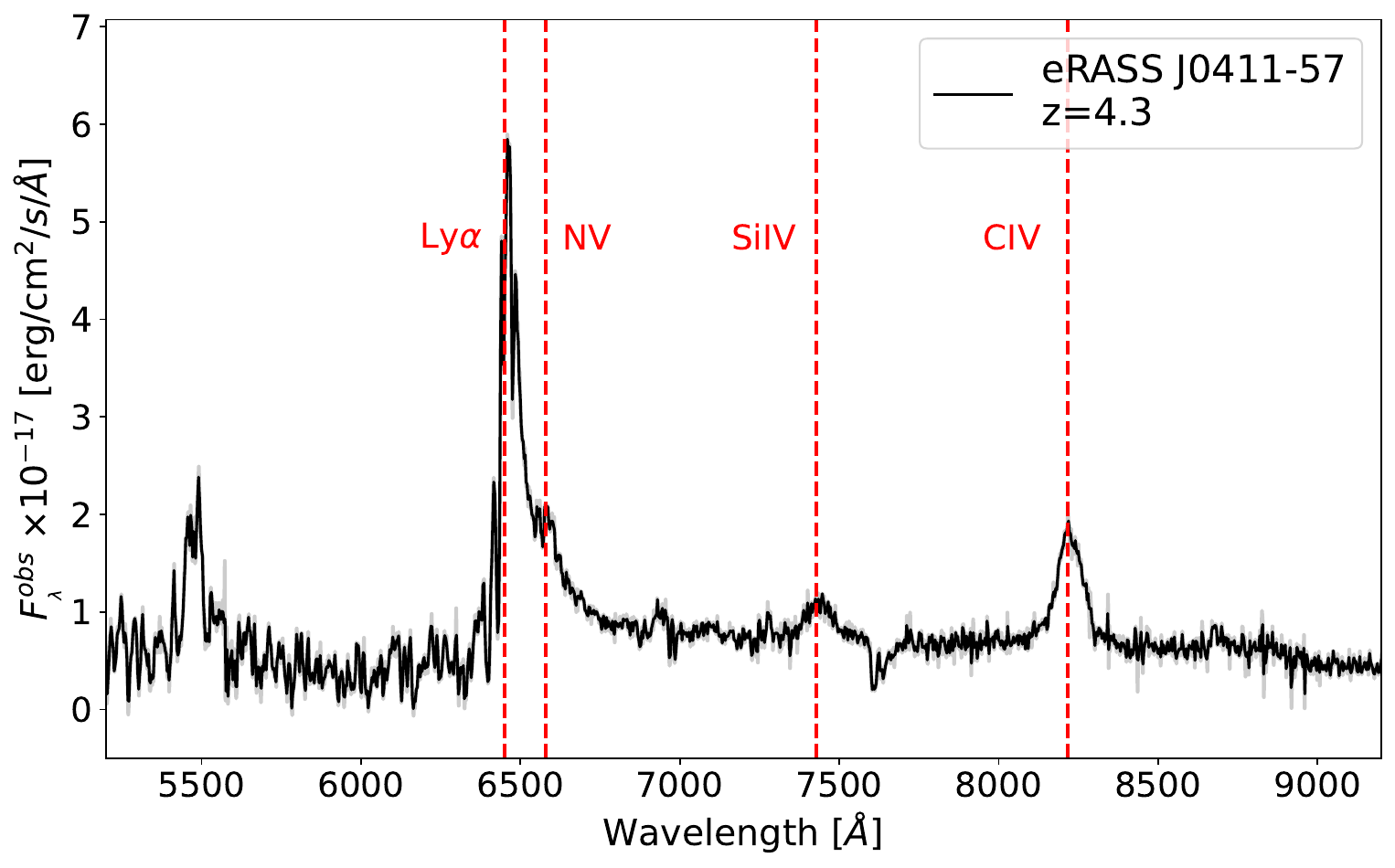}
	\includegraphics[width=0.49\hsize]{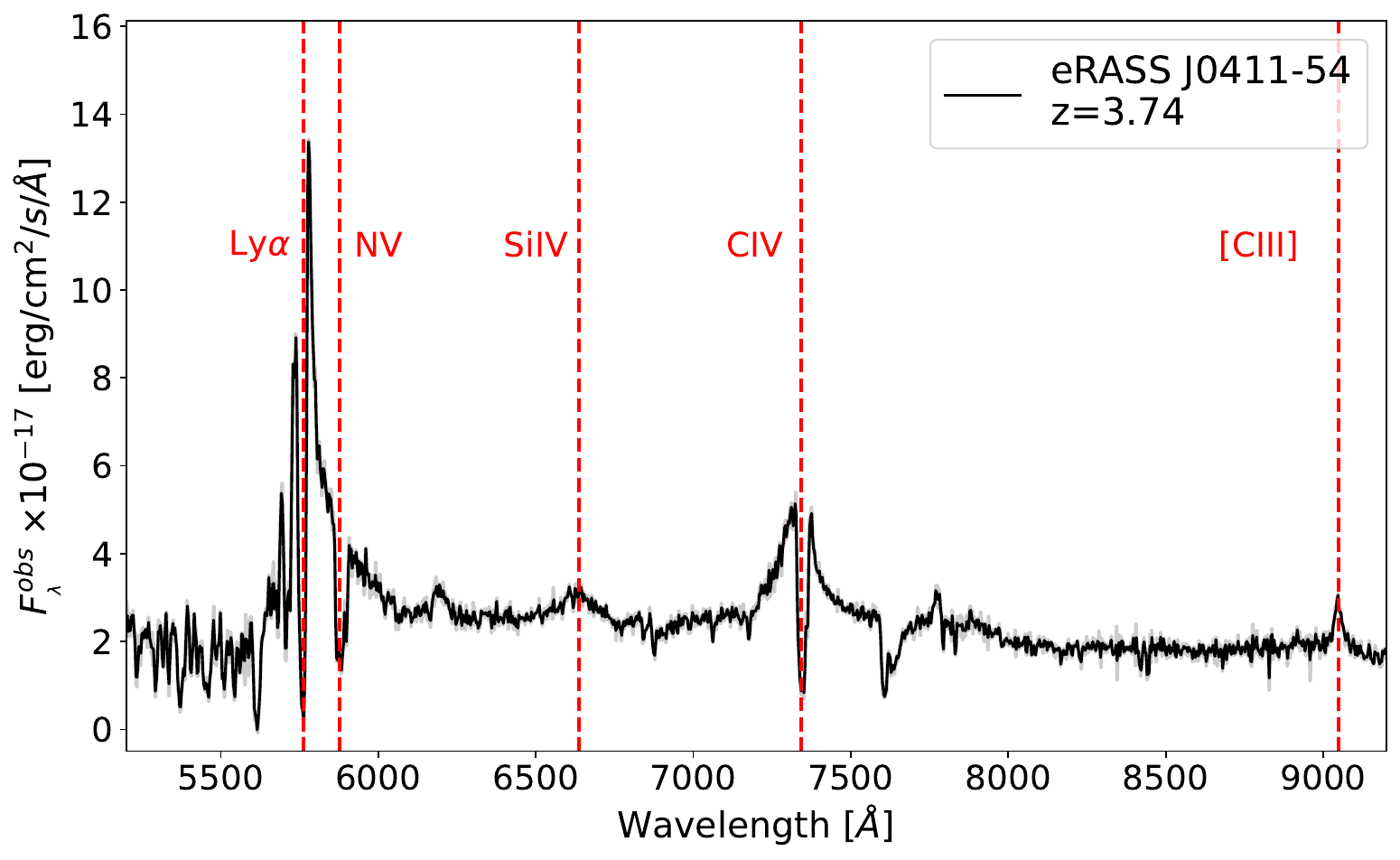}

    \caption{Continued.}
    \label{fig:optical_spectra2}
\end{figure*}

\begin{figure*}
	\includegraphics[width=0.49\hsize]{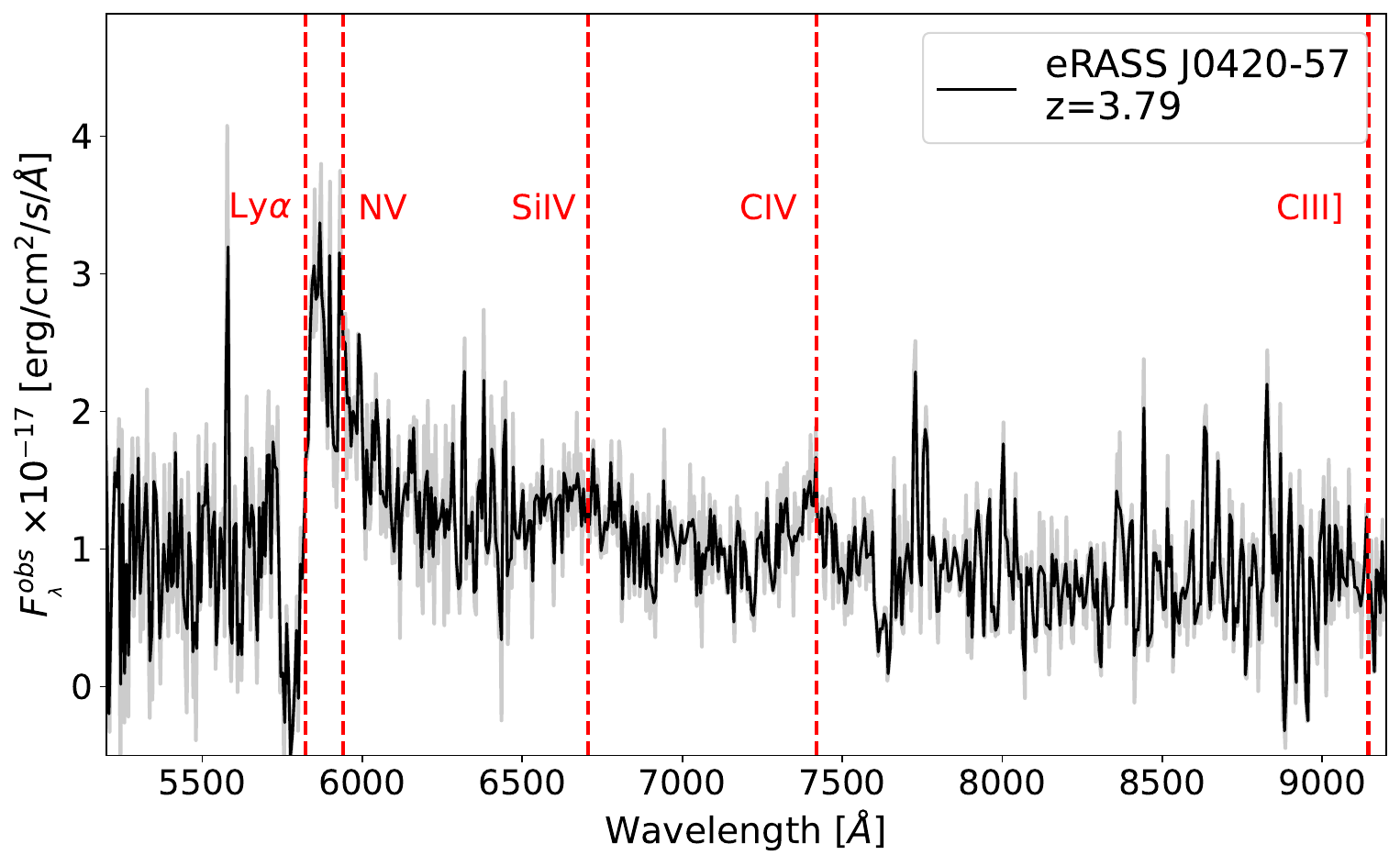}
	\includegraphics[width=0.49\hsize]{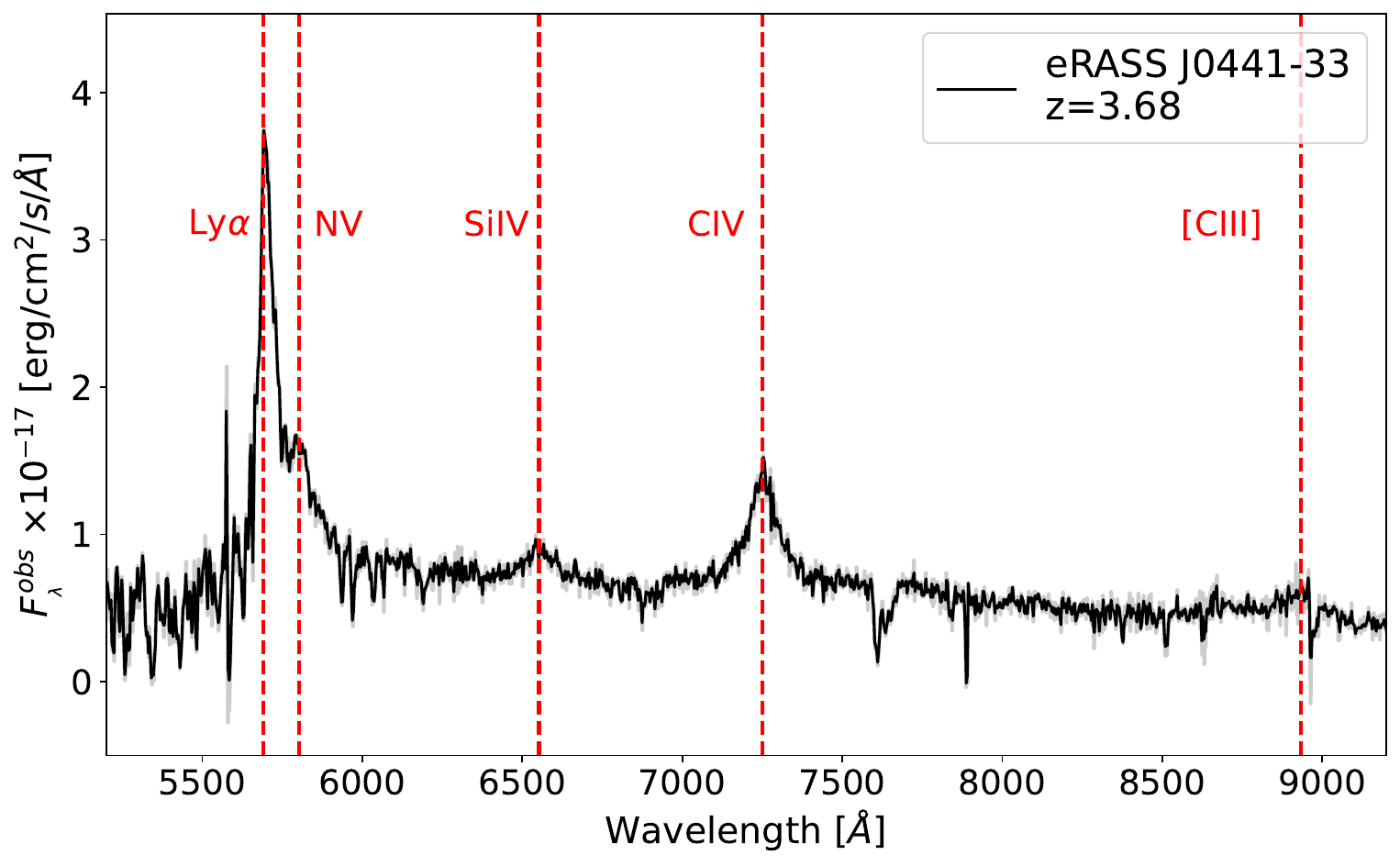}
	\includegraphics[width=0.49\hsize]{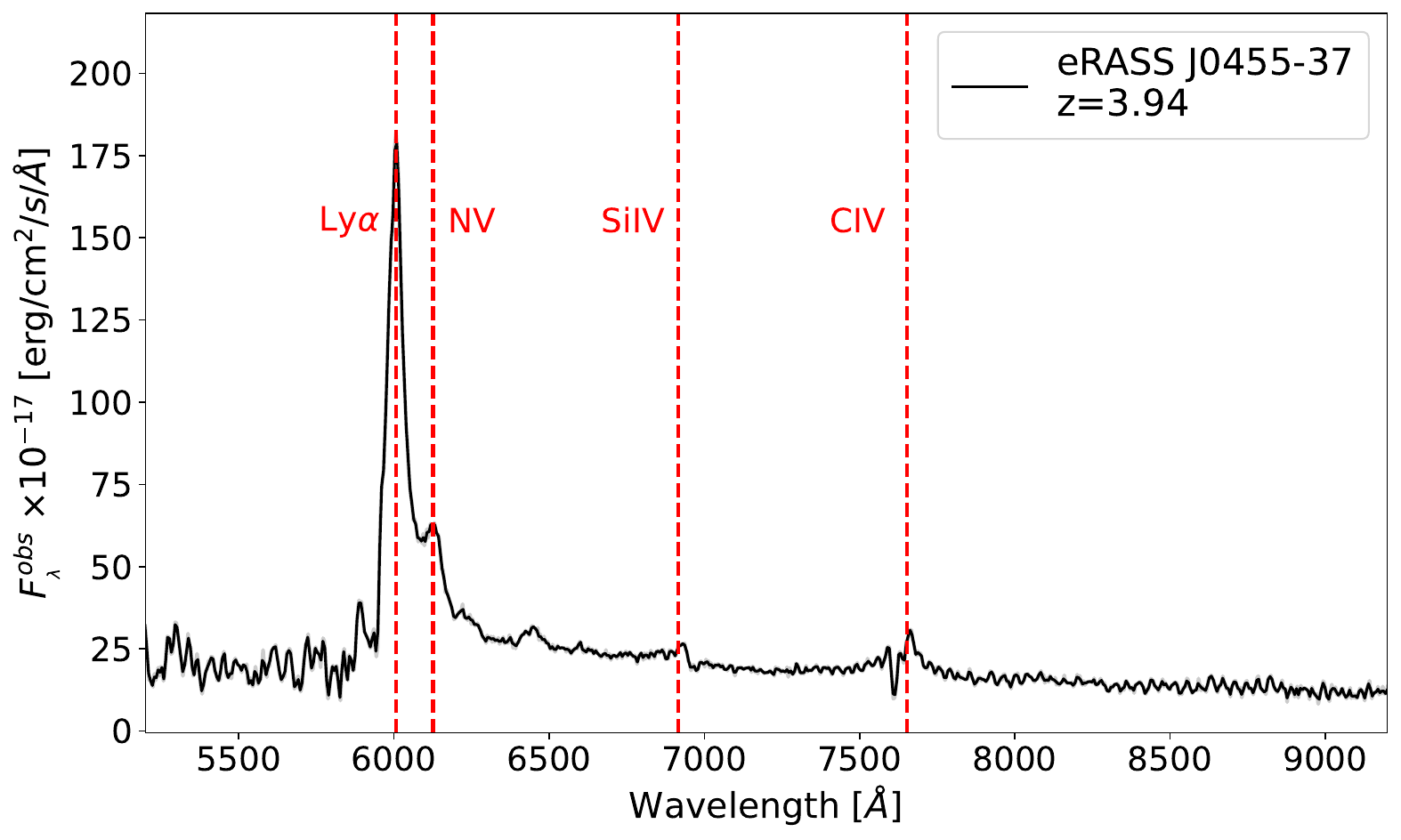}
	\includegraphics[width=0.49\hsize]{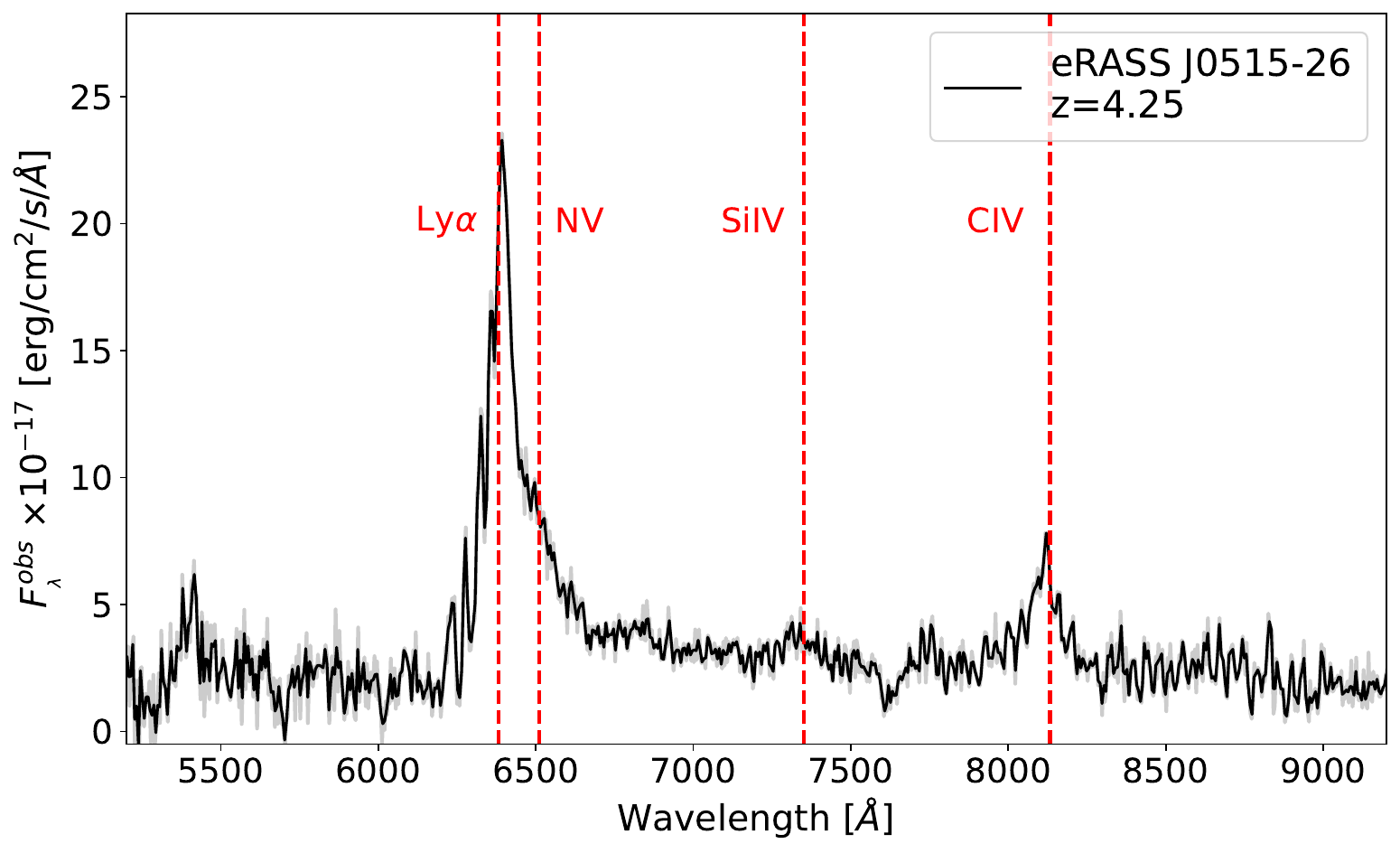}
	\includegraphics[width=0.49\hsize]{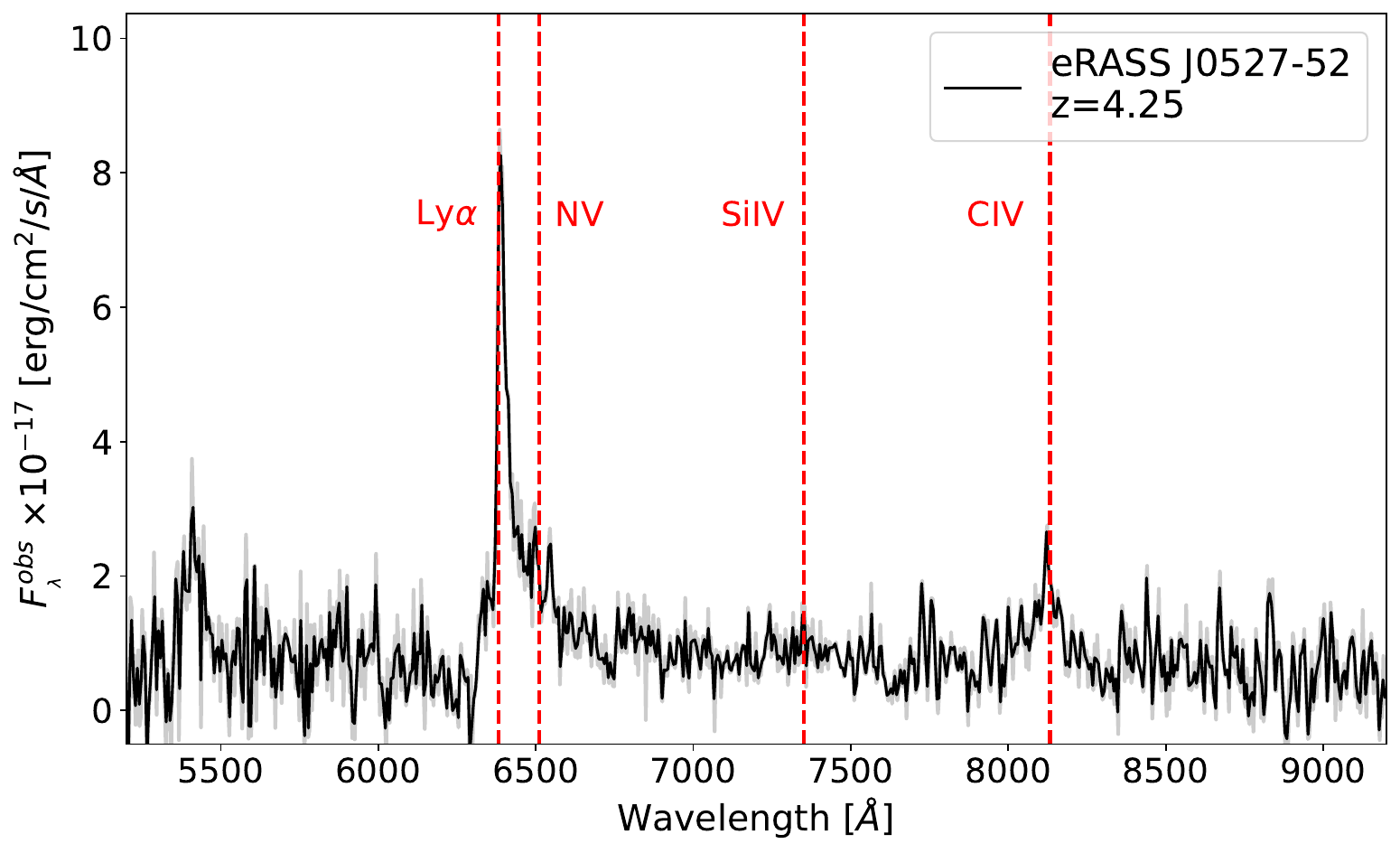}
	\includegraphics[width=0.49\hsize]{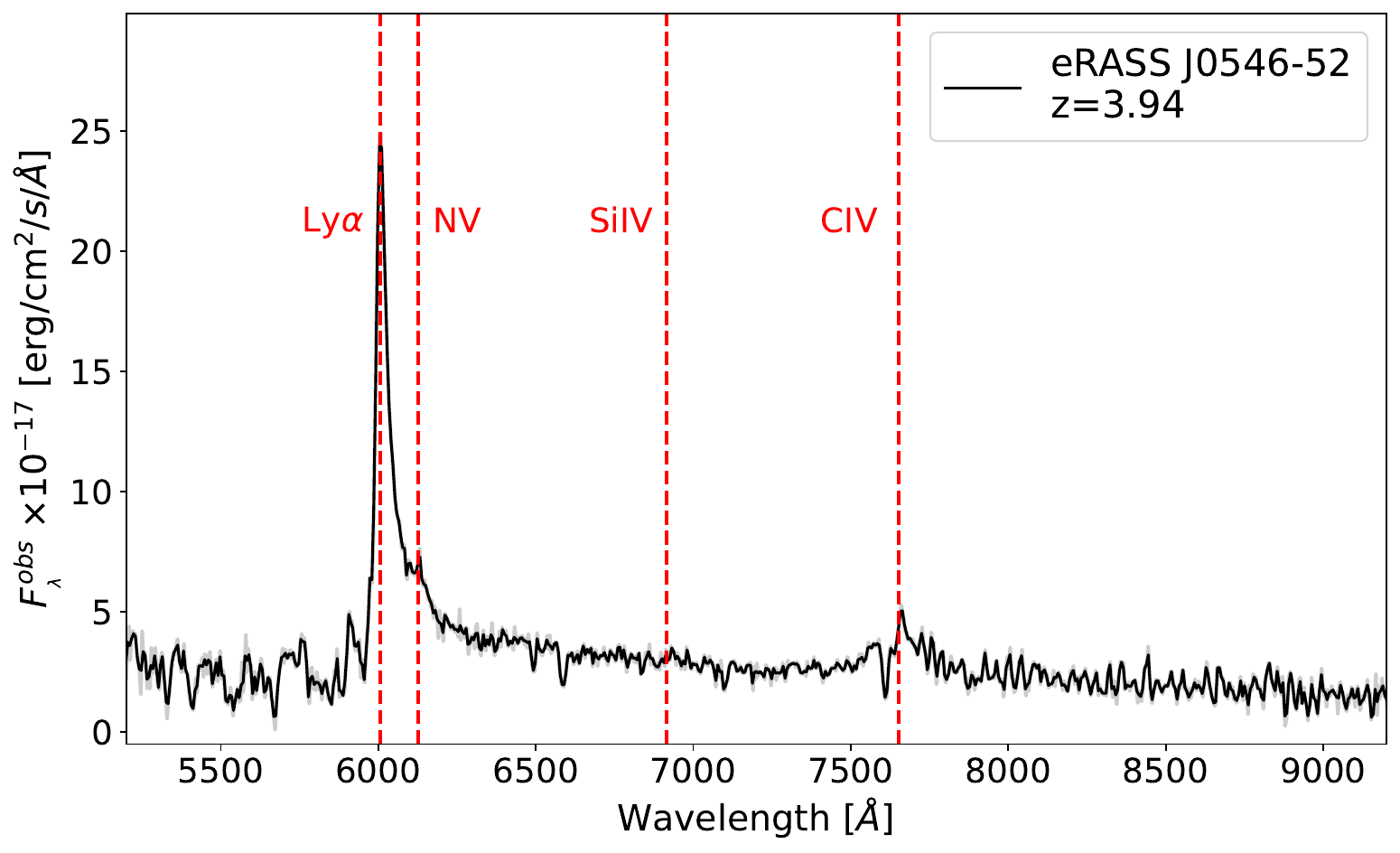}
	\includegraphics[width=0.49\hsize]{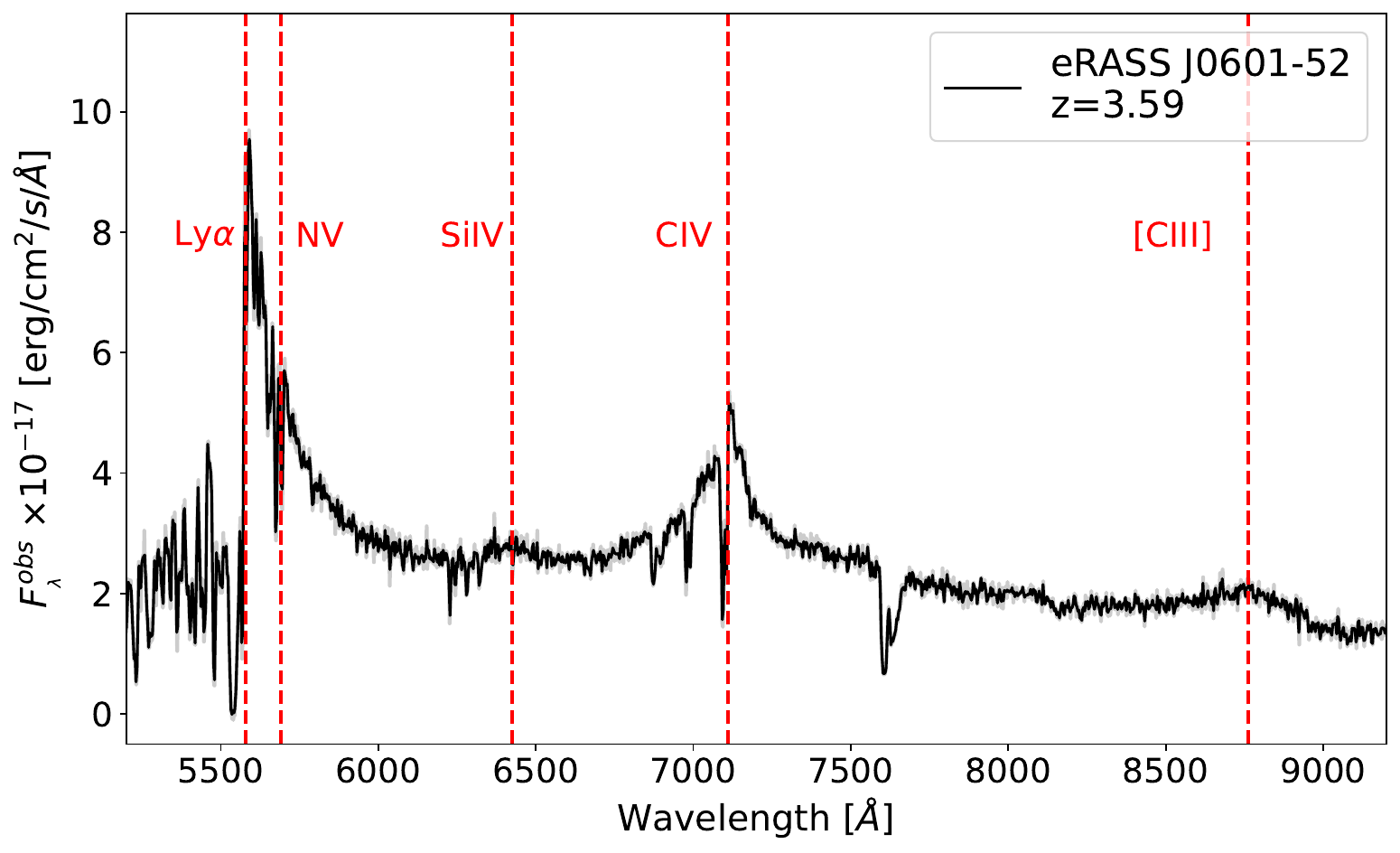}
	\includegraphics[width=0.49\hsize]{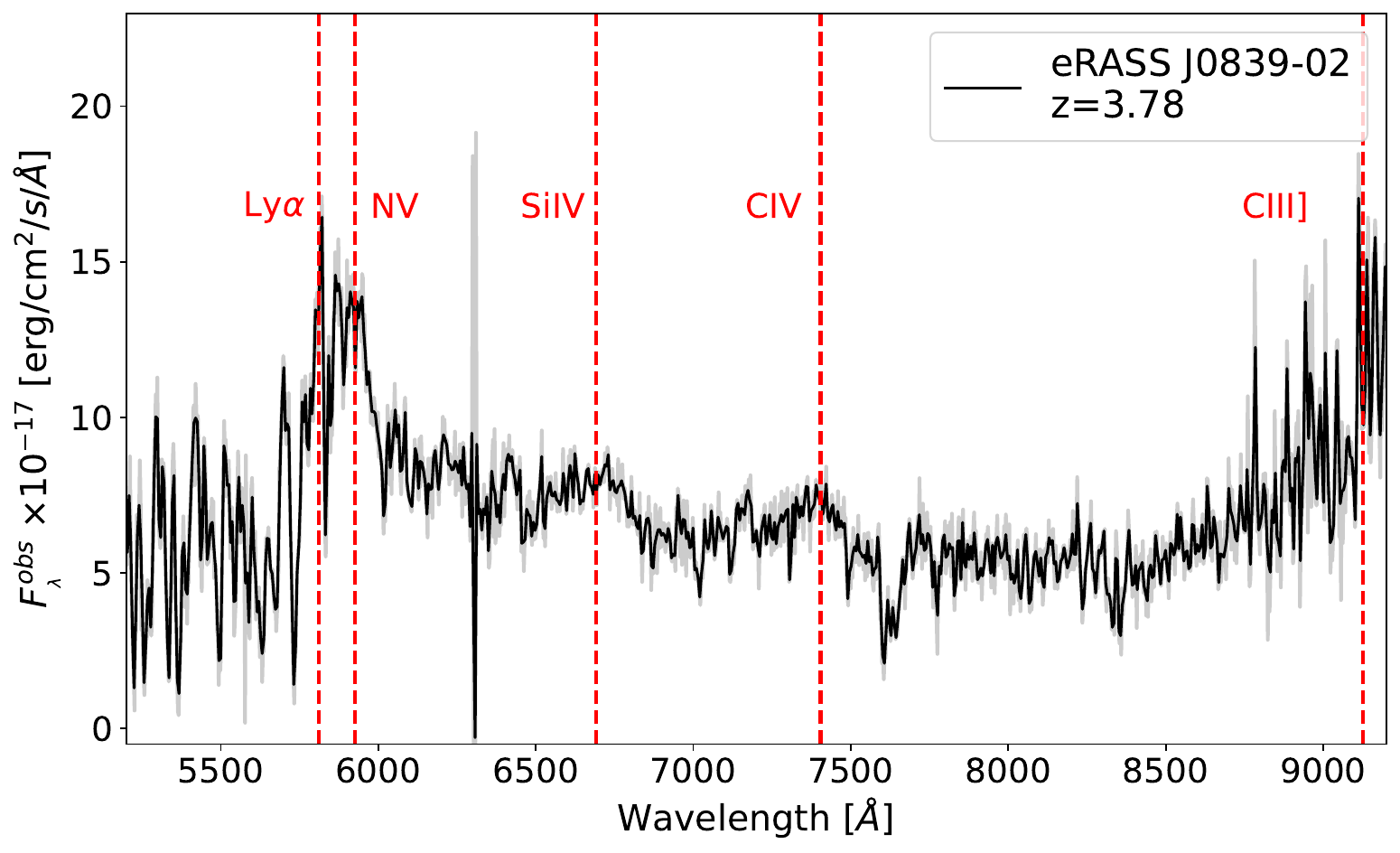}

    \caption{Continued.}
    \label{fig:optical_spectra3}
\end{figure*}

\begin{figure*}
	\includegraphics[width=0.49\hsize]{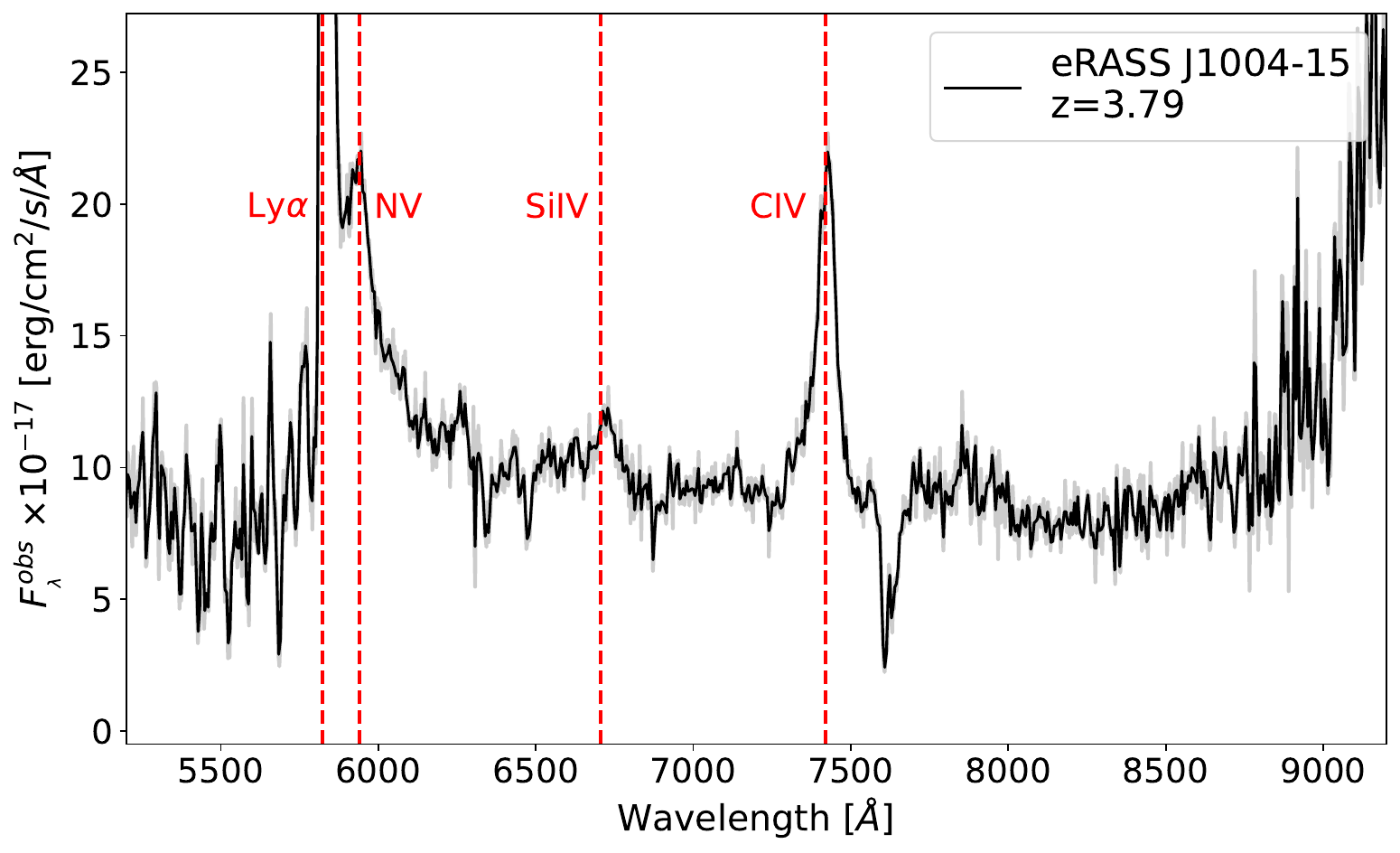}
	\includegraphics[width=0.49\hsize]{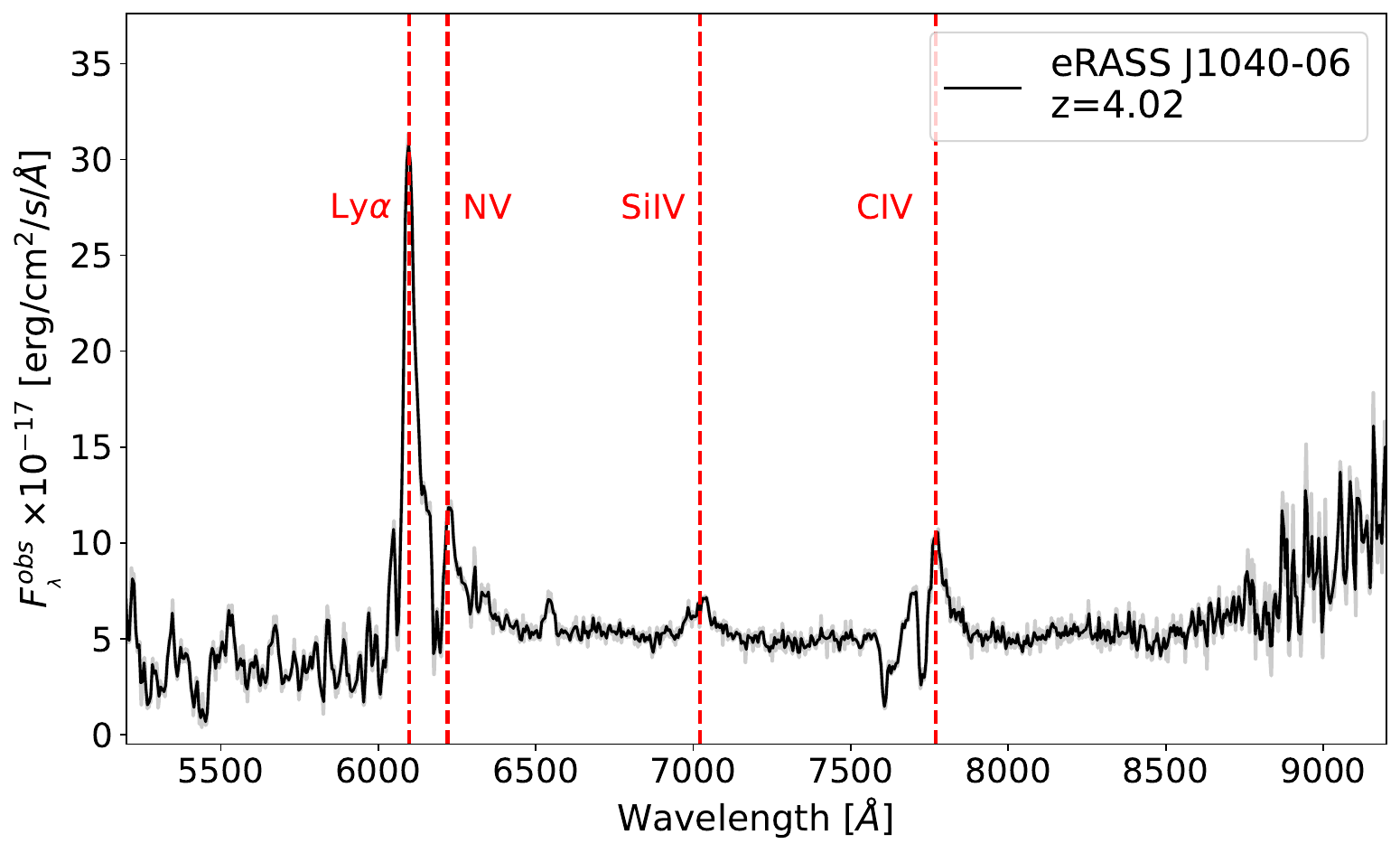}
	\includegraphics[width=0.49\hsize]{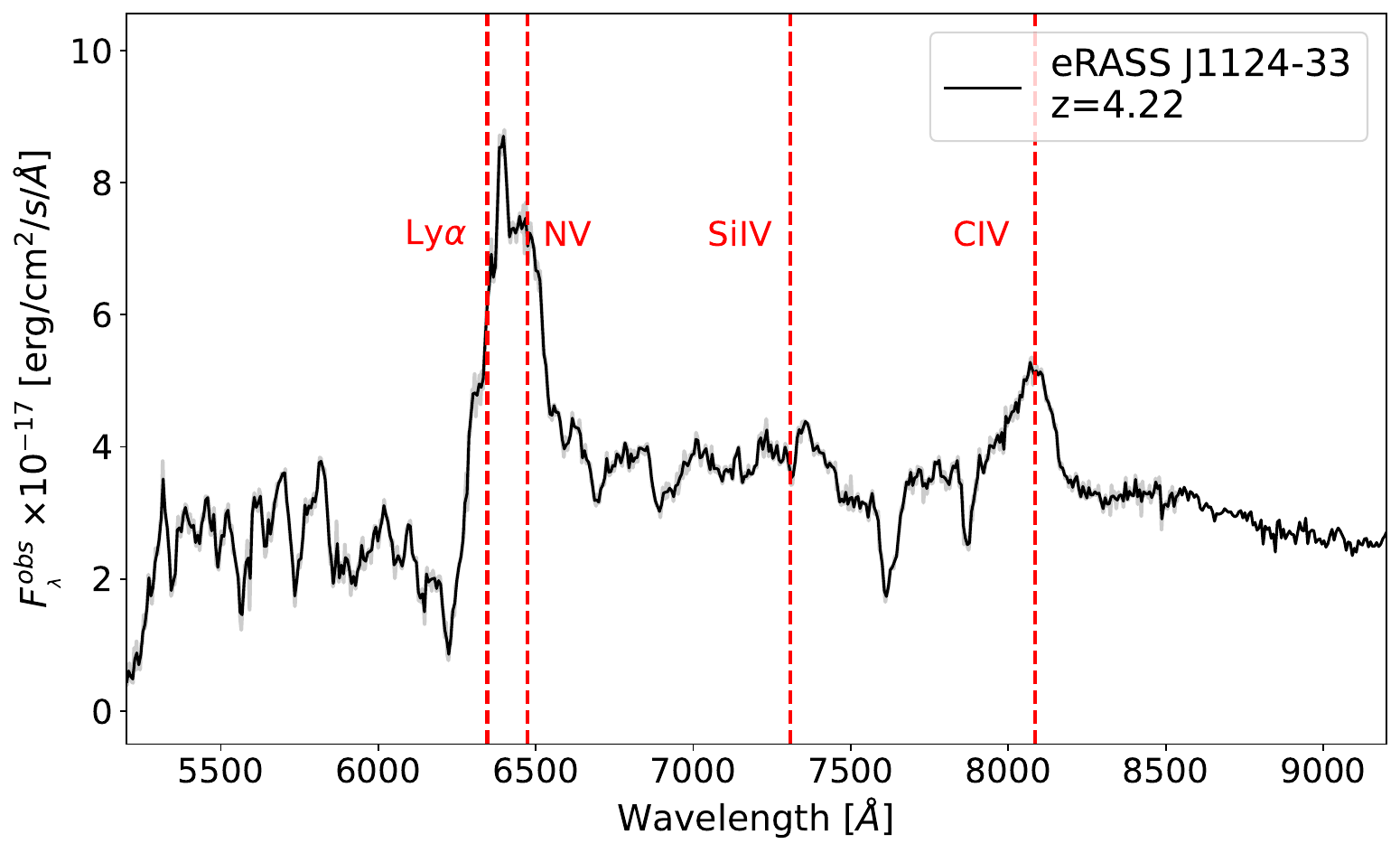}
	\includegraphics[width=0.49\hsize]{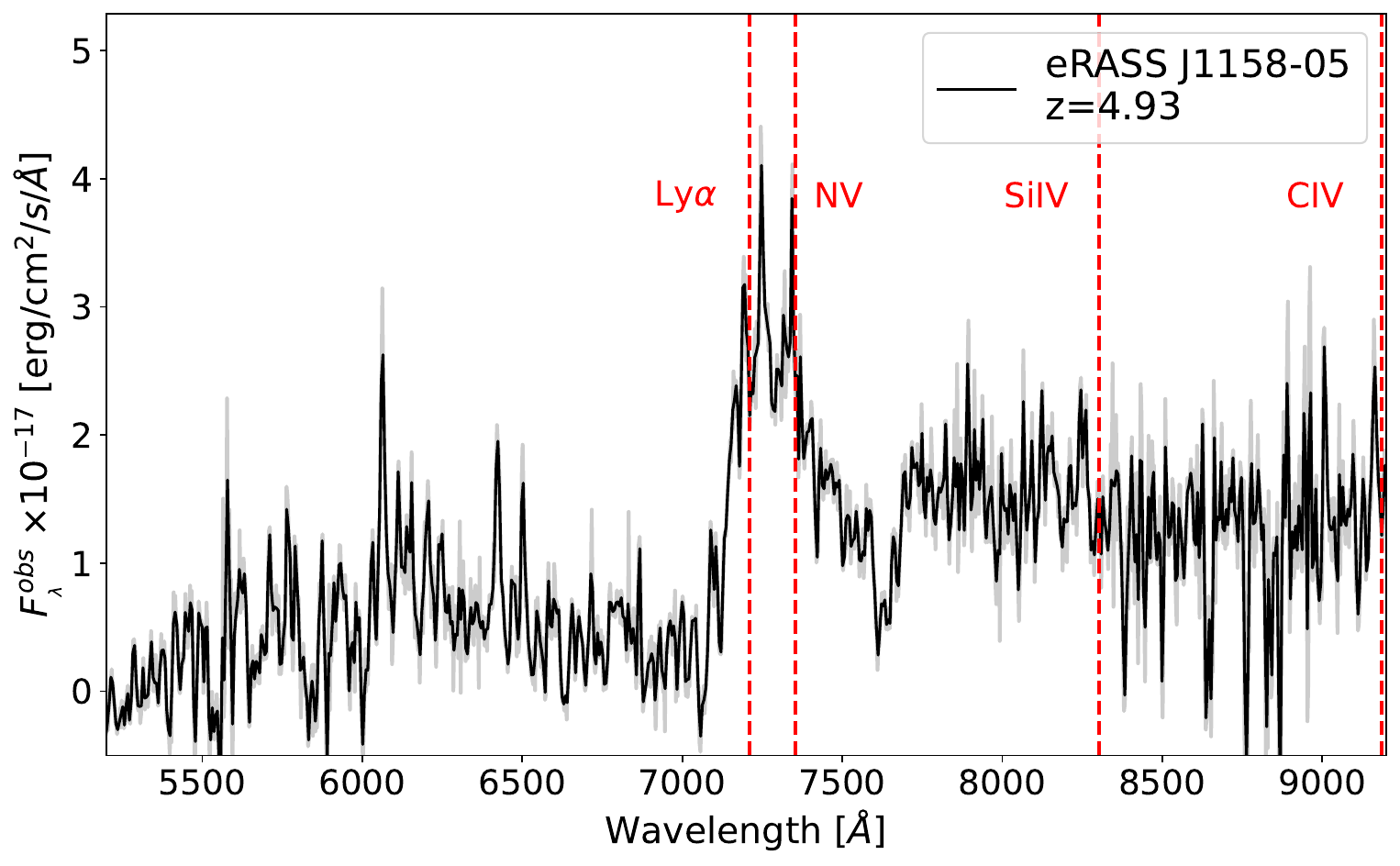}
	\includegraphics[width=0.49\hsize]{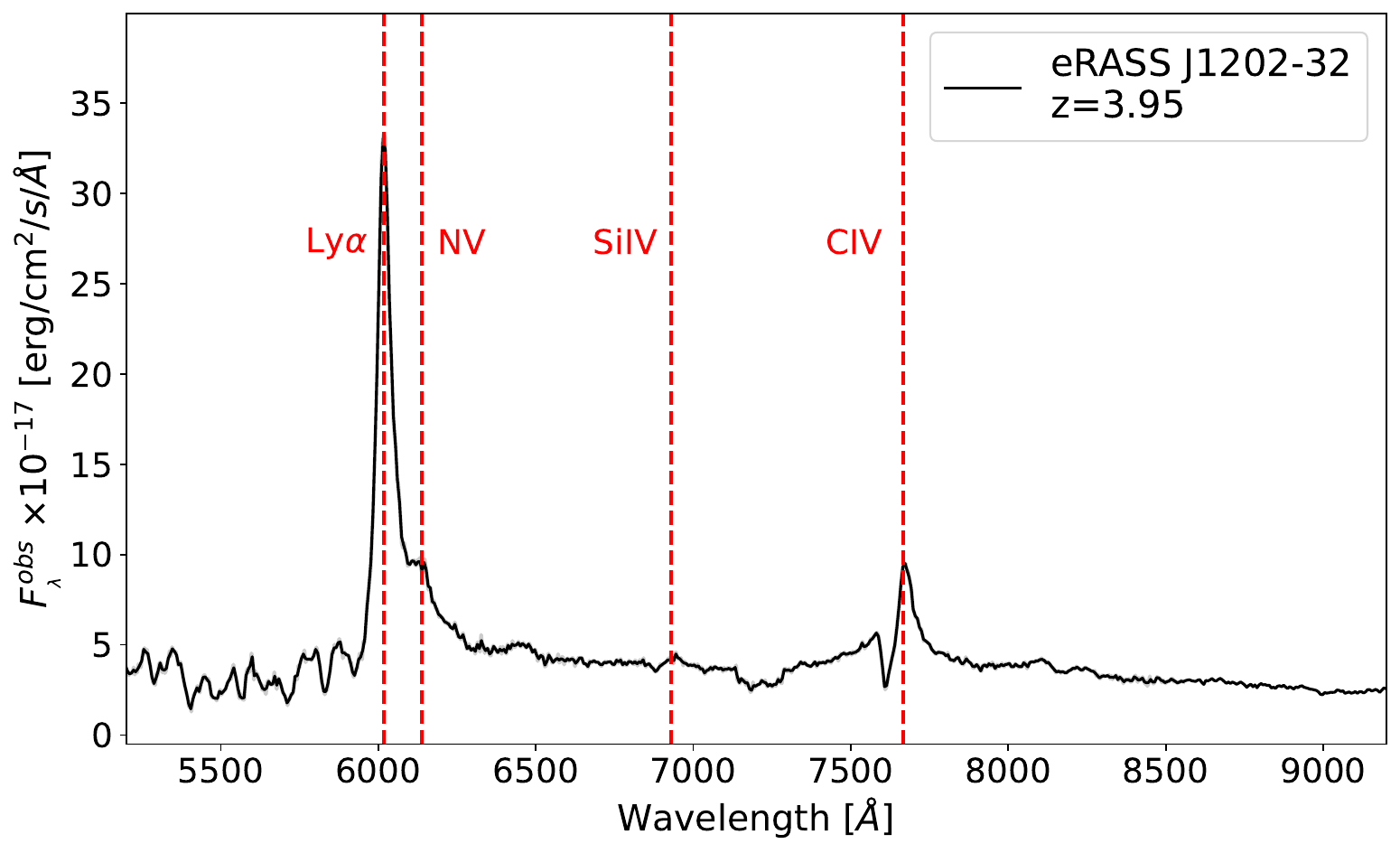}
	\includegraphics[width=0.49\hsize]{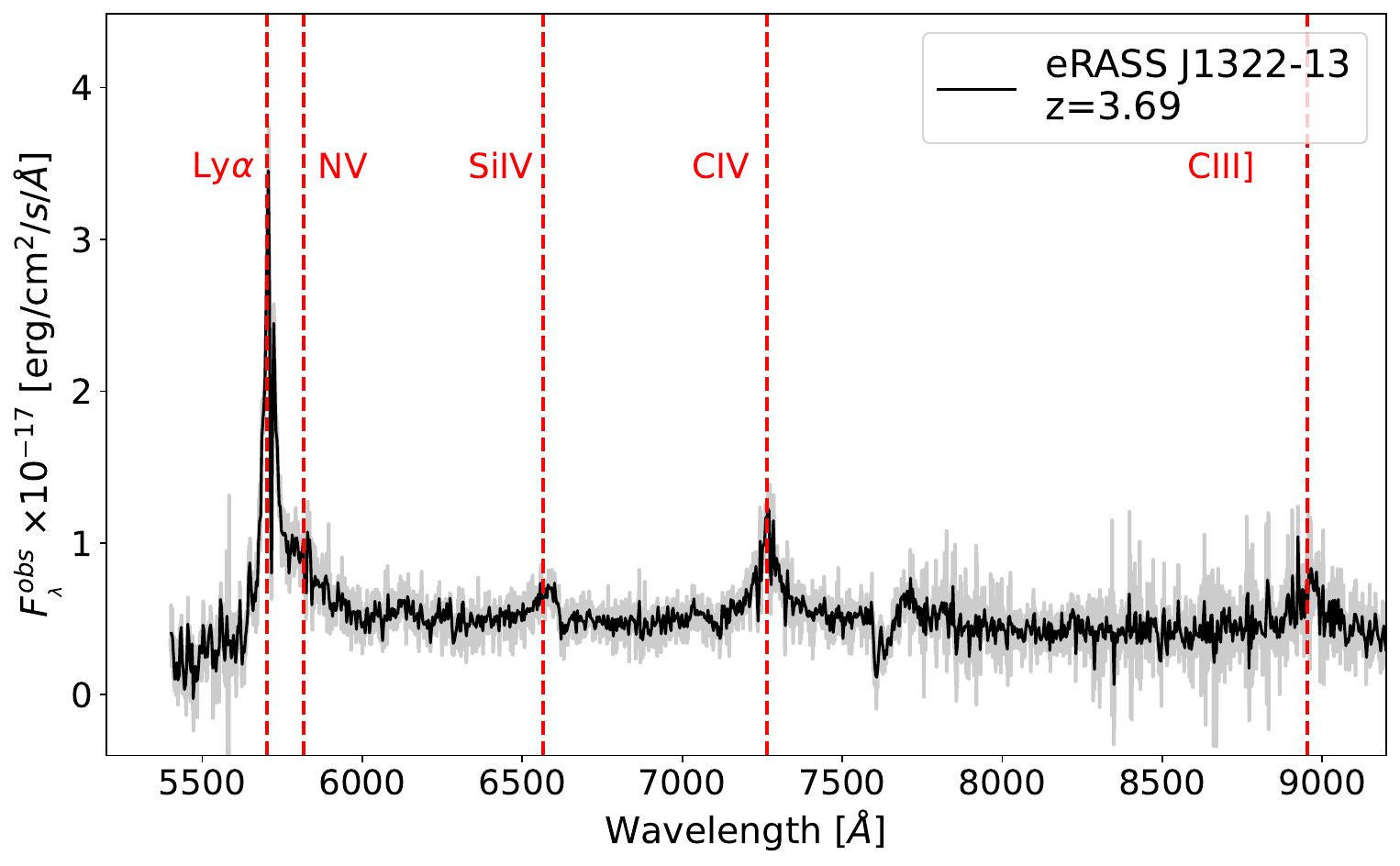}
    \includegraphics[width=0.49\hsize]{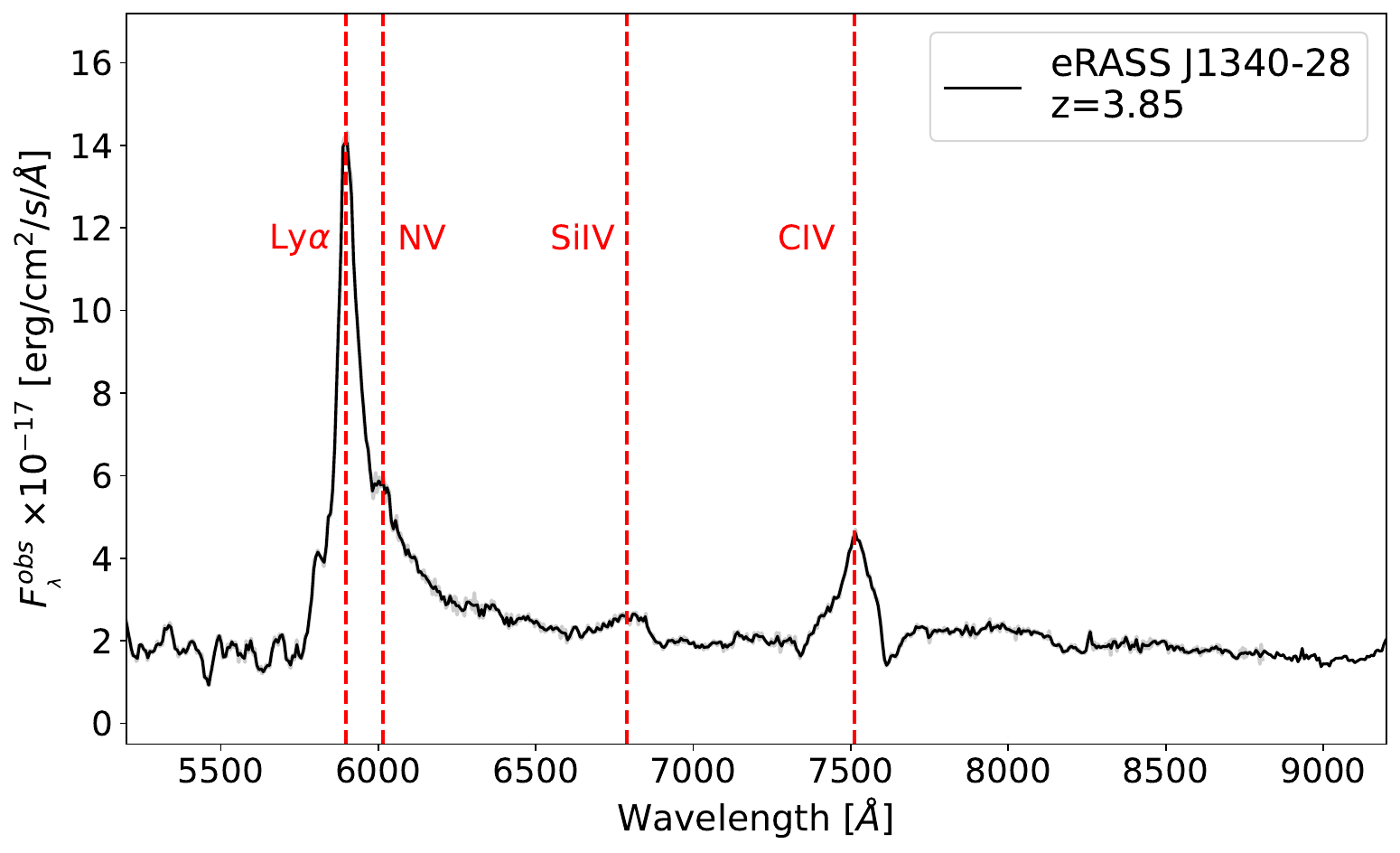}
	\includegraphics[width=0.49\hsize]{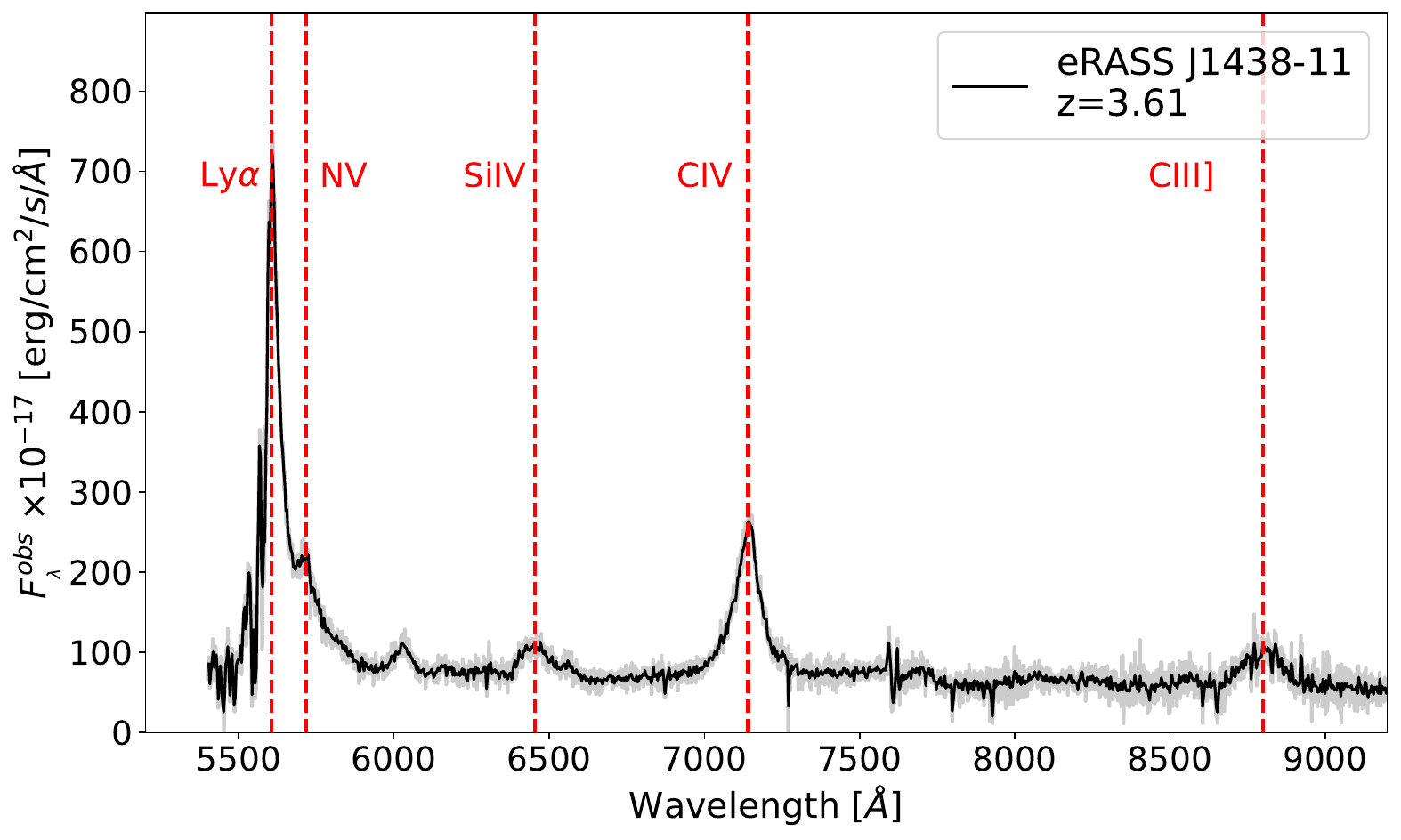}
    \caption{Continued.}
    \label{fig:optical_spectra4}
\end{figure*}

\begin{figure*}
    \includegraphics[width=0.49\hsize]{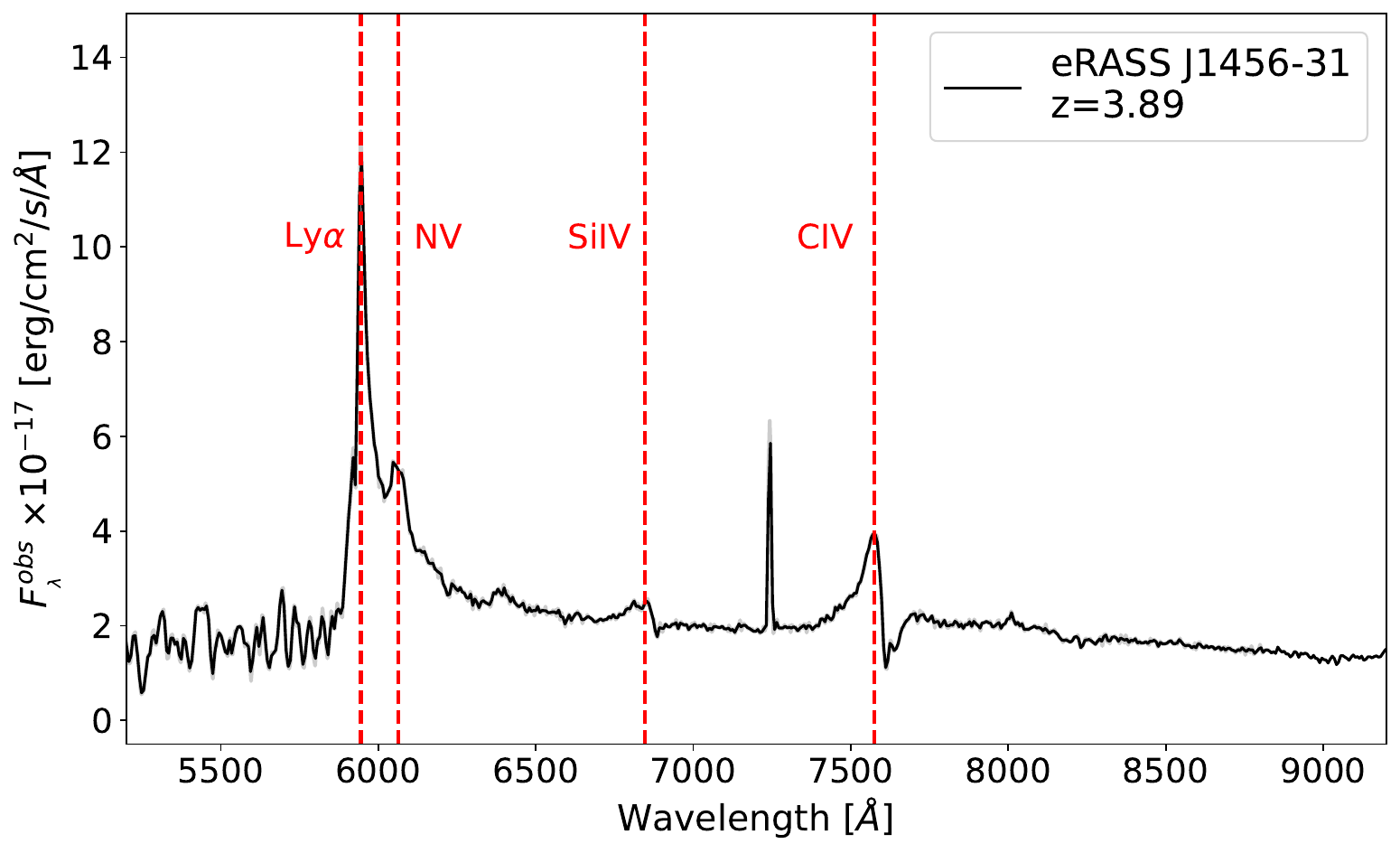}
    \includegraphics[width=0.49\hsize]{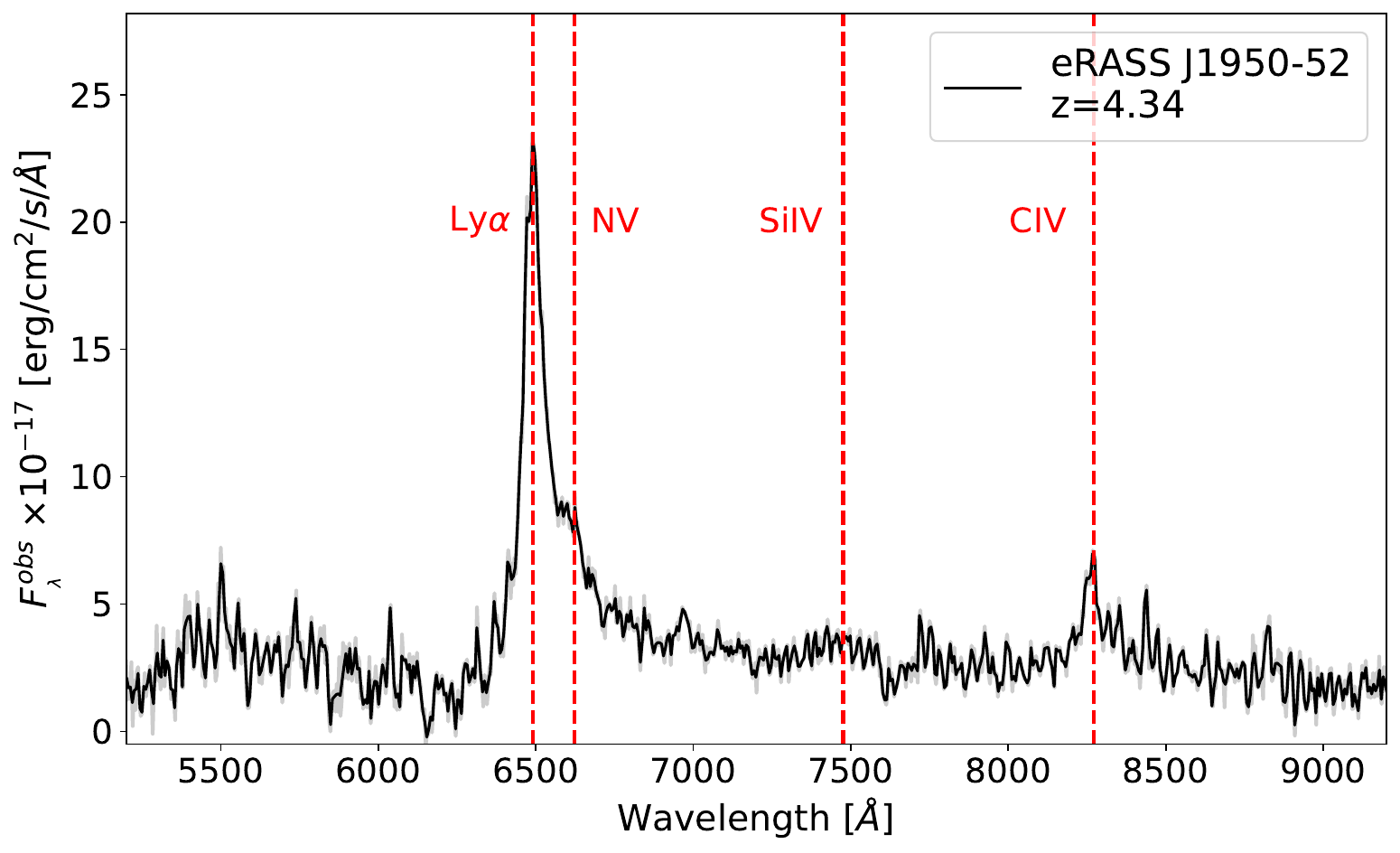}
    \includegraphics[width=0.49\hsize]{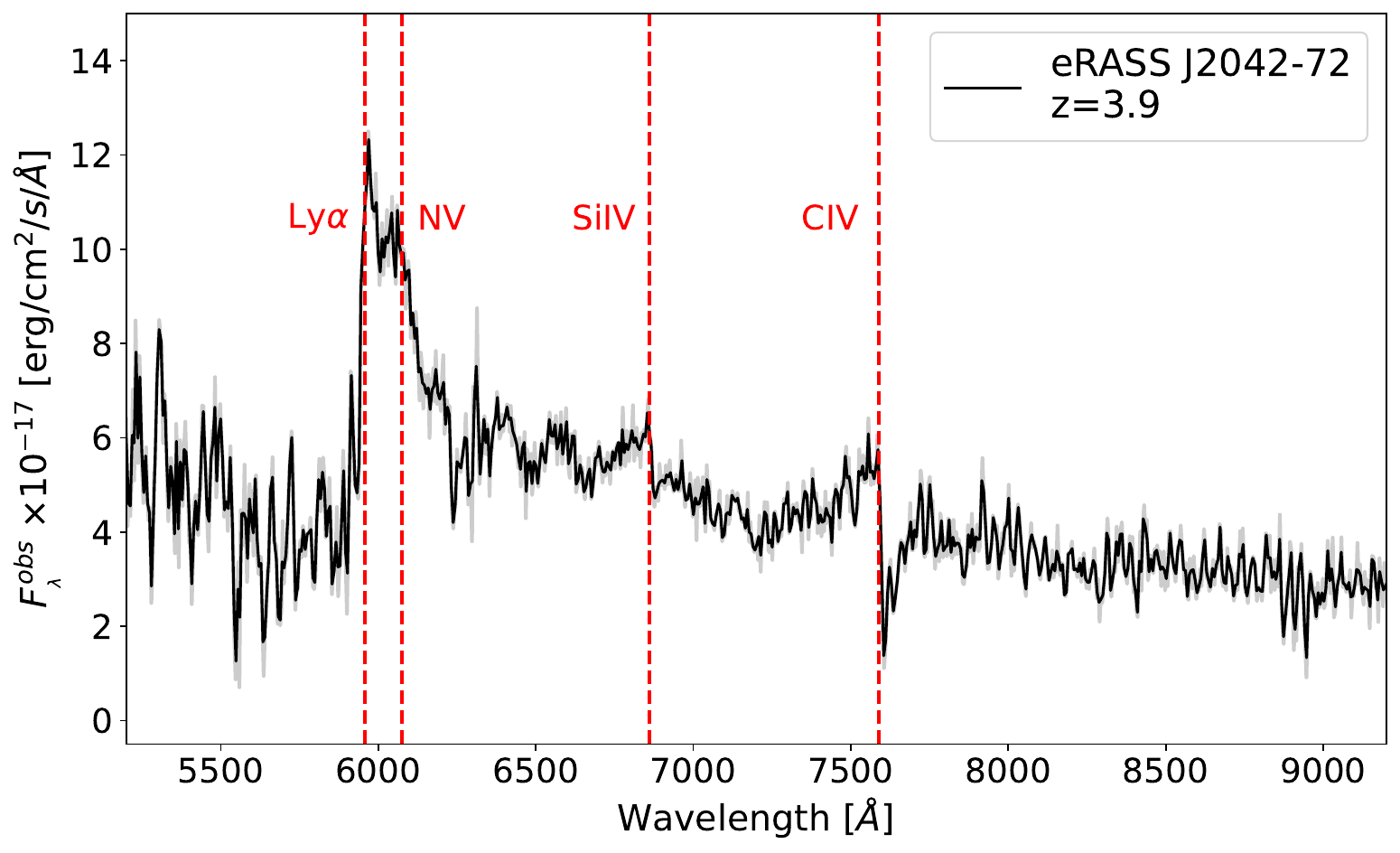}
	\includegraphics[width=0.49\hsize]{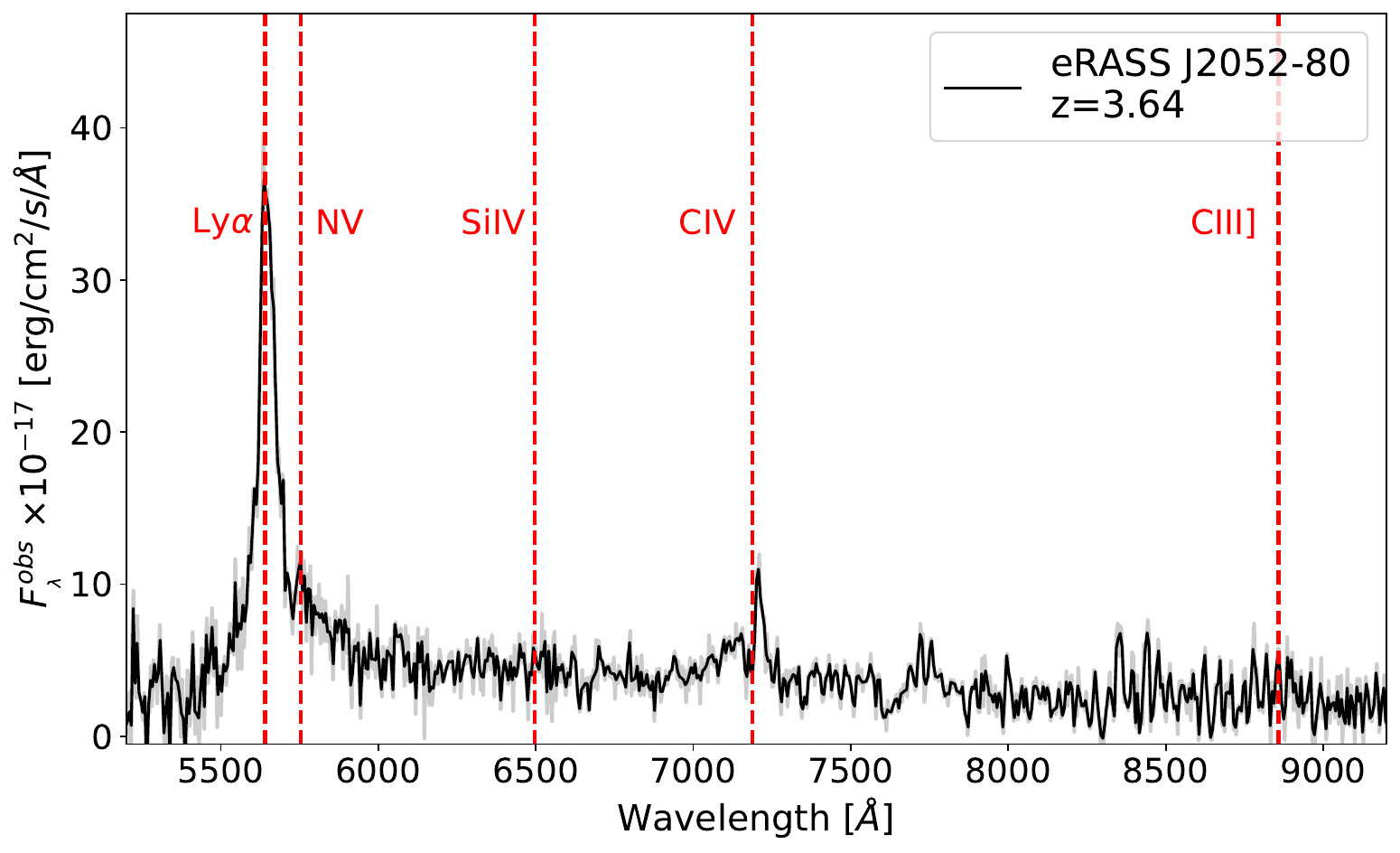}
    \includegraphics[width=0.49\hsize]{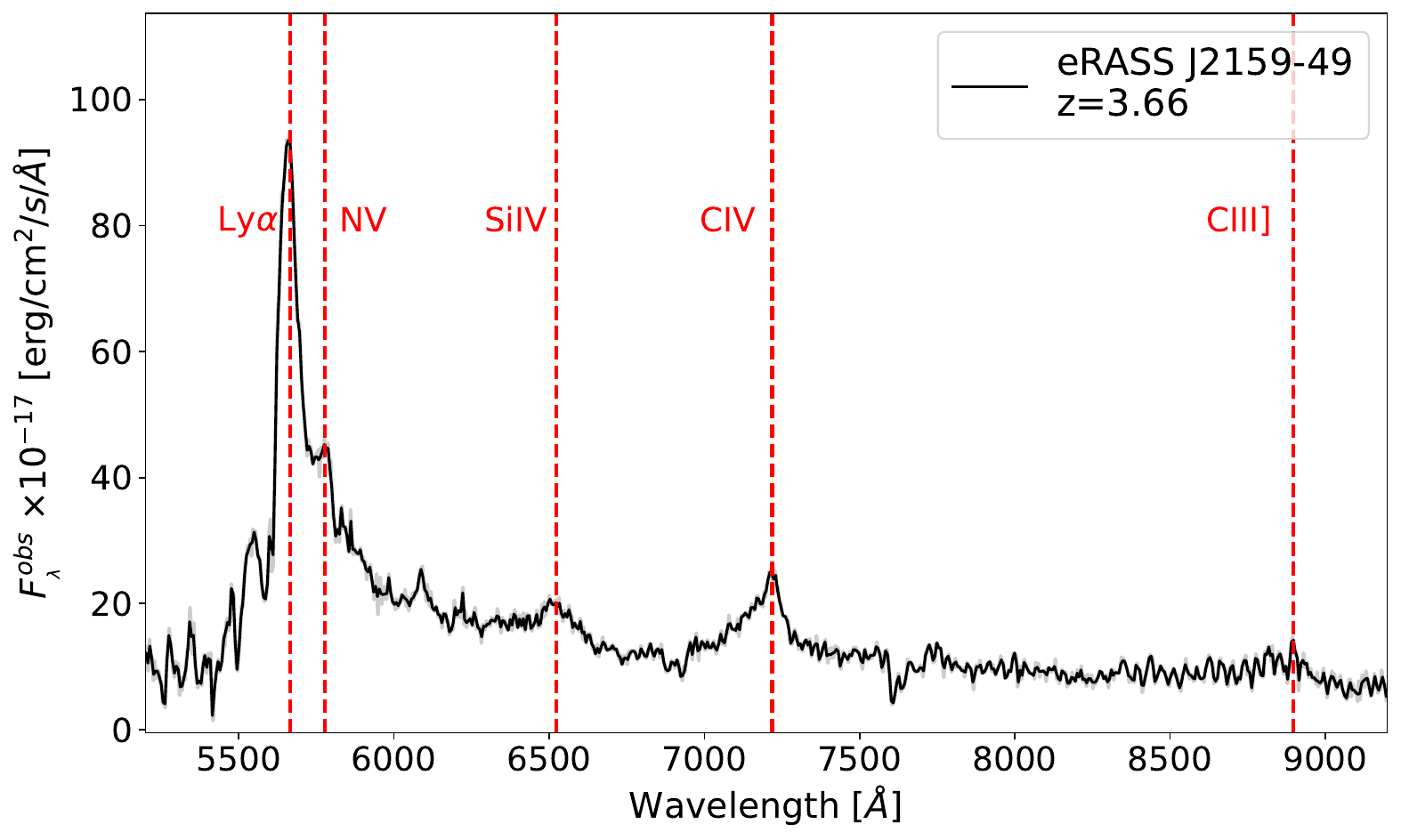}
	\includegraphics[width=0.49\hsize]{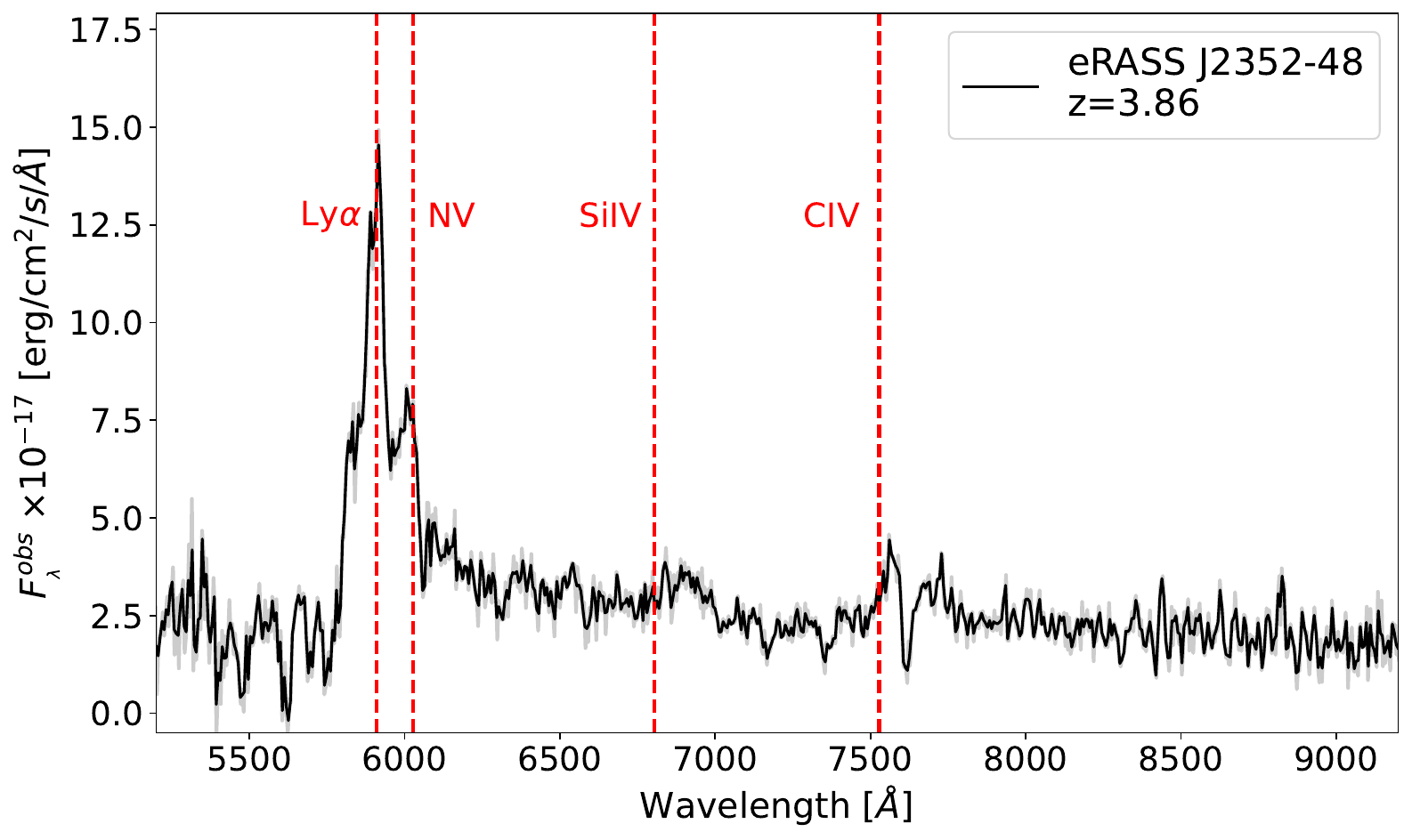}
	\includegraphics[width=0.49\hsize]{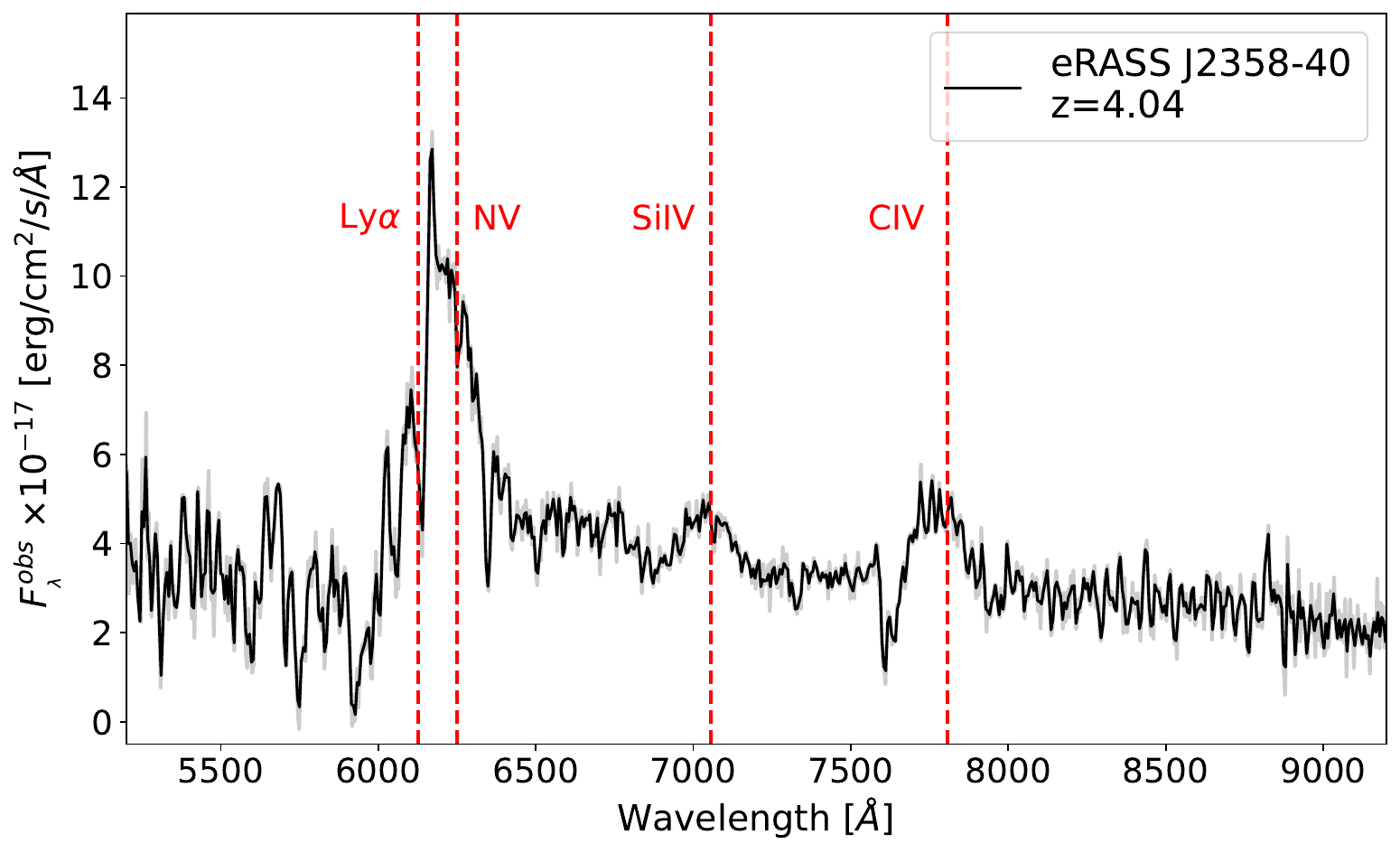}

    \caption{Continued.}
    \label{fig:optical_spectra5}
\end{figure*}

\begin{figure*}
	\includegraphics[width=0.505\hsize]{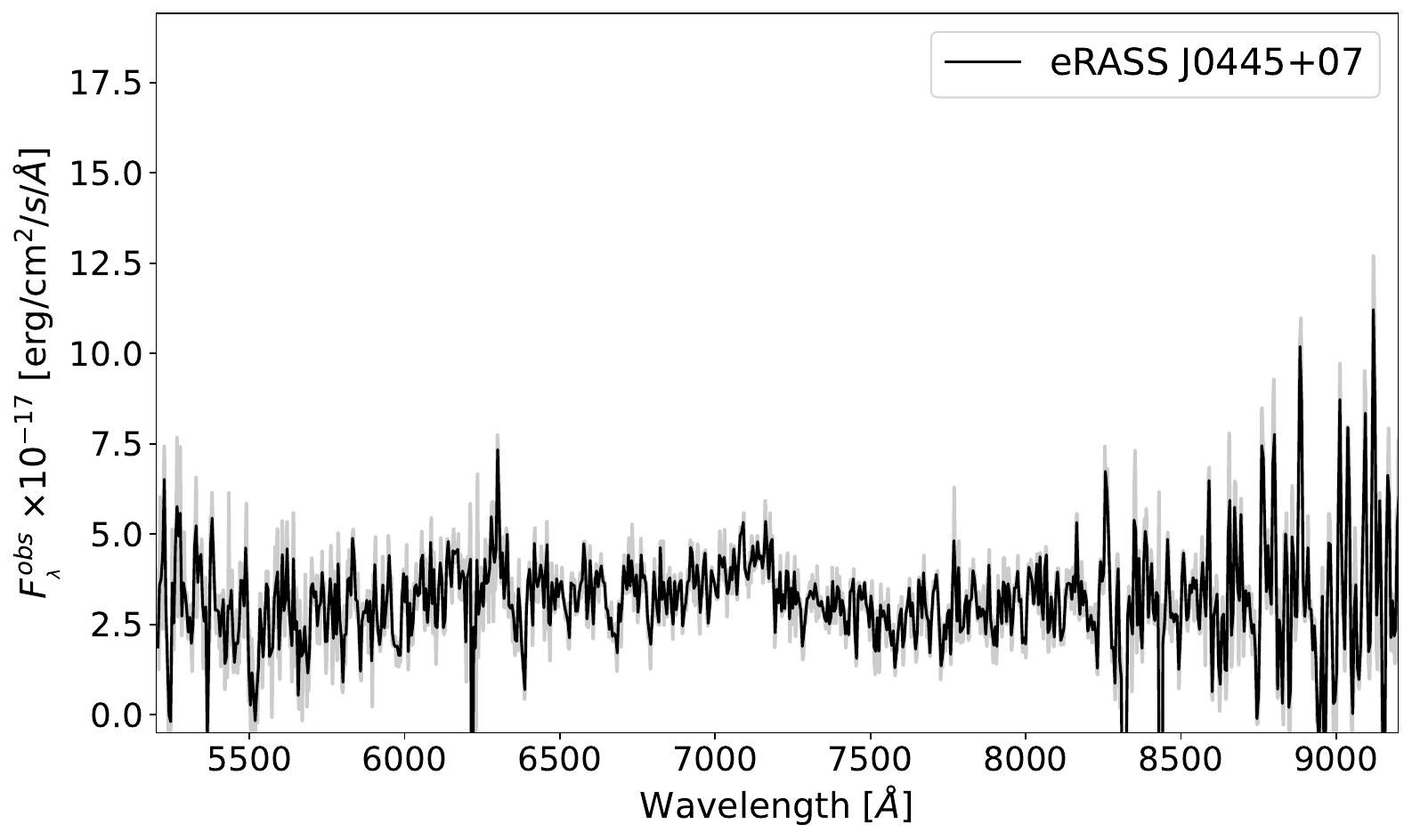}
	\includegraphics[width=0.49\hsize]{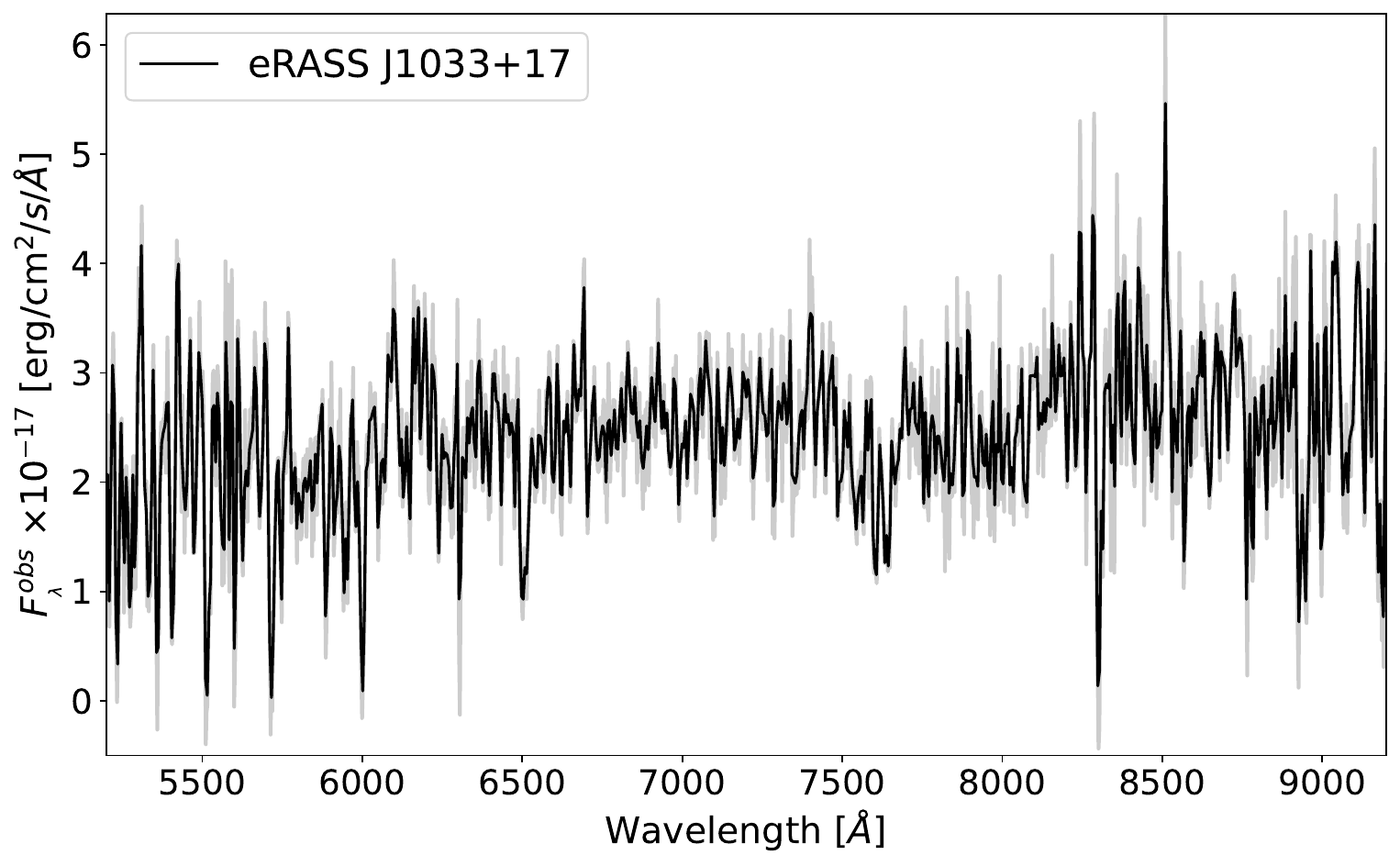}
	\includegraphics[width=0.49\hsize]{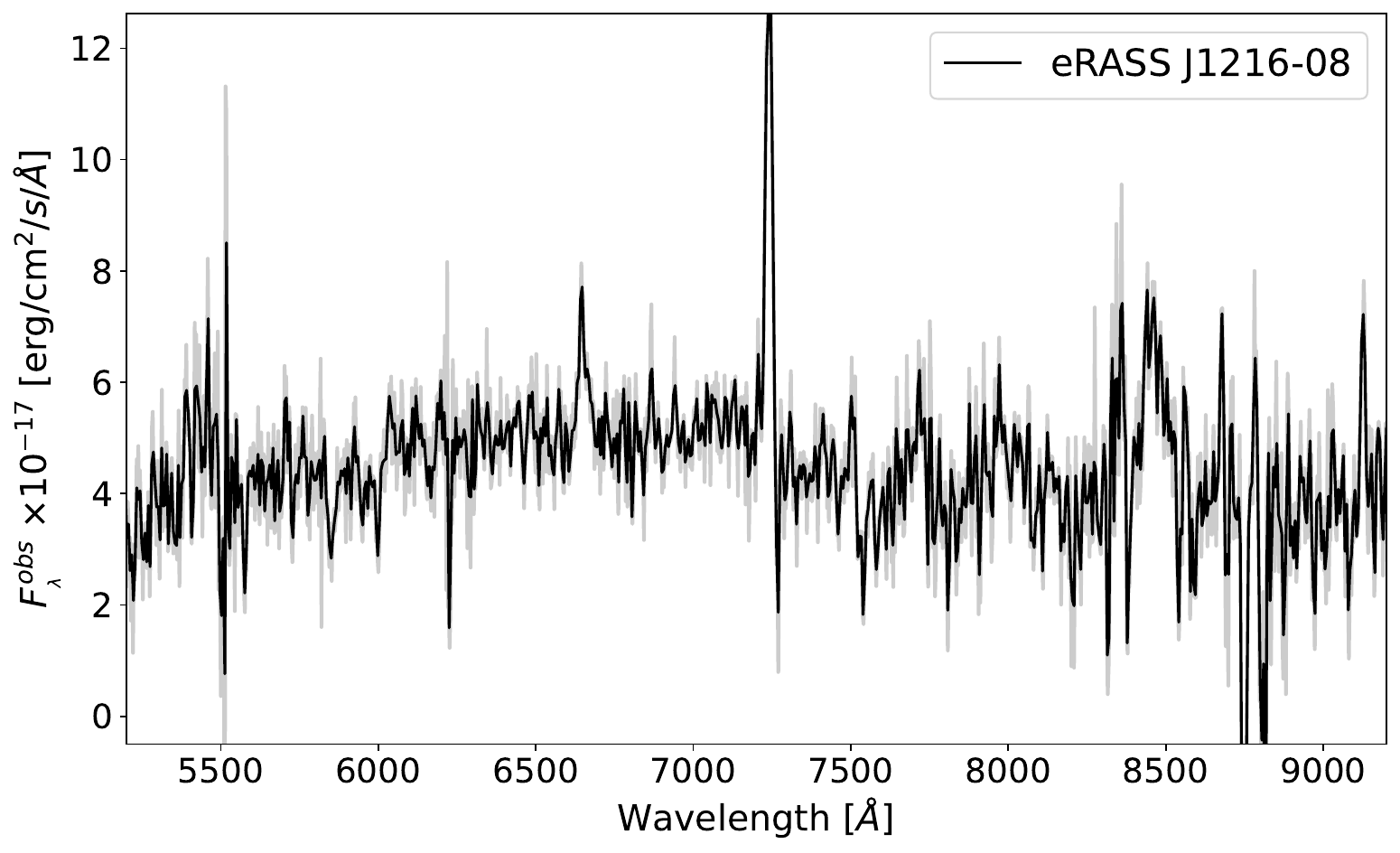}
	\includegraphics[width=0.49\hsize]{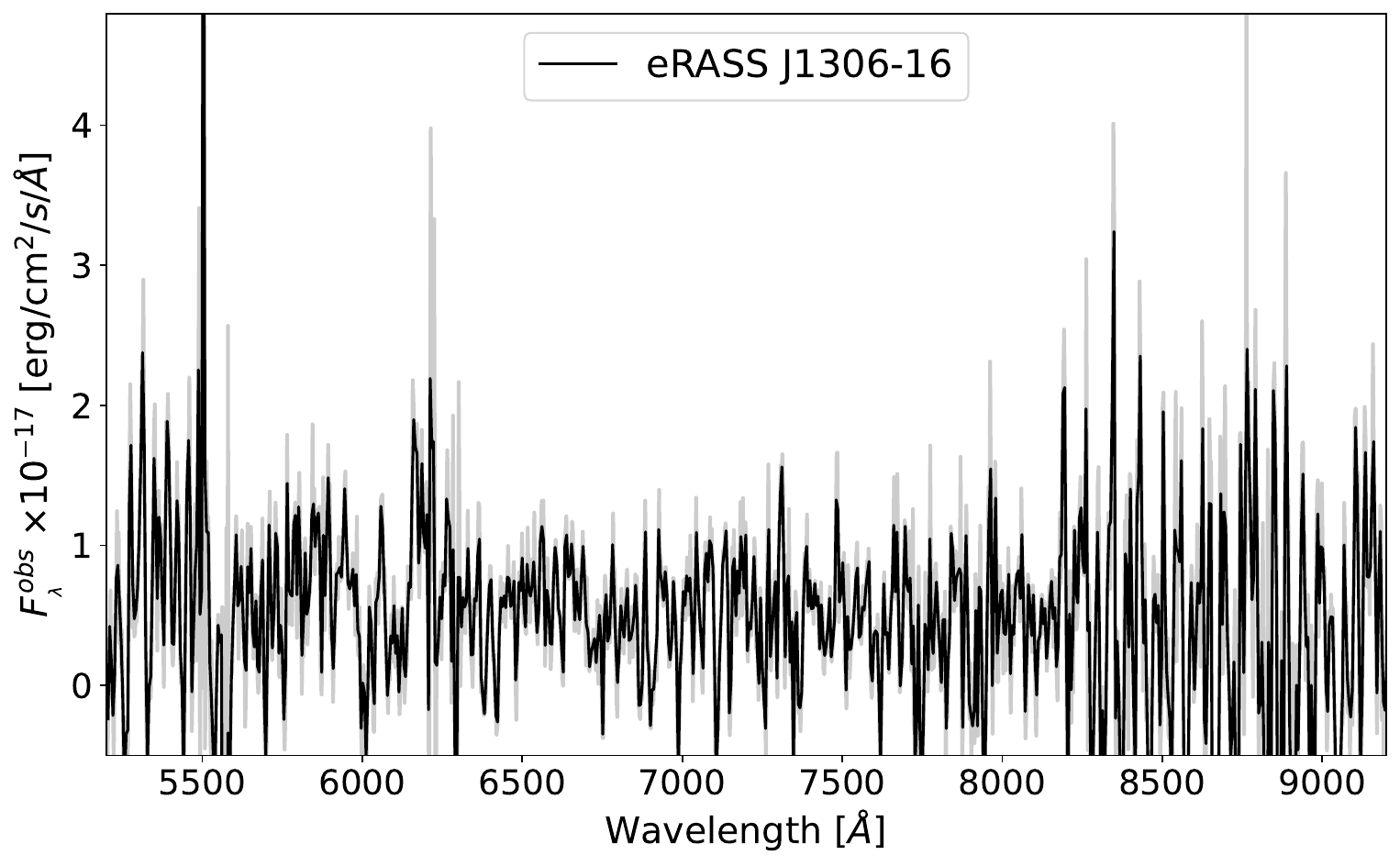}
	\includegraphics[width=0.49\hsize]{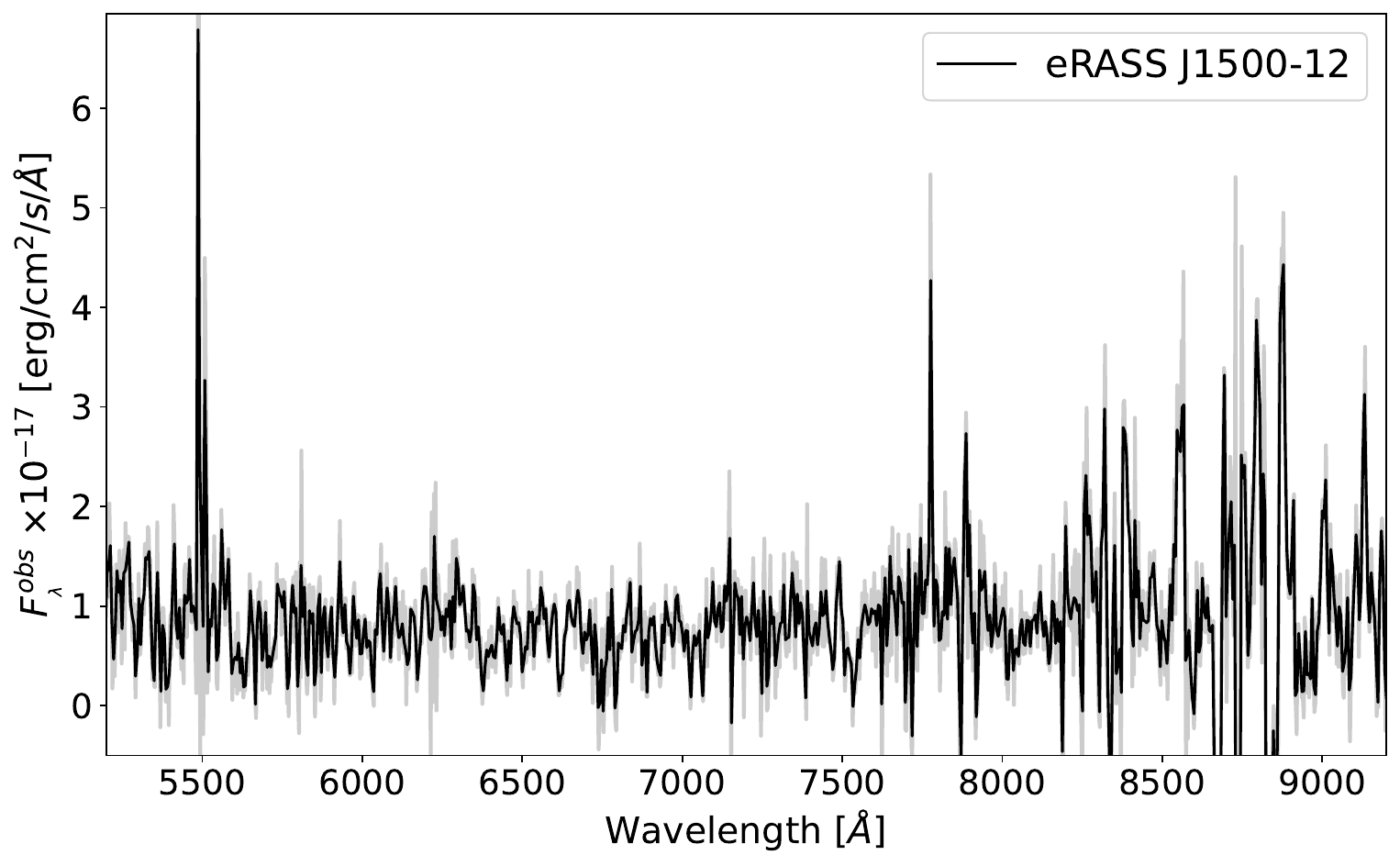}
	\includegraphics[width=0.49\hsize]{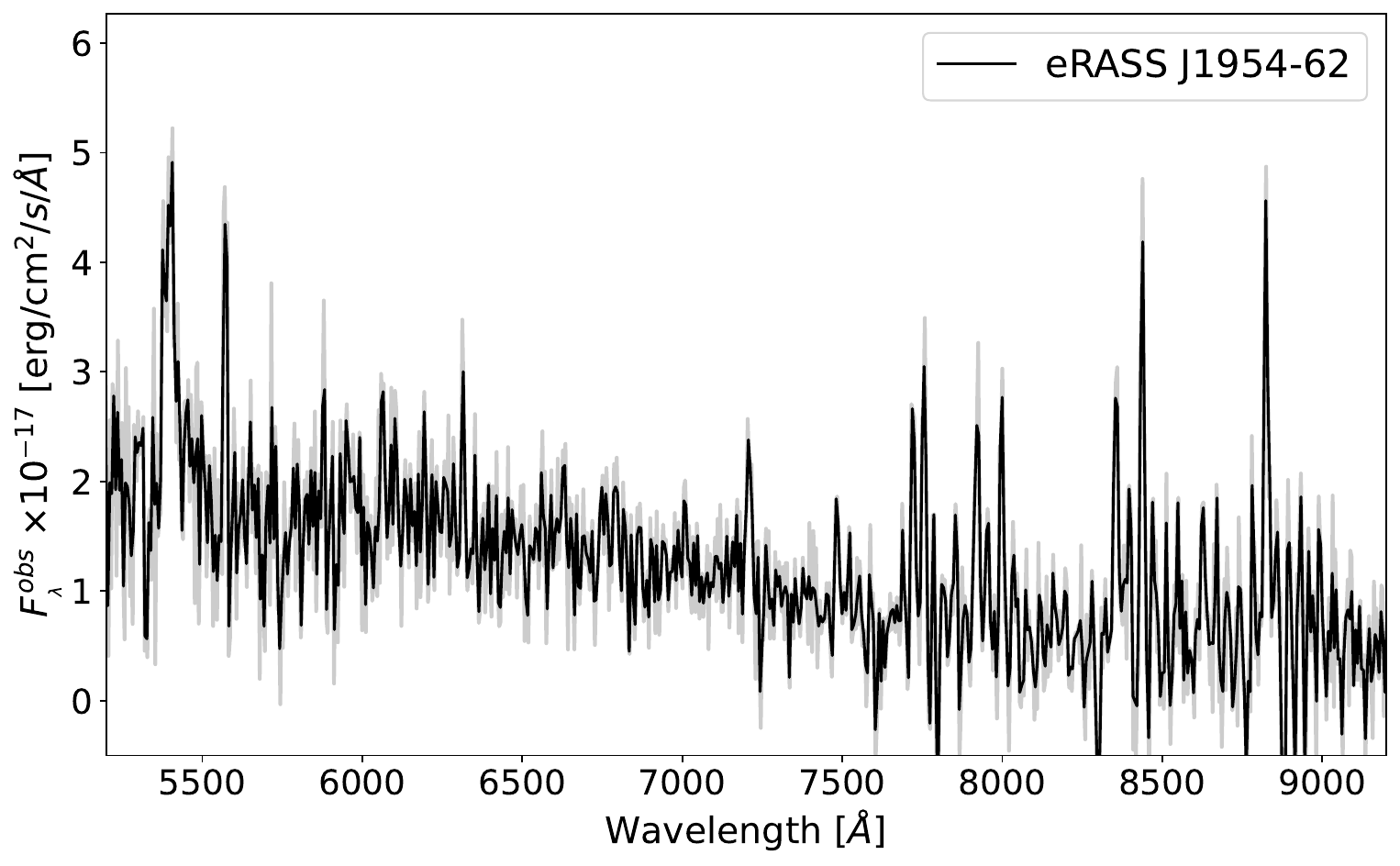}
	\includegraphics[width=0.49\hsize]{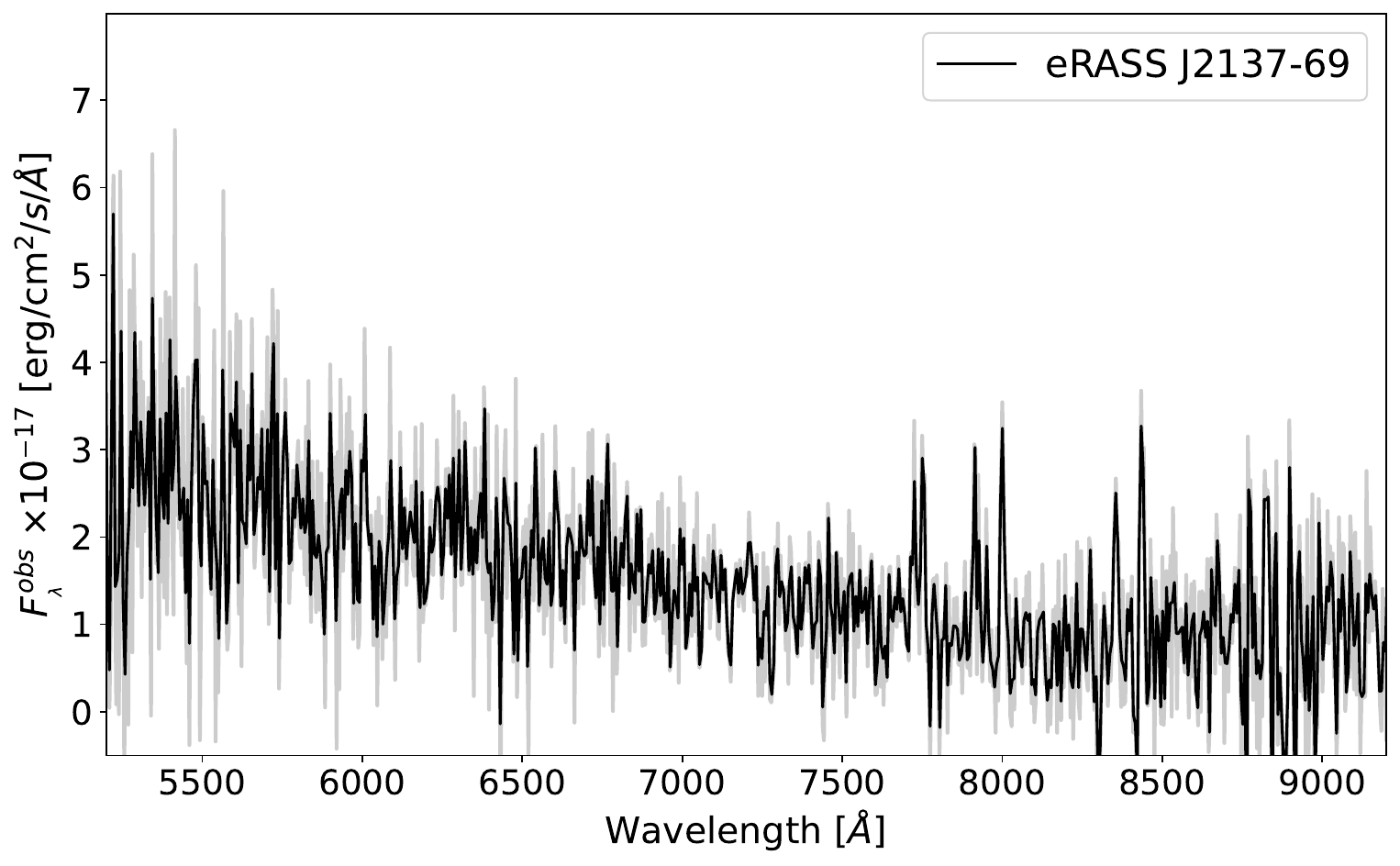}
    \caption{Optical spectra of candidate objects not confirmed to be high-redshift sources.}
    \label{fig:contaminants}
\end{figure*}

\section{Properties of candidates and high-\texorpdfstring{\lowercase{$z$}}{z} sample}
In this section, we list the coordinates, fluxes, magnitudes and exposure time of the candidates observed as part of this project, see Tab. \ref{tab:obs_targets}.\\
In Tab. \ref{tab:lum_sample} we also report the luminosities and R, $\tilde{\alpha}_{\rm ox}$ parameters of the high-$z$ quasars selected in work. For objects already known in the literature, we also include a reference to their discovery paper.

\begin{landscape}
\begin{table}
   \centering
    
    \begin{tabular}{lcccccccccccc}
    Name & RA & Dec & mag\_g & mag\_r & mag\_W2 & F$_{\rm0.2-2.3keV}$ & S$^{\rm peak}_{\rm 1.37GHz}$ & redshift & Exp. Time & Telescope\\
    & (deg) & (deg) & (AB) & (AB) & (Vega) & 10$^{-14}$~(erg~s$^{-1}$~cm$^{-2}$) & (mJy~beam$^{-1}$) & & (sec) & \\
    \hline
    \hline

1eRASS J000246.5$-$481557 	 & 	  0.6928211  	 & 	 $-$48.265888 	 & 	 20.62$\pm$0.01 	 & 	 19.23$\pm$0.01 	 & 	 15.89$\pm$0.05 	 & 	 5.9$\pm$2.3 	 & 	 32.3$\pm$1.9   	 & 	 4.25    	 & 	1200 	 & 	 NTT \\
1eRASS J005156.1$-$322644 	 & 	 12.9827681 	 & 	 $-$32.445111 	 & 	 21.76$\pm$0.03 	 & 	 19.94$\pm$0.01 	 & 	 15.72$\pm$0.05 	 & 	 4.4$\pm$2.1 	 & 	 13.8$\pm$0.8   	 & 	 4.04    	 & 	2400 	 & 	 NTT \\
1eRASS J020511.5$-$642105 	 & 	 31.3010603 	 & 	 $-$64.3505934 	 & 	 21.37$\pm$0.02 	 & 	 20.32$\pm$0.01 	 & 	 16.50$\pm$0.08 	 & 	 3.3$\pm$1.4 	 & 	 5.7$\pm$0.4    	 & 	 3.96    	 & 	3600 	 & 	 NTT \\
1eRASS J020520.9$-$220935 	 & 	 31.3384093 	 & 	 $-$22.1600903 	 & 	 20.43$\pm$0.02 	 & 	 19.29$\pm$0.01 	 & 	 15.61$\pm$0.05 	 & 	 6.5$\pm$2.2 	 & 	 10.9$\pm$0.7   	 & 	 3.57    	 & 	1500 	 & 	 NTT \\
1eRASS J023732.6$-$434820 	 & 	 39.3848765 	 & 	 $-$43.8063663 	 & 	 20.09$\pm$0.01 	 & 	 18.64$\pm$0.01 	 & 	 14.97$\pm$0.02 	 & 	 5.4$\pm$1.8 	 & 	 348.6$\pm$20.9 	 & 	 3.65    	 & 	1500 	 & 	 NTT \\
1eRASS J025903.8$-$694531 	 & 	 44.7638041 	 & 	 $-$69.7575884 	 & 	 20.85$\pm$0.01 	 & 	 19.55$\pm$0.01 	 & 	 14.82$\pm$0.02 	 & 	 3.0$\pm$1.1 	 & 	 12.6$\pm$0.8   	 & 	 3.44    	 & 	1500 	 & 	 NTT \\
1eRASS J030025.4$-$044644 	 & 	 45.1068359 	 & 	 $-$4.7807305 	 & 	 20.36$\pm$0.01 	 & 	 18.76$\pm$0.01 	 & 	 14.76$\pm$0.02 	 & 	 3.8$\pm$1.6 	 & 	 4.4$\pm$0.3    	 & 	 3.79    	 & 	1800 	 & 	 TNG \\
1eRASS J031010.4$-$350515 	 & 	 47.5432909 	 & 	 $-$35.087383 	 & 	 24.08$\pm$0.29 	 & 	 21.41$\pm$0.03 	 & 	 17.09$\pm$0.14 	 & 	 5.6$\pm$1.6 	 & 	 12.1$\pm$0.7   	 & 	 4.68    	 & 	4800 	 & 	 NTT \\
1eRASS J031226.4$-$131306 	 & 	 48.1105292 	 & 	 $-$13.2188854 	 & 	 22.28$\pm$0.06 	 & 	 20.99$\pm$0.03 	 & 	 16.50$\pm$0.09 	 & 	 3.2$\pm$1.5 	 & 	 114.8$\pm$6.9   	 & 	 3.69    	 & 	8280 	 & 	 LBT \\
1eRASS J031522.9$-$113314 	 & 	 48.8475552 	 & 	 $-$11.555209 	 & 	 20.28$\pm$0.01 	 & 	 19.11$\pm$0.01 	 & 	 14.69$\pm$0.02 	 & 	 4.2$\pm$1.6 	 & 	 7.3$\pm$0.5    	 & 	 3.79    	 & 	3150 	 & 	 LBT \\
1eRASS J034609.3$-$335933 	 & 	 56.5421750 	 & 	 $-$33.991950 	 & 	 21.76$\pm$0.03 	 & 	 20.47$\pm$0.02 	 & 	 16.67$\pm$0.10 	 & 	 2.7$\pm$1.1 	 & 	 17.7$\pm$1.1   	 & 	 4.19    	 & 	2100 	 & 	 NTT \\
1eRASS J034926.0$-$345434 	 & 	 57.3605886 	 & 	 $-$34.9116605 	 & 	 22.32$\pm$0.06 	 & 	 20.91$\pm$0.02 	 & 	 16.48$\pm$0.08 	 & 	 3.6$\pm$1.2 	 & 	 22.0$\pm$1.3   	 & 	 3.80    	 & 	1800 	 & 	 CLAY \\
1eRASS J035611.0$-$460756 	 & 	 59.0441182 	 & 	 $-$46.1332813 	 & 	 21.79$\pm$0.04 	 & 	 20.47$\pm$0.02 	 & 	 17.03$\pm$0.12 	 & 	 2.3$\pm$0.9 	 & 	 38.5$\pm$2.3   	 & 	 3.74    	 & 	3600 	 & 	 NTT \\
1eRASS J040453.6$-$333203 	 & 	 61.2222249 	 & 	 $-$33.5337843 	 & 	 22.79$\pm$0.12 	 & 	 20.81$\pm$0.03 	 & 	 16.51$\pm$0.09 	 & 	 3.9$\pm$1.5 	 & 	 17.5$\pm$1.1   	 & 	 4.62    	 & 	3000 	 & 	 NTT \\
1eRASS J041105.2$-$021450 	 & 	 62.7712502 	 & 	 $-$2.2474116 	 & 	 22.19$\pm$0.04 	 & 	 20.51$\pm$0.01 	 & 	 16.26$\pm$0.09 	 & 	 10.2$\pm$3.2 	 & 	 55.0$\pm$3.3   	 & 	 4.13    	 & 	7200 	 & 	 LBT \\
1eRASS J041138.1$-$570106 	 & 	 62.9071838 	 & 	 $-$57.0182188 	 & 	 22.71$\pm$0.08 	 & 	 21.06$\pm$0.02 	 & 	 16.28$\pm$0.05 	 & 	 1.6$\pm$0.7 	 & 	 52.0$\pm$3.1   	 & 	 4.30    	 & 	1800 	 & 	 CLAY \\
1eRASS J041146.9$-$540515 	 & 	 62.9442444 	 & 	 $-$54.0877494 	 & 	 21.82$\pm$0.04 	 & 	 20.44$\pm$0.02 	 & 	 15.95$\pm$0.04 	 & 	 4.1$\pm$1.0 	 & 	 59.6$\pm$3.6   	 & 	 3.74    	 & 	1200 	 & 	 CLAY \\
1eRASS J042048.9$-$570304 	 & 	 65.2044899 	 & 	 $-$57.0506008 	 & 	 22.03$\pm$0.04 	 & 	 21.01$\pm$0.03 	 & 	 16.67$\pm$0.07 	 & 	 8.8$\pm$1.4 	 & 	 55.8$\pm$3.4   	 & 	 3.80    	 & 	3000 	 & 	 NTT \\
1eRASS J044117.7$-$335111 	 & 	 70.3242198 	 & 	 $-$33.8527567 	 & 	 22.36$\pm$0.05 	 & 	 21.00$\pm$0.02 	 & 	 16.71$\pm$0.10 	 & 	 3.3$\pm$1.5 	 & 	 62.1$\pm$3.7   	 & 	 3.68    	 & 	2200 	 & 	 CLAY \\
1eRASS J045513.9$-$371157 	 & 	 73.8078434 	 & 	 $-$37.1996787 	 & 	 18.60$\pm$0.01 	 & 	 17.46$\pm$0.01 	 & 	 13.83$\pm$0.02 	 & 	 6.0$\pm$1.7 	 & 	 4.7$\pm$0.3    	 & 	 3.94    	 & 	900 	 & 	 NTT \\
1eRASS J051559.6$-$261006 	 & 	 78.9976636 	 & 	 $-$26.1682215 	 & 	 21.94$\pm$0.04 	 & 	 19.76$\pm$0.01 	 & 	 15.29$\pm$0.08 	 & 	 4.4$\pm$1.7 	 & 	 115.7$\pm$6.9   	 & 	 4.24    	 & 	1800 	 & 	 NTT \\
1eRASS J052733.7$-$523901 	 & 	 81.8941508 	 & 	 $-$52.6506556 	 & 	 22.82$\pm$0.07 	 & 	 21.13$\pm$0.02 	 & 	 17.49$\pm$0.15 	 & 	 1.2$\pm$0.6 	 & 	 2.0$\pm$0.2    	 & 	 4.25    	 & 	2800 	 & 	 NTT \\
1eRASS J054609.1$-$523426 	 & 	 86.5386685 	 & 	 $-$52.5735398 	 & 	 20.93$\pm$0.02 	 & 	 19.69$\pm$0.01 	 & 	 15.28$\pm$0.02 	 & 	 1.8$\pm$0.7 	 & 	 19.0$\pm$1.1   	 & 	 3.94    	 & 	3000 	 & 	 NTT \\
1eRASS J060057.9$-$521526 	 & 	 90.2400900 	 & 	 $-$52.256608 	 & 	 21.26$\pm$0.02 	 & 	 19.97$\pm$0.01 	 & 	 16.37$\pm$0.05 	 & 	 1.8$\pm$0.8 	 & 	 2.2$\pm$0.2    	 & 	 3.59    	 & 	600 	 & 	 CLAY \\
1eRASS J083910.9$-$020232 	 & 	 129.7963081 	 & 	 $-$2.0413959 	 & 	 20.11$\pm$0.01 	 & 	 19.00$\pm$0.01 	 & 	 15.27$\pm$0.04 	 & 	 37.1$\pm$6.6 	 & 	 209.3$\pm$12.6 	 & 	 3.75    	 & 	2100 	 & 	 TNG \\
1eRASS J100448.0$-$153202 	 & 	 151.2000094 	 & 	 $-$15.5332302 	 & 	 19.87$\pm$0.01 	 & 	 18.49$\pm$0.01 	 & 	 14.75$\pm$0.03 	 & 	 15.6$\pm$4.4 	 & 	 2.4$\pm$0.2    	 & 	 3.80    	 & 	1800 	 & 	 TNG \\
1eRASS J104018.5$-$063636 	 & 	 160.0784126 	 & 	 $-$6.6119352 	 & 	 20.45$\pm$0.01 	 & 	 19.06$\pm$0.01 	 & 	 14.27$\pm$0.02 	 & 	 6.6$\pm$3.1 	 & 	 18.5$\pm$1.1   	 & 	 4.01    	 & 	2700 	 & 	 TNG \\
1eRASS J112415.7$-$331735 	 & 	 171.0657772 	 & 	 $-$33.2935816 	 & 	 21.24$\pm$0.02 	 & 	 19.44$\pm$0.02 	 & 	 15.44$\pm$0.04 	 & 	 5.6$\pm$2.2 	 & 	 60.8$\pm$3.7   	 & 	 4.22    	 & 	1600 	 & 	 Gemini \\
1eRASS J115839.3$-$052221 	 & 	 179.6623472 	 & 	 $-$5.3739697 	 & 	     ---        	 & 	 21.49$\pm$0.06 	 & 	 16.15$\pm$0.09 	 & 	 7.6$\pm$2.4 	 & 	 62.7$\pm$3.8   	 & 	 4.96    	 & 	1320 	 & 	 TNG \\
1eRASS J120222.0$-$320757 	 & 	 180.5900608 	 & 	 $-$32.1313261 	 & 	 20.51$\pm$0.02 	 & 	 19.23$\pm$0.01 	 & 	 14.94$\pm$0.03 	 & 	 2.9$\pm$1.5 	 & 	 41.7$\pm$2.5   	 & 	 3.95    	 & 	1320 	 & 	 Gemini \\
1eRASS J134030.6$-$282556 	 & 	 205.1251442 	 & 	 $-$28.4333186 	 & 	 21.34$\pm$0.03 	 & 	 19.99$\pm$0.01 	 & 	 15.93$\pm$0.07 	 & 	 4.9$\pm$1.9 	 & 	 2.4$\pm$0.2    	 & 	 3.85    	 & 	2680 	 & 	 Gemini \\
1eRASS J143844.0$-$111851 	 & 	 219.6834211 	 & 	 $-$11.3142684 	 & 	 20.56$\pm$0.02 	 & 	 19.50$\pm$0.01 	 & 	 15.47$\pm$0.04 	 & 	 4.5$\pm$2.0 	 & 	 21.6$\pm$2.2   	 & 	 3.61    	 & 	2700  	 & 	 LBT \\
1eRASS J145635.5$-$314933 	 & 	 224.1510724 	 & 	 $-$31.8254278 	 & 	 21.19$\pm$0.03 	 & 	 19.98$\pm$0.01 	 & 	 16.22$\pm$0.08 	 & 	 6.3$\pm$2.5 	 & 	 34.6$\pm$2.1   	 & 	 3.89    	 & 	2760 	 & 	 Gemini \\
1eRASS J195040.7$-$523939 	 & 	 297.6679830 	 & 	 $-$52.6609784 	 & 	 21.63$\pm$0.02 	 & 	 19.63$\pm$0.01 	 & 	 16.35$\pm$0.09 	 & 	 4.9$\pm$2.6 	 & 	 28.1$\pm$1.7   	 & 	 4.43    	 & 	3000 	 & 	 NTT \\
1eRASS J204223.4$-$721207 	 & 	 310.5940819 	 & 	 $-$72.201895 	 & 	 20.22$\pm$0.02 	 & 	 19.12$\pm$0.01 	 & 	 15.64$\pm$0.04 	 & 	 5.1$\pm$2.2 	 & 	 5.8$\pm$0.4    	 & 	 3.90    	 & 	2400 	 & 	 NTT \\
1eRASS J205253.1$-$801247 	 & 	 313.2269879 	 & 	 $-$80.2117198 	 & 	 20.73$\pm$0.01 	 & 	 19.49$\pm$0.01 	 & 	 14.49$\pm$0.02 	 & 	 6.6$\pm$2.2 	 & 	 149.1$\pm$8.9   	 & 	 3.64    	 & 	3900 	 & 	 NTT \\
1eRASS J215950.0$-$494435 	 & 	 329.9585773 	 & 	 $-$49.7428258 	 & 	 19.26$\pm$0.01 	 & 	 18.04$\pm$0.01 	 & 	 13.74$\pm$0.01 	 & 	 3.9$\pm$2.0 	 & 	 17.8$\pm$1.1   	 & 	 3.66    	 & 	3000 	 & 	 NTT \\
1eRASS J235242.2$-$483954 	 & 	 358.1761584 	 & 	 $-$48.6651304 	 & 	 21.41$\pm$0.03 	 & 	 19.82$\pm$0.01 	 & 	 15.08$\pm$0.03 	 & 	 3.8$\pm$1.8 	 & 	 0.9$\pm$0.2    	 & 	 3.94    	 & 	2100 	 & 	 NTT \\
1eRASS J235828.7$-$401434 	 & 	 359.6196190 	 & 	 $-$40.239828 	 & 	 21.00$\pm$0.02 	 & 	 19.56$\pm$0.01 	 & 	 16.05$\pm$0.06 	 & 	 4.2$\pm$2.3 	 & 	 83.6$\pm$5.0   	 & 	 4.04    	 & 	2400 	 & 	 NTT \\

\hline
\hline

1eRASS J044500.9+071554 	 & 	 71.2559487 	 & 	 7.26498540 	 & 	     ---        	 & 	 19.52$\pm$0.01 	 & 	 13.60$\pm$0.02 	 & 	 5.7$\pm$2.3 	 & 	 292.8$\pm$17.6   	 & 	 ---    	 & 1800	& 	TNG  \\
1eRASS J103300.8+175659 	 & 	 158.2528123 	 & 	 17.9506654 	 & 	 20.93$\pm$0.02 	 & 	 19.77$\pm$0.01 	 & 	 15.07$\pm$0.03 	 & 	 9.5$\pm$3.6 	 & 	 3.5$\pm$0.3    	 & 	 ---    	 & 1800	& 	TNG  \\
1eRASS J130641.1$-$164126 	 & 	 196.6710923 	 & 	 $-$16.6901593 	 & 	     ---        	 & 	 21.21$\pm$0.04 	 & 	 15.61$\pm$0.04 	 & 	 7.7$\pm$2.5 	 & 	 58.2$\pm$3.5   	 & 	 ---    	 & 2100	& 	TNG  \\
1eRASS J121630.8$-$080640 	 & 	 184.1288333 	 & 	 $-$8.1108180 	 & 	 20.30$\pm$0.01 	 & 	 19.24$\pm$0.01 	 & 	 13.42$\pm$0.01 	 & 	 32.1$\pm$5.3 	 & 	 6.8$\pm$0.4    	 & 	 ---    	 & 7000	& 	TNG \\
1eRASS J150046.8$-$125146 	 & 	 225.1955377 	 & 	 $-$12.8643865 	 & 	 21.89$\pm$0.05 	 & 	 20.41$\pm$0.02 	 & 	 14.68$\pm$0.03 	 & 	 4.6$\pm$1.9 	 & 	 78.9$\pm$4.7   	 & 	 ---    	 & 7200	& 	TNG \\
1eRASS J195415.0$-$620901 	 & 	 298.5621543 	 & 	 $-$62.1522444 	 & 	 22.06$\pm$0.04 	 & 	 20.76$\pm$0.03 	 & 	 16.19$\pm$0.07 	 & 	 4.7$\pm$2.3 	 & 	 6.5$\pm$0.4    	 & 	 ---    	 & 4800 & 	NTT \\
1eRASS J213734.9$-$693425 	 & 	 324.3927552 	 & 	 $-$69.5734872 	 & 	     ---        	 & 	 20.52$\pm$0.03 	 & 	 16.15$\pm$0.06 	 & 	 6.0$\pm$-- 	 & 	 1.1$\pm$0.1    	 & 	 ---    	 & 2400	&  NTT \\

    \hline
    \hline

    \end{tabular}
    \caption{Targets observed as part of this project. We report the main optical, radio, and X-ray properties of all the sources, as well as their redshift (if confirmed), the telescope used, and the total on-source time.}
    \label{tab:obs_targets}
\end{table}
\end{landscape}

\begin{table*}
    \caption{Rest-frame properties of the new high-$z$ quasars discovered in this work. Col. (1): Name; col. (2) redshift; col. (3, 4, 5) monochromatic luminosities at 5~GHz, 4400\AA \, and 2500\AA; col. (6) integrated X-ray luminosity in the 2--10~keV energy band; col. (7, 8) radio loudness (R) and $\tilde{\alpha}_{\rm ox}$ parameters; col. (9) discovery references: (0) \protect\cite{Yang2023}\\
    }
   \centering
    
    \begin{tabular}{lcccccccc}
    Name & $z$ & log(L$_{\rm 5~GHz}$) & log$\left( {\rm L}_{\rm 4400~\textup{\footnotesize \AA}}\right)$ & log$\left( {\rm L}_{\rm 2500~\textup{\footnotesize \AA}}\right)$ & log$\left( \nu {\rm L}_{\rm 2-10~keV}\right)$ & log(R) & $\tilde{\alpha}_{\rm ox}$ & Reference\\
    && (erg~s$^{-1}$~Hz$^{-1}$) & (erg~s$^{-1}$~Hz$^{-1}$) & (erg~s$^{-1}$~Hz$^{-1}$) & (erg~s$^{-1}$) &  &  &\\
    (1) & (2) & (3) & (4) & (5) & (6) & (7) & (8) & (9)\\
    \hline
    \hline

1eRASS J000246.5-481557 	 & 	 4.25 	 & 	 34.11$\pm$0.04 	 & 	 31.49$\pm$0.06 	 & 	 31.56$\pm$0.05 	 & 	 46.20$\pm$0.22 	 & 	 2.63$\pm$0.07 	 & 	 1.31$\pm$0.07 	 & 	 This work \\ 
1eRASS J005156.1-322644 	 & 	 4.04 	 & 	 33.66$\pm$0.04 	 & 	 31.38$\pm$0.05 	 & 	 31.21$\pm$0.05 	 & 	 46.02$\pm$0.25 	 & 	 2.28$\pm$0.07 	 & 	 1.25$\pm$0.08 	 & 	 This work  \\ 
1eRASS J020511.5-642105 	 & 	 3.96 	 & 	 33.31$\pm$0.05 	 & 	 31.08$\pm$0.06 	 & 	 31.01$\pm$0.06 	 & 	 45.88$\pm$0.23 	 & 	 2.23$\pm$0.08 	 & 	 1.24$\pm$0.07 	 & 	 This work  \\ 
1eRASS J020413.2-325124 	 & 	 3.80 	 & 	 33.27$\pm$0.05 	 & 	 31.02$\pm$0.06 	 & 	 32.26$\pm$0.05 	 & 	 46.47$\pm$0.17 	 & 	 2.26$\pm$0.08 	 & 	 1.44$\pm$0.05 	 & 	 This work  \\ 
1eRASS J020520.9-220935 	 & 	 3.57 	 & 	 33.42$\pm$0.06 	 & 	 31.39$\pm$0.05 	 & 	 31.26$\pm$0.05 	 & 	 46.07$\pm$0.20 	 & 	 2.03$\pm$0.08 	 & 	 1.26$\pm$0.06 	 & 	 This work  \\ 
1eRASS J023732.6-434820 	 & 	 3.65 	 & 	 35.08$\pm$0.04 	 & 	 31.68$\pm$0.05 	 & 	 31.58$\pm$0.05 	 & 	 46.01$\pm$0.20 	 & 	 3.40$\pm$0.06 	 & 	 1.37$\pm$0.06 	 & 	 This work  \\ 
1eRASS J025903.8-694531 	 & 	 3.44 	 & 	 33.61$\pm$0.04 	 & 	 31.49$\pm$0.05 	 & 	 31.29$\pm$0.05 	 & 	 45.69$\pm$0.21 	 & 	 2.12$\pm$0.06 	 & 	 1.38$\pm$0.06 	 & 	 This work  \\ 
1eRASS J030025.4-044644 	 & 	 3.79 	 & 	 33.18$\pm$0.10 	 & 	 31.76$\pm$0.05 	 & 	 31.56$\pm$0.05 	 & 	 45.89$\pm$0.23 	 & 	 1.42$\pm$0.11 	 & 	 1.40$\pm$0.08 	 & 	 This work  \\ 
1eRASS J031010.4-350515 	 & 	 4.68 	 & 	 33.72$\pm$0.04 	 & 	 30.99$\pm$0.08 	 & 	 30.99$\pm$0.07 	 & 	 46.28$\pm$0.20 	 & 	 2.73$\pm$0.09 	 & 	 1.11$\pm$0.06 	 & 	 This work  \\ 
1eRASS J031226.4-131306 	 & 	 3.69 	 & 	 34.59$\pm$0.03 	 & 	 30.97$\pm$0.07 	 & 	 30.73$\pm$0.07 	 & 	 45.79$\pm$0.24 	 & 	 3.62$\pm$0.07 	 & 	 1.18$\pm$0.07 	 & 	 This work  \\ 
1eRASS J031522.9-113314 	 & 	 3.79 	 & 	 33.20$\pm$0.08 	 & 	 31.72$\pm$0.05 	 & 	 31.52$\pm$0.05 	 & 	 45.93$\pm$0.22 	 & 	 1.48$\pm$0.09 	 & 	 1.37$\pm$0.07 	 & 	 This work  \\ 
1eRASS J034609.3-335933 	 & 	 4.19 	 & 	 33.88$\pm$0.03 	 & 	 31.14$\pm$0.06 	 & 	 31.13$\pm$0.06 	 & 	 45.85$\pm$0.23 	 & 	 2.74$\pm$0.07 	 & 	 1.28$\pm$0.07 	 & 	 This work  \\ 
1eRASS J034926.0-345434 	 & 	 3.80 	 & 	 33.91$\pm$0.03 	 & 	 30.98$\pm$0.06 	 & 	 30.78$\pm$0.06 	 & 	 45.87$\pm$0.20 	 & 	 2.93$\pm$0.07 	 & 	 1.17$\pm$0.06 	 & 	 This work  \\ 
1eRASS J035611.0-460756 	 & 	 3.74 	 & 	 34.13$\pm$0.04 	 & 	 30.93$\pm$0.06 	 & 	 30.83$\pm$0.06 	 & 	 45.67$\pm$0.21 	 & 	 3.20$\pm$0.07 	 & 	 1.24$\pm$0.07 	 & 	 This work  \\ 
1eRASS J040453.6-333203 	 & 	 4.62 	 & 	 34.01$\pm$0.04 	 & 	 31.29$\pm$0.06 	 & 	 31.21$\pm$0.06 	 & 	 46.11$\pm$0.23 	 & 	 2.72$\pm$0.07 	 & 	 1.23$\pm$0.07 	 & 	 This work  \\ 
1eRASS J041105.2-021450 	 & 	 4.13 	 & 	 34.32$\pm$0.03 	 & 	 31.24$\pm$0.06 	 & 	 31.12$\pm$0.06 	 & 	 46.41$\pm$0.20 	 & 	 3.08$\pm$0.07 	 & 	 1.11$\pm$0.06 	 & 	 This work  \\ 
1eRASS J041138.1-570106 	 & 	 4.30 	 & 	 34.38$\pm$0.15 	 & 	 31.15$\pm$0.07 	 & 	 30.92$\pm$0.07 	 & 	 45.65$\pm$0.24 	 & 	 3.22$\pm$0.17 	 & 	 1.28$\pm$0.09 	 & 	 This work  \\ 
1eRASS J041146.9-540515 	 & 	 3.74 	 & 	 34.45$\pm$0.04 	 & 	 31.19$\pm$0.05 	 & 	 31.00$\pm$0.06 	 & 	 45.91$\pm$0.17 	 & 	 3.26$\pm$0.07 	 & 	 1.22$\pm$0.05 	 & 	 This work  \\ 
1eRASS J042048.9-570304 	 & 	 3.80 	 & 	 34.29$\pm$0.04 	 & 	 30.88$\pm$0.06 	 & 	 30.81$\pm$0.06 	 & 	 46.26$\pm$0.15 	 & 	 3.41$\pm$0.07 	 & 	 1.06$\pm$0.05 	 & 	 This work  \\ 
1eRASS J044117.7-335111 	 & 	 3.68 	 & 	 34.26$\pm$0.03 	 & 	 30.86$\pm$0.06 	 & 	 30.76$\pm$0.06 	 & 	 45.80$\pm$0.24 	 & 	 3.40$\pm$0.07 	 & 	 1.18$\pm$0.07 	 & 	 This work  \\ 
1eRASS J045513.9-371157 	 & 	 3.94 	 & 	 33.26$\pm$0.11 	 & 	 32.21$\pm$0.05 	 & 	 32.09$\pm$0.05 	 & 	 46.13$\pm$0.18 	 & 	 1.06$\pm$0.12 	 & 	 1.49$\pm$0.06 	 & 	 This work  \\ 
1eRASS J051559.6-261006 	 & 	 4.24 	 & 	 34.52$\pm$0.06 	 & 	 31.62$\pm$0.07 	 & 	 31.43$\pm$0.06 	 & 	 46.07$\pm$0.22 	 & 	 2.90$\pm$0.09 	 & 	 1.31$\pm$0.07 	 & 	 This work  \\ 
1eRASS J052733.7-523901 	 & 	 4.25 	 & 	 33.04$\pm$0.15 	 & 	 30.77$\pm$0.07 	 & 	 30.74$\pm$0.07 	 & 	 45.51$\pm$0.25 	 & 	 2.27$\pm$0.17 	 & 	 1.27$\pm$0.09   & 	 This work  \\ 
1eRASS J054609.1-523426 	 & 	 3.94 	 & 	 33.74$\pm$0.17 	 & 	 31.49$\pm$0.05 	 & 	 31.31$\pm$0.05 	 & 	 45.60$\pm$0.22 	 & 	 2.24$\pm$0.17 	 & 	 1.41$\pm$0.08 	 & 	 This work  \\ 
1eRASS J060057.9-521526 	 & 	 3.59 	 & 	 32.88$\pm$0.34 	 & 	 31.14$\pm$0.05 	 & 	 31.14$\pm$0.05 	 & 	 45.51$\pm$0.24 	 & 	 1.73$\pm$0.34 	 & 	 1.39$\pm$0.13 	 & 	 This work  \\ 
1eRASS J083910.9-020232 	 & 	 3.75 	 & 	 34.86$\pm$0.03 	 & 	 31.56$\pm$0.05 	 & 	 31.52$\pm$0.05 	 & 	 46.87$\pm$0.16 	 & 	 3.30$\pm$0.06 	 & 	 1.09$\pm$0.05 	 & 	 This work  \\ 
1eRASS J100448.0-153202 	 & 	 3.80 	 & 	 32.84$\pm$0.16 	 & 	 31.82$\pm$0.05 	 & 	 31.70$\pm$0.05 	 & 	 46.51$\pm$0.18 	 & 	 1.03$\pm$0.17 	 & 	 1.25$\pm$0.07 	 & 	 This work  \\ 
1eRASS J104018.5-063636 	 & 	 4.01 	 & 	 33.89$\pm$0.04 	 & 	 31.91$\pm$0.05 	 & 	 31.68$\pm$0.05 	 & 	 46.19$\pm$0.25 	 & 	 1.98$\pm$0.06 	 & 	 1.34$\pm$0.08 	 & 	 This work  \\ 
1eRASS J115839.3-052221 	 & 	 4.96 	 & 	 34.52$\pm$0.04 	 & 	 31.32$\pm$0.07 	 & 	 31.23$\pm$0.07 	 & 	 46.47$\pm$0.21 	 & 	 3.19$\pm$0.08 	 & 	 1.12$\pm$0.06 	 & 	 This work, (0)  \\ 
1eRASS J112415.7-331735 	 & 	 4.22 	 & 	 34.35$\pm$0.04 	 & 	 31.54$\pm$0.05 	 & 	 31.46$\pm$0.05 	 & 	 46.17$\pm$0.22 	 & 	 2.81$\pm$0.07 	 & 	 1.29$\pm$0.07 	 & 	 This work  \\ 
1eRASS J120222.0-320757 	 & 	 3.95 	 & 	 34.14$\pm$0.06 	 & 	 31.68$\pm$0.05 	 & 	 31.49$\pm$0.05 	 & 	 45.82$\pm$0.27 	 & 	 2.47$\pm$0.08 	 & 	 1.40$\pm$0.08 	 & 	 This work  \\ 
1eRASS J134030.6-282556 	 & 	 3.85 	 & 	 32.98$\pm$0.17 	 & 	 31.32$\pm$0.05 	 & 	 31.18$\pm$0.05 	 & 	 46.02$\pm$0.22 	 & 	 1.66$\pm$0.18 	 & 	 1.24$\pm$0.08 	 & 	 This work  \\ 
1eRASS J143844.0-111851 	 & 	 3.61 	 & 	 33.84$\pm$0.06 	 & 	 31.42$\pm$0.05 	 & 	 31.26$\pm$0.05 	 & 	 45.92$\pm$0.23 	 & 	 2.43$\pm$0.08 	 & 	 1.30$\pm$0.07 	 & 	 This work  \\ 
1eRASS J145635.5-314933 	 & 	 3.86 	 & 	 34.11$\pm$0.06 	 & 	 31.23$\pm$0.06 	 & 	 31.13$\pm$0.06 	 & 	 46.13$\pm$0.22 	 & 	 2.88$\pm$0.08 	 & 	 1.20$\pm$0.07 	 & 	 This work  \\ 
1eRASS J195040.7-523939 	 & 	 4.43 	 & 	 34.06$\pm$0.17 	 & 	 31.41$\pm$0.06 	 & 	 31.35$\pm$0.06 	 & 	 46.16$\pm$0.27 	 & 	 2.65$\pm$0.18 	 & 	 1.25$\pm$0.10 	 & 	 This work  \\ 
1eRASS J204223.4-721207 	 & 	 3.90 	 & 	 33.34$\pm$0.26 	 & 	 31.49$\pm$0.05 	 & 	 31.47$\pm$0.05 	 & 	 46.05$\pm$0.23 	 & 	 1.85$\pm$0.26 	 & 	 1.32$\pm$0.11 	 & 	 This work  \\ 
1eRASS J205253.1-801247 	 & 	 3.64 	 & 	 34.63$\pm$0.04 	 & 	 31.81$\pm$0.05 	 & 	 31.35$\pm$0.05 	 & 	 46.09$\pm$0.20 	 & 	 2.81$\pm$0.06 	 & 	 1.27$\pm$0.06 	 & 	 This work  \\ 
1eRASS J215950.0-494435 	 & 	 3.66 	 & 	 33.64$\pm$0.17 	 & 	 32.07$\pm$0.05 	 & 	 31.90$\pm$0.05 	 & 	 45.87$\pm$0.26 	 & 	 1.57$\pm$0.17 	 & 	 1.51$\pm$0.09 	 & 	 This work  \\ 
1eRASS J235242.2-483954 	 & 	 3.94 	 & 	 32.34$\pm$0.37 	 & 	 31.54$\pm$0.05 	 & 	 31.33$\pm$0.05 	 & 	 45.94$\pm$0.25 	 & 	 0.80$\pm$0.37 	 & 	 1.32$\pm$0.14 	 & 	 This work  \\ 
1eRASS J235828.7-401434 	 & 	 4.40 	 & 	 34.52$\pm$0.04 	 & 	 31.40$\pm$0.06 	 & 	 31.37$\pm$0.06 	 & 	 46.09$\pm$0.28 	 & 	 3.12$\pm$0.07 	 & 	 1.28$\pm$0.09 	 & 	 This work  \\ 

    \hline
    \hline

    \end{tabular}
    \label{tab:lum_sample}
\end{table*}

\begin{table*}
    \caption{Continuation of Table \protect\ref{tab:lum_sample} including sources known from the literature. References for the discovery: (1) \protect\cite{Wolf2020}, (2) \protect\cite{Ighina2025}, (3) \protect\cite{Onken2022}, (4) \protect\cite{Peroux2001}, (5) \protect\cite{Schindler2019}, (6) \protect\cite{Storrie2001}, (7) \protect\cite{Paris2014}, (8) \protect\cite{Caccianiga2019}, (9) \protect\cite{Lyke2020}, (10) \protect\cite{Schneider2010}, (11) \protect\cite{Sbarrato2012}, (12) \protect\cite{Zickgraf1997}, (13) \protect\cite{Mahony2011}, (14) \protect\cite{Schneider2007}, (15) \protect\cite{Shaver1996}, (16) \protect\cite{Sbarrato2013}, (17) \protect\cite{Hook2002}, (18) \protect\cite{White1991}.}

   \centering
    
    \begin{tabular}{lcccccccc}
    Name & $z$ & log(L$_{\rm 5~GHz}$) & log$\left( {\rm L}_{\rm 4400~\textup{\footnotesize \AA}}\right)$ & log$\left( {\rm L}_{\rm 2500~\textup{\footnotesize \AA}}\right)$ & log($\nu$L$_{\rm 2-10~keV}$) & log(R) & $\tilde{\alpha}_{\rm ox}$ & Reference\\
    & & (erg~s$^{-1}$~Hz$^{-1}$) & (erg~s$^{-1}$~Hz$^{-1}$) & (erg~s$^{-1}$~Hz$^{-1}$) & (erg~s$^{-1}$) &  &  &\\
    (1) & (2) & (3) & (4) & (5) & (6) & (7) & (8) & (9)\\

    \hline
    \hline

1eRASS J013539.6-212630 	 & 	 4.94 	 & 	 34.19$\pm$0.03 	 & 	 32.38$\pm$0.06 	 & 	 32.22$\pm$0.06 	 & 	 46.31$\pm$0.23 	 & 	 1.81$\pm$0.07 	 & 	 1.47$\pm$0.07  & (1) \\ 
1eRASS J012713.9-445454 	 & 	 4.92 	 & 	 33.90$\pm$0.04 	 & 	 30.99$\pm$0.08 	 & 	 31.00$\pm$0.08 	 & 	 46.21$\pm$0.24 	 & 	 2.91$\pm$0.09 	 & 	 1.13$\pm$0.07  & (2) \\ 
1eRASS J021217.8-394544 	 & 	 4.13 	 & 	 34.06$\pm$0.03 	 & 	 32.30$\pm$0.05 	 & 	 32.17$\pm$0.05 	 & 	 46.05$\pm$0.23 	 & 	 1.76$\pm$0.06 	 & 	 1.53$\pm$0.07  & (3) \\ 
1eRASS J032233.5-594328 	 & 	 4.42 	 & 	 33.70$\pm$0.19 	 & 	 31.92$\pm$0.05 	 & 	 31.71$\pm$0.05 	 & 	 45.82$\pm$0.22 	 & 	 1.78$\pm$0.19 	 & 	 1.46$\pm$0.09  & (3) \\ 
1eRASS J033951.4-474000 	 & 	 4.45 	 & 	 34.10$\pm$0.16 	 & 	 31.74$\pm$0.05 	 & 	 31.66$\pm$0.05 	 & 	 45.94$\pm$0.21 	 & 	 2.36$\pm$0.17 	 & 	 1.41$\pm$0.08  & (3) \\ 
1eRASS J035505.0-381141 	 & 	 4.54 	 & 	 32.98$\pm$0.10 	 & 	 32.39$\pm$0.05 	 & 	 32.25$\pm$0.05 	 & 	 46.11$\pm$0.21 	 & 	 0.59$\pm$0.11 	 & 	 1.54$\pm$0.07  & (4) \\ 
1eRASS J043623.9-000356 	 & 	 3.85 	 & 	 34.42$\pm$0.03 	 & 	 32.22$\pm$0.05 	 & 	 32.11$\pm$0.05 	 & 	 45.96$\pm$0.24 	 & 	 2.21$\pm$0.06 	 & 	 1.55$\pm$0.07  & (5) \\ 
1eRASS J052506.2-334305 	 & 	 4.41 	 & 	 34.69$\pm$0.03 	 & 	 32.01$\pm$0.05 	 & 	 31.88$\pm$0.05 	 & 	 47.04$\pm$0.15 	 & 	 2.68$\pm$0.06 	 & 	 1.15$\pm$0.05  & (6) \\ 
1eRASS J073631.1+284035 	 & 	 3.62 	 & 	 34.24$\pm$0.04 	 & 	 31.11$\pm$0.06 	 & 	 31.04$\pm$0.06 	 & 	 46.16$\pm$0.23 	 & 	 3.13$\pm$0.07 	 & 	 1.16$\pm$0.07  & (7) \\ 
1eRASS J083548.8+182517 	 & 	 4.41 	 & 	 34.29$\pm$0.04 	 & 	 31.09$\pm$0.07 	 & 	 31.01$\pm$0.07 	 & 	 46.51$\pm$0.22 	 & 	 3.19$\pm$0.08 	 & 	 1.04$\pm$0.07  & (8) \\ 
1eRASS J085111.2+142347 	 & 	 4.17 	 & 	 33.78$\pm$0.05 	 & 	 31.43$\pm$0.06 	 & 	 31.30$\pm$0.06 	 & 	 46.21$\pm$0.26 	 & 	 2.34$\pm$0.07 	 & 	 1.23$\pm$0.08  & (9) \\ 
1eRASS J085943.8+212511 	 & 	 3.69 	 & 	 33.85$\pm$0.06 	 & 	 31.93$\pm$0.05 	 & 	 31.74$\pm$0.05 	 & 	 46.12$\pm$0.24 	 & 	 1.92$\pm$0.08 	 & 	 1.39$\pm$0.08  & (10) \\ 
1eRASS J091824.4+063655 	 & 	 4.15 	 & 	 34.15$\pm$0.06 	 & 	 31.74$\pm$0.05 	 & 	 31.56$\pm$0.05 	 & 	 46.23$\pm$0.25 	 & 	 2.41$\pm$0.08 	 & 	 1.30$\pm$0.08  & (11) \\ 
1eRASS J095736.7-000426 	 & 	 3.86 	 & 	 33.68$\pm$0.05 	 & 	 31.24$\pm$0.06 	 & 	 31.13$\pm$0.05 	 & 	 46.53$\pm$0.18 	 & 	 2.44$\pm$0.07 	 & 	 1.08$\pm$0.06  & (7) \\ 
1eRASS J102107.9+220923 	 & 	 4.26 	 & 	 34.80$\pm$0.04 	 & 	 30.94$\pm$0.08 	 & 	 30.73$\pm$0.08 	 & 	 46.17$\pm$0.26 	 & 	 3.86$\pm$0.09 	 & 	 1.06$\pm$0.08  & (5) \\ 
1eRASS J101540.9-032741 	 & 	 3.84 	 & 	 34.36$\pm$0.03 	 & 	 31.99$\pm$0.05 	 & 	 31.89$\pm$0.05 	 & 	 46.26$\pm$0.22 	 & 	 2.37$\pm$0.06 	 & 	 1.39$\pm$0.07  & (9) \\ 
1eRASS J102838.7-084434 	 & 	 4.27 	 & 	 34.75$\pm$0.03 	 & 	 31.65$\pm$0.05 	 & 	 31.66$\pm$0.05 	 & 	 46.87$\pm$0.17 	 & 	 3.10$\pm$0.06 	 & 	 1.14$\pm$0.05  & (12) \\ 
1eRASS J104742.6+094737 	 & 	 4.23 	 & 	 33.81$\pm$0.05 	 & 	 31.35$\pm$0.06 	 & 	 31.14$\pm$0.06 	 & 	 46.22$\pm$0.25 	 & 	 2.46$\pm$0.08 	 & 	 1.17$\pm$0.08  & (8) \\ 
1eRASS J111403.0-050233 	 & 	 3.82 	 & 	 34.01$\pm$0.06 	 & 	 32.20$\pm$0.05 	 & 	 32.09$\pm$0.05 	 & 	 46.36$\pm$0.21 	 & 	 1.81$\pm$0.08 	 & 	 1.42$\pm$0.06  & (5) \\ 
1eRASS J115503.5-310759 	 & 	 4.30 	 & 	 34.44$\pm$0.06 	 & 	 31.57$\pm$0.05 	 & 	 31.42$\pm$0.05 	 & 	 46.24$\pm$0.21 	 & 	 2.87$\pm$0.08 	 & 	 1.25$\pm$0.07  & (13)\\ 
1eRASS J121547.4+155637 	 & 	 3.63 	 & 	 33.70$\pm$0.06 	 & 	 31.02$\pm$0.06 	 & 	 30.89$\pm$0.06 	 & 	 45.92$\pm$0.26 	 & 	 2.68$\pm$0.09 	 & 	 1.19$\pm$0.08  & (14) \\ 
1eRASS J125359.0-405931 	 & 	 4.46 	 & 	 34.82$\pm$0.05 	 & 	 31.38$\pm$0.06 	 & 	 31.26$\pm$0.06 	 & 	 46.33$\pm$0.20 	 & 	 3.44$\pm$0.08 	 & 	 1.18$\pm$0.06  & (15) \\ 
1eRASS J132512.1+112333 	 & 	 4.41 	 & 	 34.50$\pm$0.06 	 & 	 31.68$\pm$0.06 	 & 	 31.45$\pm$0.06 	 & 	 46.19$\pm$0.23 	 & 	 2.83$\pm$0.08 	 & 	 1.28$\pm$0.07  & (16) \\ 
1eRASS J141209.5+062411 	 & 	 4.46 	 & 	 34.29$\pm$0.04 	 & 	 31.23$\pm$0.06 	 & 	 31.25$\pm$0.06 	 & 	 46.11$\pm$0.24 	 & 	 3.07$\pm$0.08 	 & 	 1.24$\pm$0.07  & (16) \\ 
1eRASS J145146.7-151218 	 & 	 4.76 	 & 	 33.92$\pm$0.07 	 & 	 32.54$\pm$0.06 	 & 	 32.41$\pm$0.05 	 & 	 46.47$\pm$0.21 	 & 	 1.38$\pm$0.09 	 & 	 1.48$\pm$0.07  & (17) \\ 
1eRASS J215202.3-780712 	 & 	 3.99 	 & 	 35.21$\pm$0.04 	 & 	 31.33$\pm$0.05 	 & 	 31.24$\pm$0.05 	 & 	 46.00$\pm$0.24 	 & 	 3.87$\pm$0.07 	 & 	 1.27$\pm$0.07  & (18) \\ 

    \hline
    \hline

    \end{tabular}
    \label{tab:lum_sample2}
\end{table*}




\bsp	
\label{lastpage}
\end{document}